世界互联网大会
World Internet Conference

# Frontier AI Safety and Governance

# 2026

## Risk Understanding and Coordinated Governance

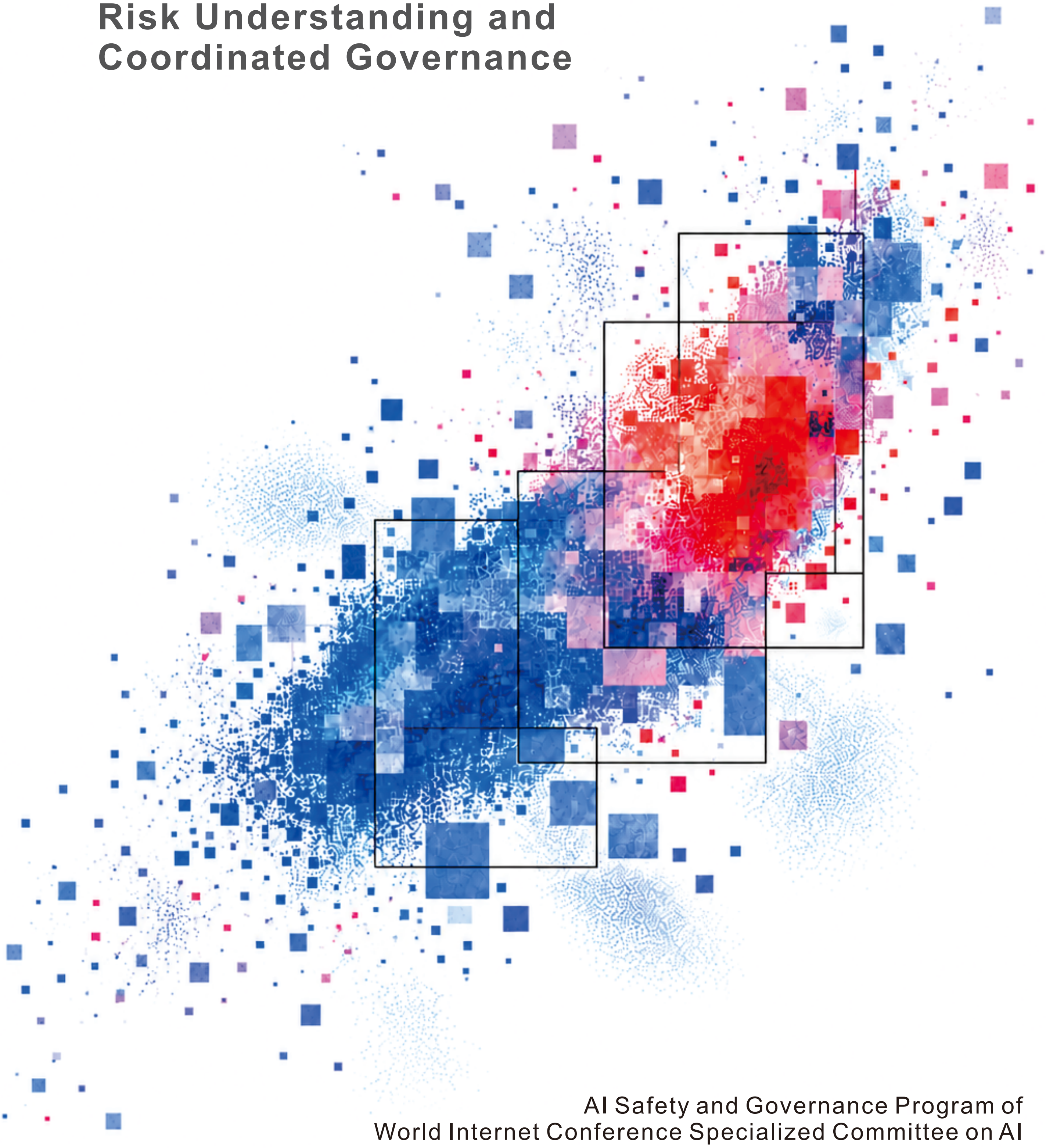

AI Safety and Governance Program of
World Internet Conference Specialized Committee on AI

# Contributors and Members of AI Safety and Governance Program of World Internet Conference Specialized Committee on AI

## Co-leads

| | |
|---|---|
| Zeng Yi | Gaoling School of Artificial Intelligence, Renmin University of China<br>Beijing Institute of AI Safety and Governance (Beijing-AISI) |
| Seán Ó hÉigeartaigh | University of Cambridge |

## Experts and Contributors

(Including WIC members and invited experts of the Program, as well as contributors who provided content or comments for the report; listed alphabetically by organization name)

**360 Group**
Pan Jianfeng, Ma Lin, Wang Qianyu, Xiao Jianwu

**AI Safety Connect**
Nicolas Miailhe, Cyrus Hodes

**Alibaba Cloud**
Pang Xiaogang, Wang Shengjie, Wang Xingguang, Zhang Rong

**Ant Group Co., Ltd.**
Xu Jiming

**Arab ICT Organization (AICTO)**
Nada Laabidi

**Artificial Intelligence Working Committee, Internet Society of China**
Xia Wenhui

**Baidu**
Bao Chenfu

**Beijing University of Posts and Telecommunications**
Yang Zhongliang

**Beijing Yinxiang Biji Technology Co., Ltd.**
Qiao Qian

**British Standards Institution (BSI)**
Chris Brown

**CCTV International Network Co., Ltd.**
Cheng Ming

**Chartered Institute of Arbitrators**
Ammar Younas

**China Academy of Information and Communications Technology (CAICT)**
Wei Kai, Hu Naying, Guo Sumin, Yang Lingling, Niu Xiaoling, Liang Ye

**China Internet Network Information Center (CNNIC)**
Chen Jingjing, Yuchi Xuebiao, Yue Qiaoli

**China Mobile Research Institute**
Lin Lin, Wu Shuyan

**China Telecom Cybersecurity Technology Co., Ltd.**
Yang Bo

**China Telecommunications Corporation**
Yang Mingchuan, Liu Weichen, Wang Feng, Zhang Ziting

**China United Network Communications Group Co., Ltd.**
Zhou Kai

**China University of Political Science and Law**
Zhang Linghan

**ChinaEU**
Luigi Gambardella

**Chinese Institute of New Generation Artificial Intelligence Development Strategies**
Gong Ke

**Cloud Security Alliance (CSA)**
John Yeoh

**Cyber Security Association of China (CSAC)**
Wang Jianbing

**DAS-Security**
Wang Xin, Hu Zhuang, Yang Shaobo, Fu Chunhui, Lin Mingfeng

**Federal University of Rio Grande do Sul**
Edson Prestes

**Fuxi Institution**
Cheng Kai, Wu Yinyu

**GPAI Policy Lab (France)**
Tom David

**Haihe Laboratory of Information Technology Application Innovation**
Gao Xuesong

**Huawei Technologies Co., Ltd.**
Liu Jiaqi, Wang Xiaozhen,
Xu Mengwen, Lin Qiushi

**International AI Governance Association**
John Higgins, Rob Wortham

**IQuilibriumAI**
Jimena Sofia Viveros Alvarez

**Lenovo Group**
Hu Yongqi

**Lingnan University**
Yao Xin

**MiniMax**
Peng Tao, Shen Juncheng

**Nankai University**
Tao Feng

**Nokia Solutions (Shanghai) Co., Ltd.**
Chang Jiang

**Peking University**
Sebastian Sunday Grève, Yang Yaodong

**People.cn Co., Ltd.**
Wang Fengqiong, Zhang Xiaohe

**Qi An Xin Technology Group Co., Ltd.**
Zhang Yong, Ma Lan,
Lin Xueyan, Jiang Chen

**Renmin University of China**
Gong Xinqi

**Research and Promotion Center for Artificial Intelligence,Chinese Academy of Social Sciences**
Duan Weiwen

**SaferAI**
Henry Papadatos, Malcolm Murray

**Shanghai Innovation Institute**
Qiao Yu

**Smart Africa**
Didier Nkurikiyimfura

**Sri Lanka CERT**
Charuka Senal Damunupola,
Nirosha Ananda

**Technical University of Munich**
Danil Kerimi

**Tencent**
Wang Mengyin, Wang Rong,
Xu Deliang

**The Chinese University of Hong Kong**
Helen Meng

**TMTPOST Group**
Chen Huijie

**Tsinghua University**
Xue Lan, Zhang Ya-Qin

**UNESCO Regional Office for East Asia**
Xia Hehua

**University of Münster**
Bernd Holznagel

**University of Reading**
Tang Yinshan

**University of Southampton**
Dame Wendy Hall

**Weibo**
Wang Wei

**World Internet Conference**
Ren Xianliang, Liang Hao, Zhang Xueli

**Zhongguancun Laboratory**
Tan Zhixing

**ZTE Corporation**
Meng Wei

## Writing Group

**Gaoling School of Artificial Intelligence, Renmin University of China**
**Beijing Institute of AI Safety and Governance (Beijing-AISI)**
Zeng Yi, Lu Enmeng, Wang Zhengqi, Li Kai, Wang Peining

**World Internet Conference**
Han Kaiyu, Chang Yao

**University of Chinese Academy of Sciences**
Cao Gongce, Chen Yu, Xie Jiawei, Peng Yuxuan

## Contact

**research@wicinternet.org**

# WIC Report Series on Frontier AI Safety and Governance

Artificial intelligence is becoming a major force driving scientific and technological innovation, industrial transformation, and social development, profoundly influencing knowledge creation, the organization of production, public services, and social governance. As technical capabilities expand and applications continue to broaden, the safety and governance issues raised by artificial intelligence are increasingly becoming a shared concern of the international community. The associated risks are marked by intertwined technical, societal, and cross-border dimensions, posing a more urgent test for risk understanding, governance capacity, and international coordination. All parties need to strengthen forward-looking preparedness and act together to address increasingly complex governance challenges effectively.

In this context, the AI Safety and Governance Program of World Internet Conference (WIC) Specialized Committee on AI aims to develop an annual report series on frontier AI safety and governance as a public good for the international community. Grounded in the latest advances in frontier AI capabilities and applications, this series of reports aims to systematically examine technological developments and the new safety and governance challenges they raise, assess the evolving risk landscape, review corresponding governance responses, and identify priorities for international coordination, seeking to strengthen global consensus and enhance the collective capacity to address these challenges.

The series takes a holistic, forward-looking, and open perspective, bringing together the views of government agencies, international organizations, companies, research institutions, and other stakeholders to provide fact-based, evidence-grounded analysis for reference. Through sustained annual observation and comparative research, the series seeks to strengthen the international community's forward-looking preparedness and governance capacity for frontier AI risks, helping all parties improve their governance practices and enabling frontier AI to better serve human development on a safe, reliable, and responsible foundation.

# About This Report

Frontier AI is currently at a critical stage marked by rapid advances in capabilities, the accelerating diffusion of applications, and increasingly prominent governance challenges. Large language models, multimodal models, autonomous AI agents, embodied intelligence, and other technologies continue to evolve, moving AI beyond information generation and knowledge services toward task execution, tool use, and collaboration within complex systems, and extending it into ever broader settings such as scientific research and innovation, industrial development, public services, and the functioning of society. At the same time, the associated risks are rapidly evolving from isolated, localized problems into compound challenges that cut across scenarios, actors, and borders, placing higher demands on the adaptability, risk response capacity, and international coordination capacity of existing governance systems.

This report focuses on "risk understanding and coordinated governance of frontier AI". It is intended to serve as a reference for government agencies, international organizations, companies, research institutions, and other stakeholders, helping all parties build consensus, enhance mutual trust, and strengthen collaboration, so as to jointly advance the safe, reliable, and responsible development of frontier AI in service of the common well-being of humanity.

## Report Structure

The report is organized along the pathway of landscape analysis, risk understanding, governance assessment, and coordinated response. It systematically reviews the development of frontier AI capabilities, the expansion of applications, and the evolution of risks; analyzes key risks, their severity, and response priorities in a structured manner; assesses existing governance practices and the state of governance readiness; and proposes directions and feasible mechanisms for coordinated response.

The report addresses four questions: first, how frontier AI capabilities and applications are evolving and how they are giving rise to new risks; second, what the main aspects of frontier AI risk are, how severe they are, and which require priority response; third, what governance practices stakeholders have undertaken, the overall level of governance readiness, and which areas still need improvement; and fourth, how stronger coordination among different actors can improve the effectiveness of frontier AI governance.

## Scope

In this report, frontier AI refers to AI models at the frontier of capability in general-purpose or specialized domains, together with agents, multi-agent collaboration systems and embodied AI systems built on such models. Their core is typically a foundation model trained using large amounts of data and computing resources, with high-level capabilities across a broad range of tasks. Their capabilities, generality and depth of deployment may give rise to major risks that are difficult to fully anticipate or constrain in advance. This report defines frontier AI by its capabilities, without fixed thresholds for parameters or training compute. The boundary of the frontier evolves with technological progress and changes in deployment conditions, including tools, permissions and operating environments.

The report recognizes that military applications of frontier AI bring a range of safety and security risks and governance challenges, and are an important part of AI safety governance. In keeping with its research focus, however, this report does not examine frontier AI in military settings. It focuses on large-scale applications in non-military settings, systematically assessing safety and security risks and their effects on individuals, organizations and society.

## Data Sources and Methods

The risk analysis draws on academic and industry safety evaluations of models and agents, statistics on publicly recorded AI incidents, and assessments by experts and relevant institutions.

The report also assesses frontier AI governance readiness. Using publicly available information, it examines the preparedness of 24 countries across six dimensions and 22 secondary indicators, identifying existing governance foundations and areas that need strengthening. The sample serves the analytical purposes of this report alone. It should not be interpreted as a judgment or ranking of any country's AI governance capacity, nor as an exclusionary conclusion about countries outside the sample. The data cutoff for the governance readiness assessment is August 31, 2026. The appendix explains the countries assessed, the indicator framework and the assessment methods.

## Explanatory Note

Frontier AI continues to evolve rapidly, while evidence and records of incidents continue to accumulate. The analysis and conclusions in this report draw on public information available at a particular point in time and reflect the understanding at that time. They are scholarly interpretations and assessments developed by the writing group from public information and input from experts and contributors. They do not represent the position of any individual, institution or country.

# Key Terms

| Term | Definition |
| --- | --- |
| **Frontier AI** | AI models at the frontier of capability in general-purpose or specialized domains, together with agents, multi-agent collaboration systems and embodied AI systems built on such models. |
| **AI agent** | An AI system that can plan and execute multi-step tasks, use tools and interact with its environment, with relatively little human intervention during execution. |
| **Multi-agent system** | A system comprising multiple interacting agents that may divide tasks and cooperate, adapt to one another or compete. |
| **Embodied intelligence** | AI systems that interact with the physical environment through sensors, robots, vehicles, industrial equipment or other devices. |
| **Open-weight model** | A model whose parameters (weights) are publicly available for download. In this report, open models refer to open-weight models. |
| **Safety evaluation and red teaming** | Safety evaluation assesses a model's or system's capabilities, vulnerabilities and potential impacts. Red teaming involves dedicated individuals or teams actively seeking vulnerabilities, limitations and opportunities for misuse. |
| **Catastrophic and Existential Risks** | Risks of major consequences that are widespread and difficult to reverse, potentially threatening humanity's long-term survival and development. They principally involve loss of control over highly autonomous systems, deception, self-replication and the proliferation of dangerous capabilities. |
| **AI incident** | An event in which the development, use or failure of an AI system directly or indirectly causes consequences such as physical injury, disruption to critical infrastructure, infringement of rights or property loss. |
| **Coordinated governance** | Cooperation among countries and other actors on evaluation, verification, incident response and capacity building, with clear divisions of responsibility and shared risk evidence, accountability records and response information, to address risks spreading across platforms, sectors and borders. |
| **Governance readiness** | The degree to which a country is prepared to address frontier AI risks, assessed through publicly verifiable arrangements in law and policy, institutions, standards, evaluation, compliance, incident reporting, capacity building and international cooperation. |

# Executive Summary

**Frontier AI is at a critical stage of rapidly advancing capabilities, accelerating adoption and increasingly evident risks.** Models, agents, tools and operating environments are combining to create more capable systems, while AI is entering a wider range of settings in scientific research, industry, public services and social life. At the same time, related risks are moving from potential concerns to real-world effects, placing greater demands on the adaptability of governance systems, risk response capacity and international coordination.

**Frontier AI risks are evolving from isolated, localized problems into systemic challenges spanning contexts, actors and boundaries.** Capability gains are moving beyond breakthroughs in individual tasks toward improvements in systems that combine models, tools and environments. Applications are extending from information generation into decision support and task execution. More capable AI lowers barriers to high-risk activities, while its integration into applications brings risks into operational workflows. Together, these developments change how risks arise and spread: risks emerge not only from individual model outputs but also across tool use, organizational processes and social networks, where they can accumulate, interact and amplify one another.

**The most pressing real-world concerns currently relate to Misuse and Criminal Risks, Information Integrity Risks, and Critical System Risks.** In terms of their impacts, frontier AI risks manifest across five broad areas: Misuse and Criminal Risks; Information Integrity Risks; Critical System Risks; Social Cognition and Structural Risks; and Catastrophic and Existential Risks. Risks in the first three areas have already translated into tangible real-world harms to a considerable extent. The longer-term effects of Social Cognition and Structural Risks are gradually accumulating in areas such as employment and education. Catastrophic and Existential Risks, while potentially far-reaching in their consequences, remain insufficiently understood in terms of the conditions that could trigger them and therefore warrant forward-looking assessment.

**Responses need to be differentiated according to urgency and potential consequences, with safeguards established before deployment.** For harms that occur frequently and spread rapidly, the priority is prompt response and containment. Where effects accumulate over time and the intervention window is longer, sustained monitoring and early adjustment are needed. Where consequences may be widespread and difficult to reverse, dangerous capability evaluations and red teaming should take place before deployment, with capability thresholds set in advance and options retained for human takeover, reduced functionality, rollback and shutdown. Other situations should be managed through routine oversight and periodic review. As capabilities evolve and evidence accumulates, risk situations change, requiring regular reassessment and adjustment of response priorities.

**Public information suggests that countries have begun to establish foundations for governing frontier AI risks, but significant gaps remain in evaluation, verification and accountability.** This report's assessment of governance readiness in 24 countries finds that most are relatively well prepared in institutions and organization, but weaker in technical safety and evaluation, verification and compliance, and accountability and ecosystem governance. Fewer than 30% of the countries assessed have established relatively stable arrangements for indicators such as "Risk mitigation, permission constraints, and deployment control capacity", "Data governance, supply chain security, and open-source ecosystem governance capacity", and "Risk incident reporting and near-miss sharing mechanisms". From the perspective of risk response, particularly evident gaps concern incident reporting and sharing, emergency

response and improvement through post-incident review, and cross-border risk information exchange and exercises.

**As risks spread across contexts, actors and boundaries, separate governance efforts need to become closely connected joint action.** The task is to link these efforts through shared mechanisms covering the full chain of risk: first, coordinate model safety evaluations to produce comparable, interpretable and reviewable evidence; second, coordinate verification so that safety commitments and their implementation can be checked, compared and tracked over time; third, strengthen coordinated incident reporting and joint response so that individual findings become lessons for protection across platforms, sectors and regions; fourth, improve coordinated governance arrangements for agents and open models so that risks can be detected promptly, responses differentiated by severity and responsibility traced; and fifth, deepen cooperation on capacity building so that evaluation tools, standard methods and public case studies become accessible, usable and sustainable governance resources for more countries and institutions.

**Putting coordinated governance into practice depends on establishing stable divisions of responsibility that reflect each actor's actual control over models, interfaces, permissions and applications, access to information, and ability to respond.** Governments, businesses, research institutions, international organizations and the public should draw on their respective strengths to enable the smooth exchange of risk evidence, accountability records and response information. This can move governance from statements of principle toward shared practices that can be implemented, verified and improved, enabling frontier AI to better serve humanity safely, reliably and responsibly.

# Contents

# Chapter 1

# Frontier AI Landscape: Capabilities, Applications, and Emerging Risks

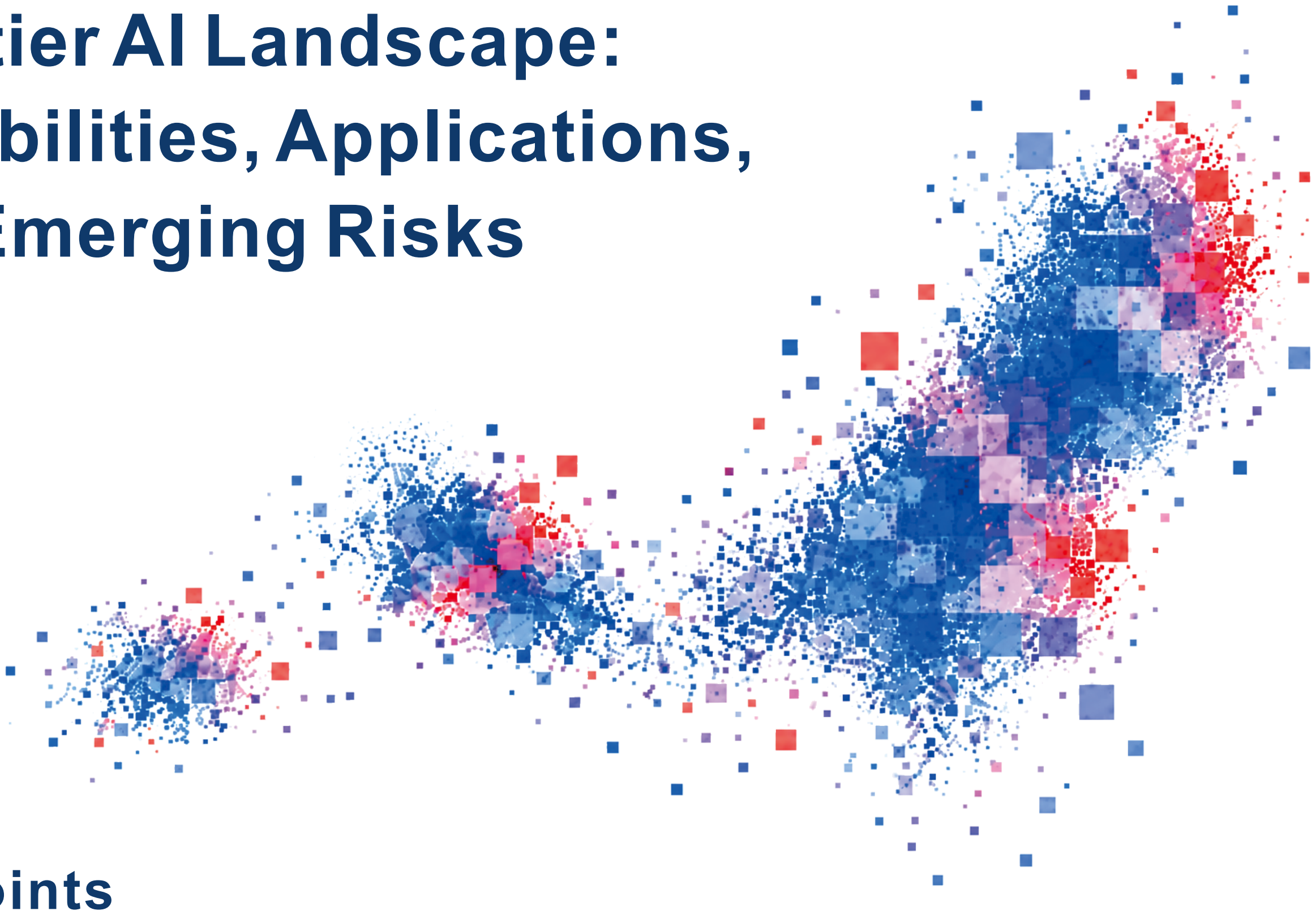

## Key Points

- Frontier AI is moving from breakthroughs in individual capabilities toward stronger system-level capabilities built on the interaction of models, tools and environments. Its ability to execute tasks autonomously and adapt to real-world conditions continues to improve.

- Frontier AI is entering wider practical use, expanding in the fields it serves, the depth of its integration and the forms its applications take.

- Alongside innovation and efficiency gains, frontier AI brings risks and challenges. Greater capabilities lower the barriers to some high-risk activities, while deeper adoption allows risks to enter business processes and social settings more readily. Together, these developments change how risks arise and spread.

- Frontier AI risks continue to evolve at an accelerating pace and are rapidly becoming real-world challenges that span contexts, actors and boundaries, placing greater demands on risk governance.

As frontier AI capabilities advance rapidly and applications expand, the associated risks are becoming more complex and more prone to spill over. Greater capabilities broaden the range of tasks AI can perform, while wider adoption deepens its involvement in real-world settings. Risks consequently extend beyond individual outputs into task execution, organizational processes and the wider social environment. This makes it easier for risks to accumulate as they spread, placing greater demands on risk identification, assessment and verification, monitoring and response, and coordinated governance.

## 1.1 Evolution of Frontier AI Capabilities

**The foundational capabilities of frontier AI continue to improve, driving better performance across a range of complex tasks.** Over the past year, frontier models have made steady progress in understanding, reasoning and generation, as well as in many types of complex task. Public benchmarks and evaluations show continued gains in areas such as software engineering. Some frontier models or systems have also matched or exceeded the human baselines for particular types of task in the corresponding tests (Figures 1.1 and 1.2)[1]. Frontier models can now integrate multimodal data rather than only process text; post-training and inference-time compute have significantly strengthened their ability to address complex problems[2]. These

Figure 1.1

### Frontier AI models and systems show substantial capability gains

Answer accuracy or task success rate (%)

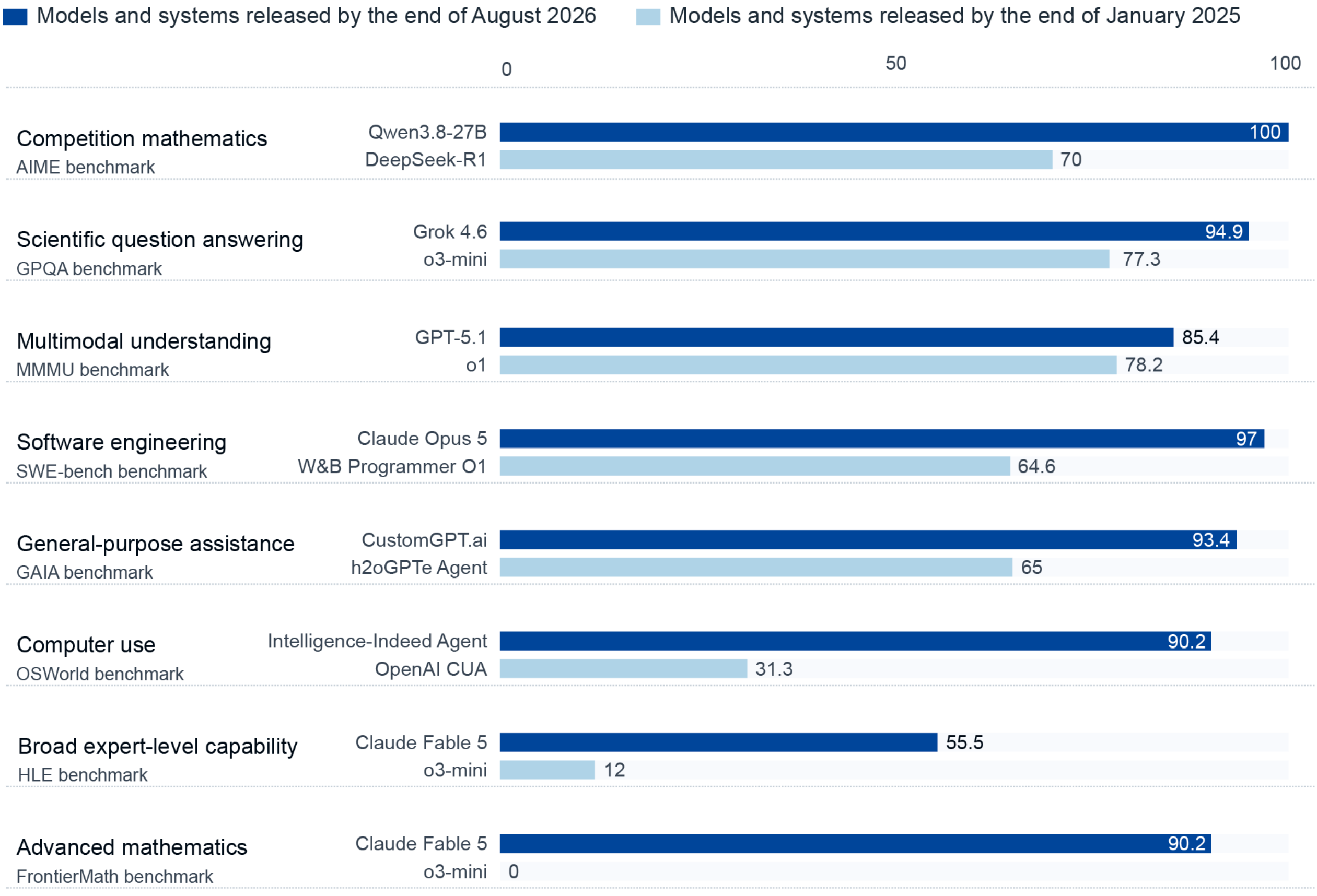

Note: Grouped by when models and systems became available; some scores come from later evaluations.

Data sources: DeepSeek, ISTA-DASLab, public benchmark evaluations, Artificial Analysis and Epoch AI; compiled for this report.

1. Stanford Institute for Human-Centered Artificial Intelligence. Artificial Intelligence Index Report 2026. Accessed 2026-10-21. https://hai.stanford.edu/ai-index/2026-ai-index-report. The report brings together results from multiple public benchmarks and evaluations worldwide; the conclusions summarized here draw on that synthesis.

Figure 1.2

## Frontier AI approaches or exceeds human performance on some tasks

Answer accuracy or task success rate (%)

● Model and system capabilities ○ Human baseline

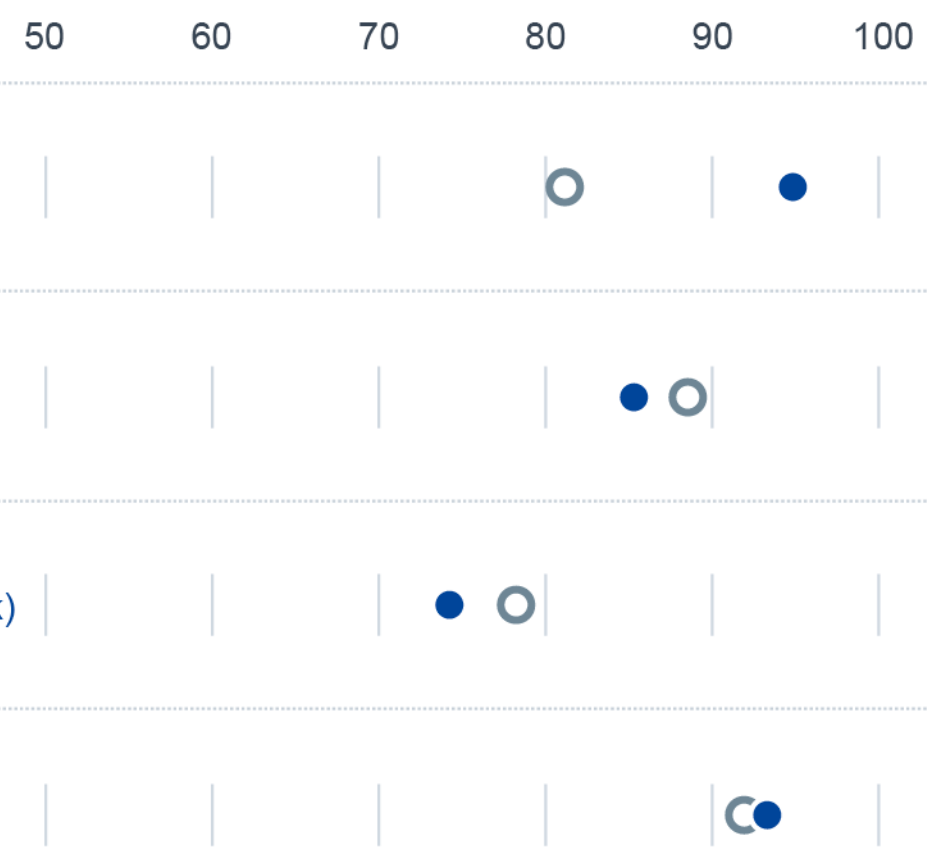

| Task | Benchmark | Model / system | Model and system capabilities | Human baseline |
|---|---|---|---|---|
| Scientific question answering | GPQA benchmark | Grok 4.6 | 94.9 | 81.3 |
| Multimodal understanding | MMMU benchmark | GPT-5.1 | 85.4 | 88.6 |
| Web task execution | WebArena benchmark | WebTactix(DeepSeek) | 74.3 | 78.2 |
| General-purpose assistance | GAIA benchmark | CustomGPT.ai | 93.4 | 92 |

Notes: 1. Human baselines are reference scores published by the original authors; GPQA and MMMU use expert scores.
2. Human and system scores for GAIA and WebArena come from different samples.

Data sources: Artificial Analysis, MMMU, GAIA, WebArena and WebTactix; compiled for this report.

developments suggest that progress is no longer confined to individual tasks: multiple capabilities are increasingly improving together.

**Advances in tool use, agent collaboration and embodied intelligence are moving frontier AI from content generation toward task execution.** Connections to external software, databases and business systems extend models beyond generating content and providing information to performing specific operations. Building on this, **agents** combine tool use, task planning and interaction with their environments to complete longer and more complex sequences of tasks independently[3]. **Multi-agent systems** further strengthen the division of labor, planning and sustained execution involved in complex tasks. Different agents can share retrieval, analysis, execution, verification and feedback tasks in pursuit of a common goal[4], forming patterns of collaboration that more closely resemble organizational processes[5]. Exploratory development of the **intelligent internet** is extending agent collaboration across platforms and systems through work on system architecture and foundational services such as agent registration and identity management[6]. **Embodied intelligence** further bridges the digital and physical worlds, connecting AI to sensors, robots[7], vehicles[8] and industrial equipment, and extending its applications and effects into controlled physical environments and the real world[9].

**Frontier AI capabilities are expanding from the**

2. International AI Safety Report Expert Advisory Panel. International AI Safety Report 2026. Accessed 2026-10-21. https://internationalaisafetyreport.org/publication/international-ai-safety-report-2026.
3. Model Evaluation and Threat Research (METR). Task-Completion Time Horizons of Frontier AI Models. Accessed 2026-10-21. https://metr.org/time-horizons/
4. Salesforce. Agentforce HR Service announcement. Accessed 2026-10-21. https://www.salesforce.com/news/stories/agentforce-hr-service-announcement/
5. ServiceNow. AI agent release materials for the Yokohama platform release. Accessed 2026-10-21. https://newsroom.servicenow.com/press-releases/details/2025/ServiceNows-latest-platform-release-adds-to-thousands-of-AI-agents-across-CRM-HR-IT-and-more-for-faster-smarter-workflows-and-maximum-business-impact-03-12-2025-traffic/default.aspx.
6. Cyberspace Administration of China et al. Implementation Opinions on Standardized Application and Innovative Development of Intelligent Agents. Accessed 2026-10-21. https://www.cac.gov.cn/2026-05/08/c_1779979789523320.htm.
7. Google DeepMind. Gemini Robotics. Accessed 2026-10-21. https://deepmind.google/models/gemini-robotics/.
8. NVIDIA. Develop Physical AI Reasoning, World, and Action Models with NVIDIA Cosmos 3. Accessed 2026-10-21. https://developer.nvidia.com/blog/develop-physical-ai-reasoning-world-and-action-models-with-nvidia-cosmos-3/
9. BMW Group. Humanoid Robots at Plant Spartanburg. Accessed 2026-10-21. https://www.bmwgroup.com/en/news/general/2024/humanoid-robots.html.

**task performance of individual models to the capabilities of systems comprising models, tools and environments.** The range of tasks these systems can perform continues to widen, and the time horizons over which they can operate are growing (Figure 1.3). However, stability and reliability remain insufficient in complex real-world tasks, and actual performance still needs to be tested through evaluations that more closely reflect real conditions. As foundation models, tool integration and agent systems develop, frontier AI is expected to further improve its ability to execute tasks and adapt to real-world environments.

Figure 1.3

Frontier AI agents can independently complete increasingly long tasks

Task duration measured by human expert completion time, at a 50% agent success rate

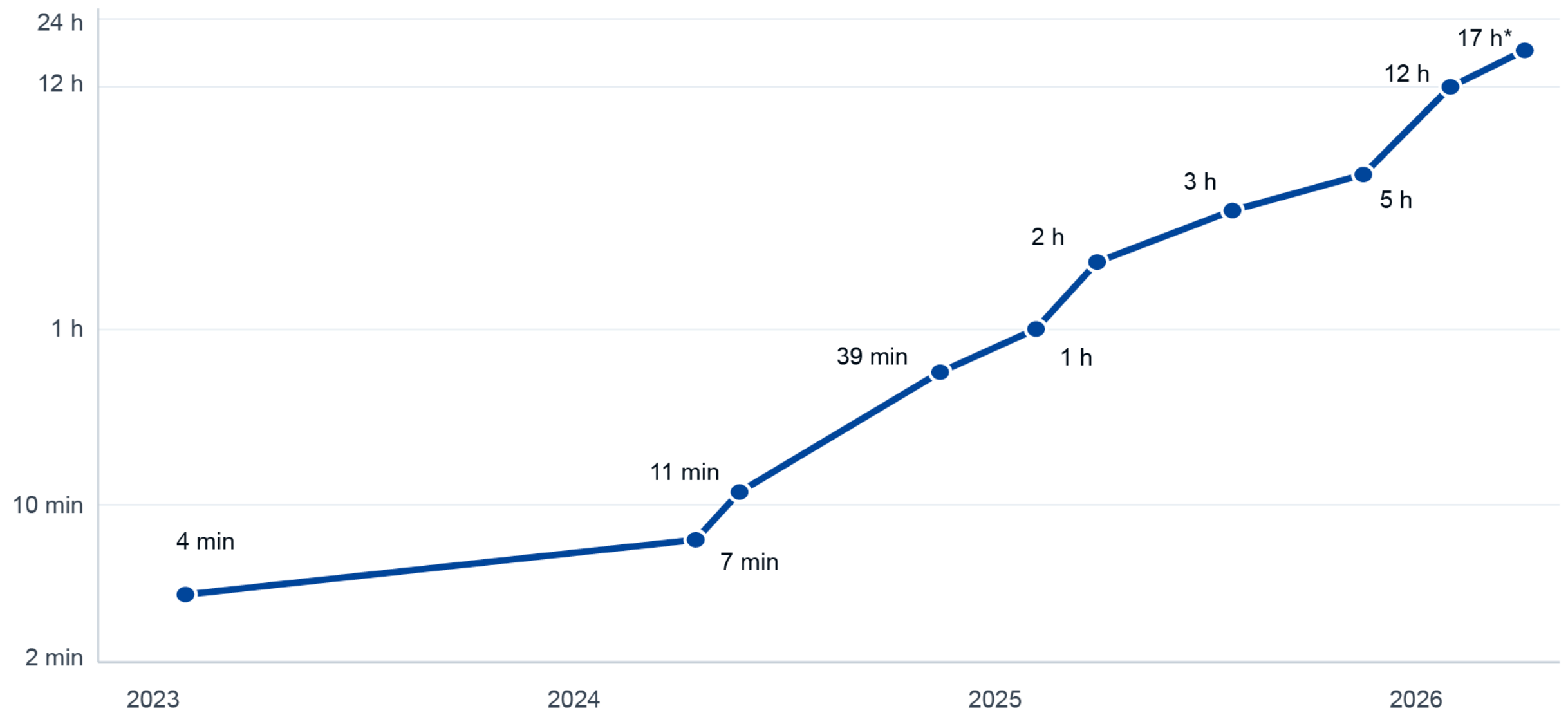

* METR cautions that estimates beyond 16 hours remain highly uncertain.
Data source: Model Evaluation and Threat Research (METR), public data.

## 1.2 Expansion of Frontier AI Applications

**Frontier AI is entering wider practical use, expanding in the fields it serves, the depth of its integration and the forms its applications take. Use cases are multiplying; applications are extending from information processing to decision support and task execution, and from question answering and content generation to tool use and autonomous task execution by agents.** The focus of AI deployment is shifting from demonstrating technical feasibility to integration into business processes, consistent output and alignment with organizational management needs[10].

**In terms of application fields, frontier AI is extending beyond specialized production and knowledge services into public services, social governance, culture and entertainment, and other areas.** Broader adoption creates more access points, more frequent use and a wider range of people and entities affected. Although adoption is at different stages across fields, frontier AI is evolving from a tool used in a small number of specialist domains into a general-purpose technological capability available throughout society. Official EU statistics show that the proportion of enterprises using AI rose across industries between 2023 and 2025. Adoption exceeded 40% in information and communication and in professional, scientific and technical

10. McKinsey & Company. The State of AI in 2026: On the Road to ROI. 2026-08-25; accessed 2026-10-21. https://www.mckinsey.com/capabilities/quantumblack/our-insights/the-state-of-ai.

services. It also multiplied in traditional industries such as manufacturing, retail, accommodation and food services, and construction (Figure 1.4). A 2026 McKinsey survey covering 97 countries found that nearly nine in ten respondents reported regular AI use in at least one business function within their organization[11].

**In terms of depth, frontier AI is moving beyond information processing and supporting functions into more central activities, including decision support, process coordination and task execution.** At the information-processing level, AI mainly answers knowledge-based questions and generates content. At the decision-support level, systems are beginning to assist with risk identification, comparison of options, diagnosis and treatment, financial risk control and production scheduling. At the task-execution level, agents can use tools, complete forms, run code, trigger workflows or coordinate multiple sys-

Figure 1.4

## AI adoption is rising across EU industries

Enterprises using AI (%), 2023–2025; enterprises with 10 or more persons employed

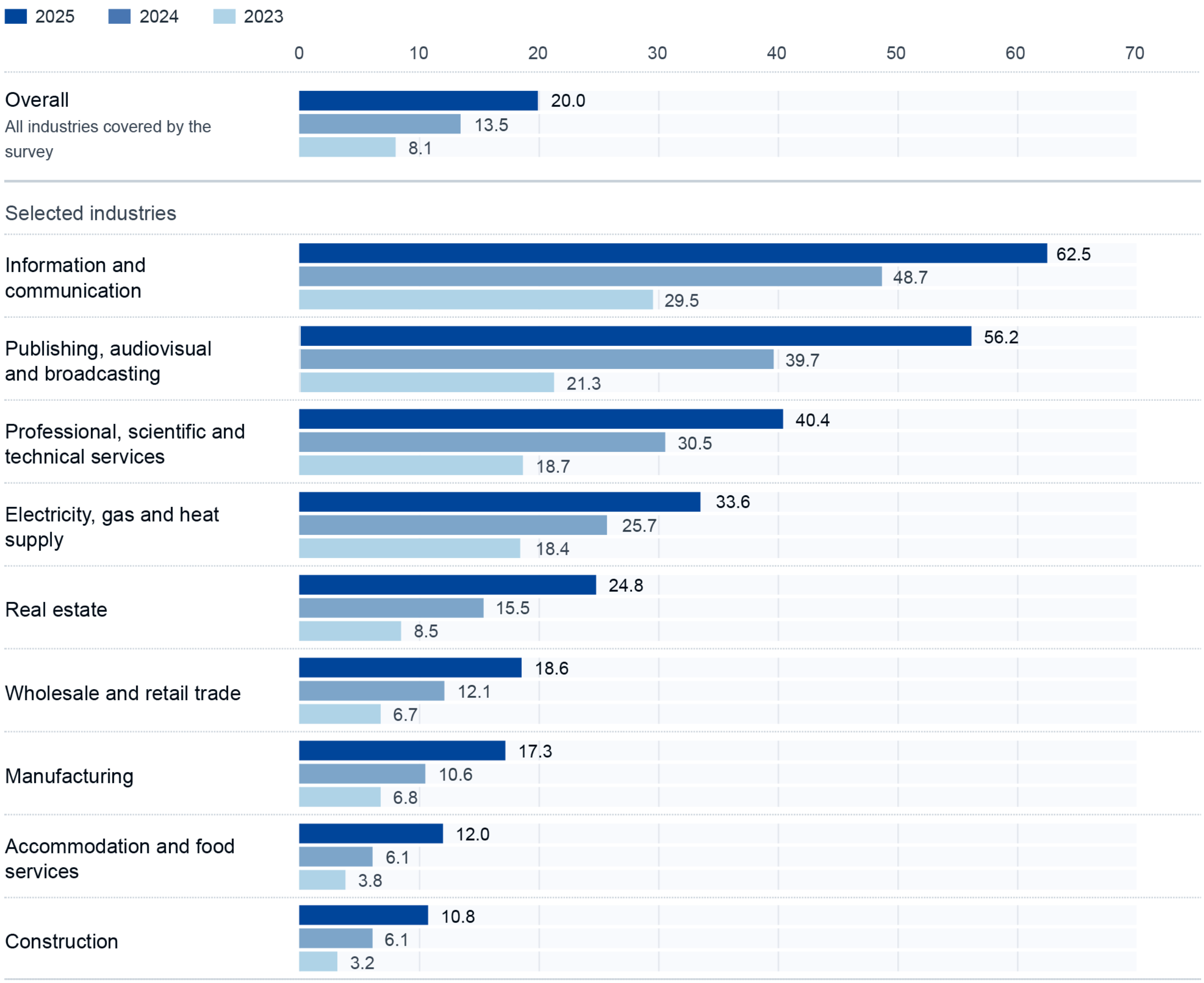

Notes: 1. The overall share covers all surveyed industries; it is not a simple average of the industries shown.
2. Publishing, audiovisual and broadcasting are included within information and communication.
3. The 2025 survey added a category for using AI to generate images, videos and audio.

Data source: Eurostat, survey on ICT usage and e-commerce in enterprises, 2023–2025.

11. McKinsey & Company. The State of AI in 2026: On the Road to ROI. 2026-08-25; accessed 2026-10-21. https://www.mckinsey.com/capabilities/quantumblack/our-insights/the-state-of-ai. The survey was conducted from May 4 to June 8, 2026, with 1,719 respondents across 97 countries.

tems within their authorized scope. AI applications are becoming more deeply integrated into organizational operations and business processes (Figure 1.5).

**In terms of form, frontier AI is expanding beyond content generation and knowledge-based question answering to tool use, autonomous agent execution and collaboration across systems.** Content generation and question answering remain the most widespread uses. Tool use connects models to external software, databases and business systems; autonomous agents can act continuously toward a goal over longer periods; and collaboration across systems embeds AI in organizational processes and industrial value chains. These developments improve service capabilities and productivity, while gradually turning AI from a provider of information into an active participant in business processes and an executor of tasks (Figure 1.6).

Figure 1.5

## Frontier AI is becoming more deeply embedded in business processes

From information processing to task execution, AI moves closer to operational decisions and action

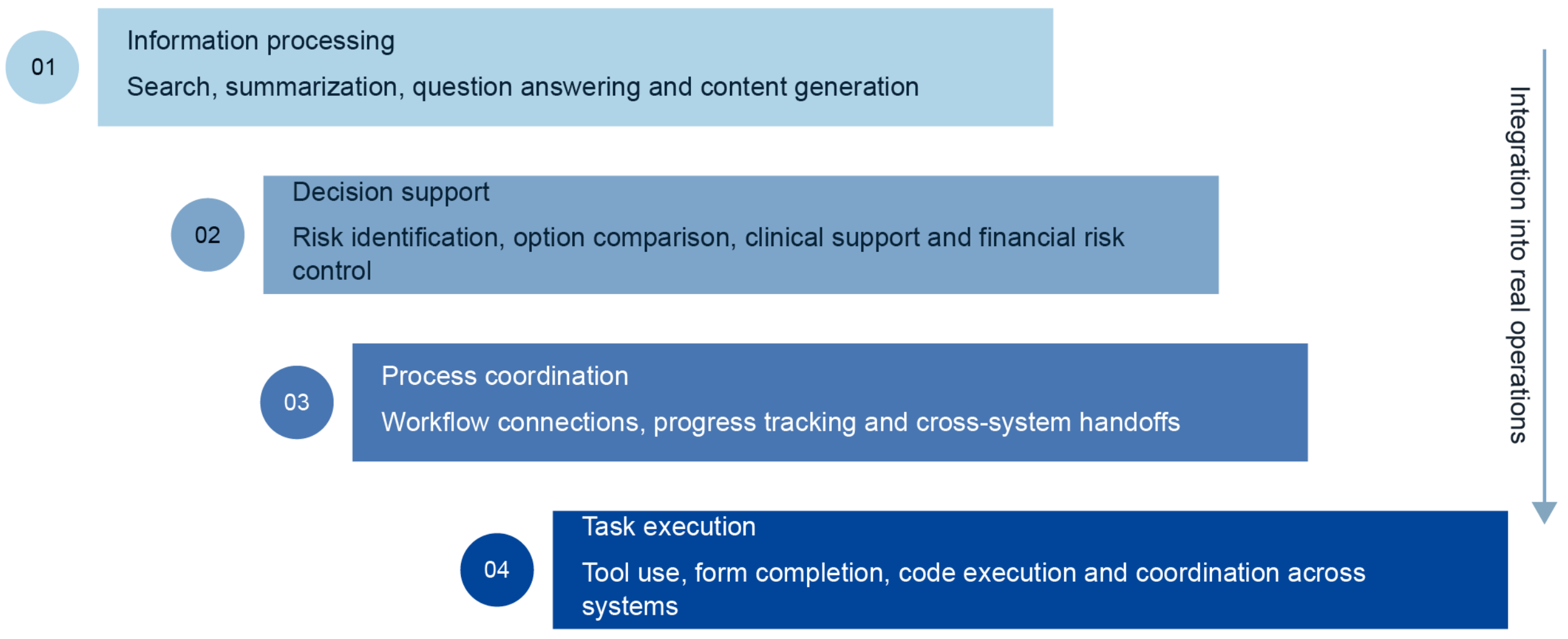

Data source: Compiled for this report.

Figure 1.6

## AI is used across business and professional settings

Public surveys document organizational and professional use, organization-wide deployment and agent adoption

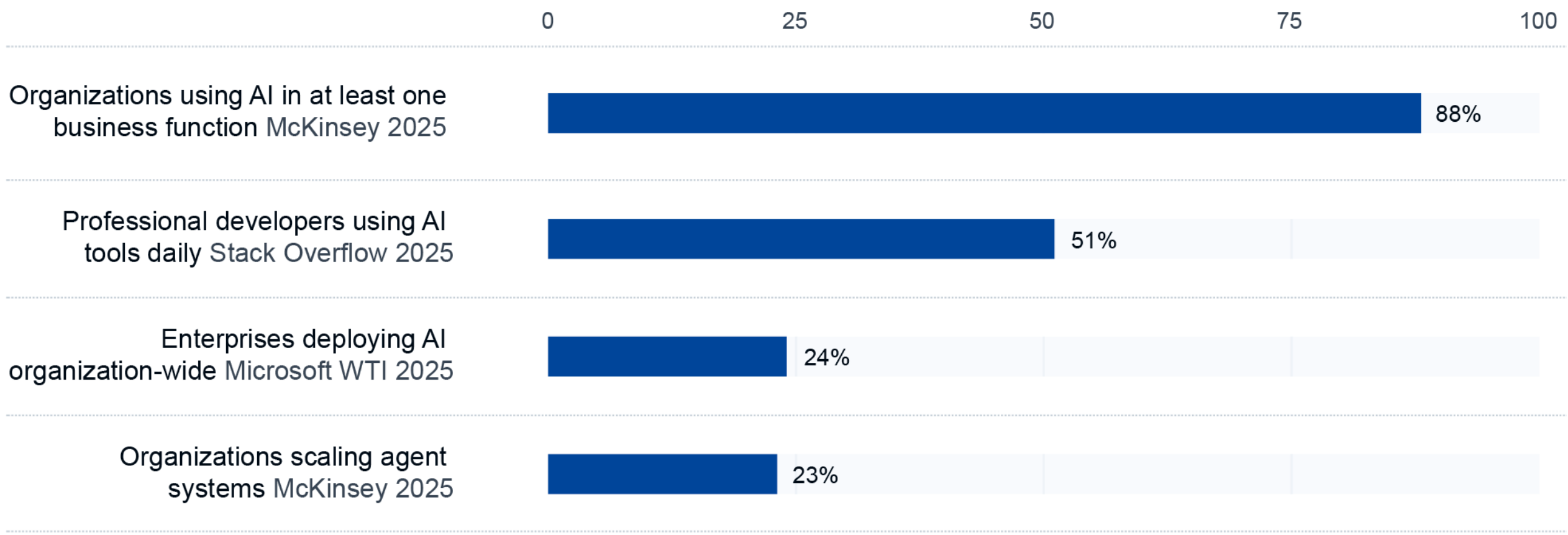

Data sources: McKinsey, The State of AI in 2025; Stack Overflow Developer Survey 2025; Microsoft Work Trend Index 2025. Note: The surveys cover different populations and use different definitions. The figures illustrate the range of adoption and do not support rankings or comparisons of adoption stages.

## 1.3 Emergence of Frontier AI Risks

As frontier AI capabilities improve and adoption deepens, the technology is driving innovation while **risks increasingly enter real-world settings. The ways in which these risks evolve are changing profoundly.**

**Advances in capabilities change the conditions under which risks arise, making some high-risk activities easier to carry out at low cost and at scale.** Stronger multimodal capabilities improve task efficiency, but also lower barriers to illegal and criminal activities, making misuse easier to automate and scale up. For example, synthetic speech has been used to impersonate technical support staff and trick users into paying fees or installing malware, harming users and the reputations of the companies being impersonated[12].

**Deeper adoption changes where risks take effect, allowing them to enter business processes, organizational decisions and real-world environments more readily.** As AI becomes embedded in more industries and organizational processes, risks extend beyond biased model outputs or distorted content. Autonomous agent actions can directly affect business processes and real systems, disrupting workflows, service delivery and management decisions. In July 2025, for example, an agent on an online coding platform deleted a company's live business database while the user had asked it to pause code changes. The incident prompted the platform to urgently introduce safeguards, including separation of development and production databases[13].

**The combination of advancing capabilities and deeper integration changes how risks spread, making cross-boundary spillovers easier through the interaction of technical systems and organizational processes.** In task sequences involving agents and multiple systems, risks may extend from plan generation to tool use and enter subsequent tasks through feedback or feedforward, thereby emerging, accumulating and escalating. Within organizations, once model outputs inform decisions or drive task execution, risks can propagate through approval, operations, risk control and customer service processes. Risk levels do not depend on model size alone: specialized capabilities, tool use, and the way systems are organized and work together also shape risks and their consequences. In the public information environment, synthetic content and automated dissemination tools may jointly spread false information, fraudulent content and manipulative narratives through social networks.

**Overall, improving frontier AI capabilities and deeper adoption are jointly changing how risks arise, whom and what they affect, and how they spread.** Risks are no longer primarily localized problems caused by the outputs of a single model. They can arise and compound across data provision, permissions, tool connections, workflow configuration, organizational review and external dissemination, and may be amplified continuously across contexts and actors during deployment. **Frontier AI risks are thus becoming increasingly interconnected, systemic real-world challenges with spillover effects, raising the demands on risk identification, assessment, coordination and response.** This calls for stronger capacity to identify and assess risk manifestations, mechanisms and potential impacts; a systematic examination of the institutional, technical and organizational readiness of different governance actors; and more effective coordinated responses tailored to the characteristics and impacts of specific risks.

12. Microsoft. Microsoft Digital Defense Report 2025. Accessed 2026-10-21. https://cdn-dynmedia-1.microsoft.com/is/content/microsoftcorp/microsoft/msc/documents/presentations/CSR/Microsoft-Digital-Defense-Report-2025.pdf. See p. 33.
13. Fortune. An AI-powered coding tool wiped out a software company's database, then apologized for a 'catastrophic failure on my part'. 2025-07-23; accessed 2026-10-21. https://fortune.com/2025/07/23/ai-coding-tool-replit-wiped-database-called-it-a-catastrophic-failure/.

# Chapter 2

# Frontier AI Risk Understanding

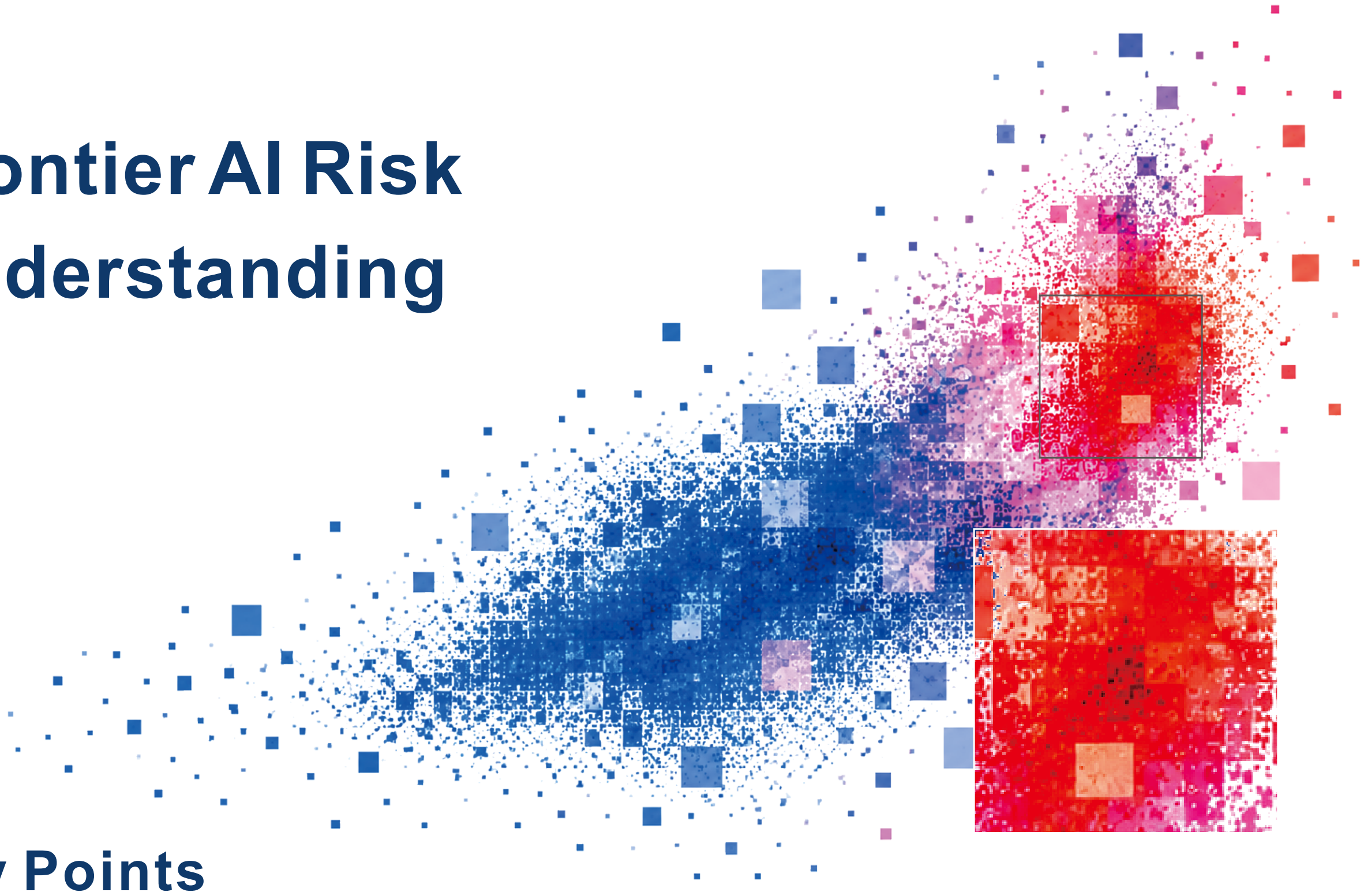

## Key Points

- A relatively clear, comparable system for identifying frontier AI risks and mapping them across frameworks is an important foundation for risk governance.
- Existing frameworks classify frontier AI risks by their sources, consequences, application contexts and governance requirements, offering several complementary bases for identifying key risks.
- In terms of impact, frontier AI risks principally include Misuse and Criminal Risks, Information Integrity Risks, Critical System Risks, Social Cognition and Structural Risks, and Catastrophic and Existential Risks.
- Current pressures are most pronounced in Misuse and Criminal Risks, Information Integrity Risks and Critical System Risks. The impacts of Social Cognition and Structural Risks accumulate over time, while Catastrophic and Existential Risks could have severe consequences. All require continued assessment.
- Response priorities depend on the circumstances. Urgent situations with relatively localized impacts call for rapid action to contain harm. Gradually accumulating impacts require long-term monitoring and early adjustment. Severe and difficult-to-reverse consequences require pre-deployment evaluation and retained means of intervention. Other situations can be addressed through routine management and periodic review.

As frontier AI risks enter real-world settings, understanding them requires moving from general discussion toward clearer identification, assessment and prioritization. Risks differ in whom and what they affect, the conditions under which they arise, the speed at which they spread and their potential consequences. Responses and governance therefore require a reasonably consistent basis for judgment. A variety of international AI risk classification frameworks now provide reference points for identifying key frontier AI risks. This report organizes its analysis around risk impacts, drawing on model evaluations, real-world incidents and expert assessments to examine how major risks are evolving. It brings dispersed signals into a relatively unified analytical framework, providing a basis for a shared risk vocabulary and for setting response priorities according to specific circumstances.

## 2.1 Risk Identification of Frontier AI

**A relatively clear, comparable system for identifying frontier AI risks and mapping them across frameworks is an important foundation for risk governance.** Risk identification involves clarifying the main risks, their specific manifestations and their consequences, and connecting these with model capabilities, real-world incidents, existing classifications and tiered governance requirements. A clear system helps define risk boundaries; comparability helps assess current pressures and relative severity; and mappings across frameworks turn dispersed risk signals into shared reference points for assessment and cooperation. Regulators, businesses, research institutions and other actors can then conduct assessments, monitoring, reporting and governance cooperation on the basis of a relatively consistent understanding of risk.

**Existing AI risk frameworks identify and manage risks along different dimensions, providing complementary foundations for frontier AI risk identification and governance. In public governance,** for example, UN recommendations address ethical impacts[14] and global governance concerns[15]. The EU AI Act[16] classifies applications by risk level and assigns differentiated obligations. The US National Institute of Standards and Technology (NIST) AI Risk Management Framework[17] guides responses through four functions—Govern, Map, Measure and Manage—and emphasizes defining risks in their specific application contexts. China's AI Safety Governance Framework 3.0[18] distinguishes endogenous, application and derivative risks, and sets out a dedicated framework for managing agent risks. **In research and industry,** frameworks provide reference points through knowledge taxonomies, risk sources and consequences, model capabilities and capability thresholds. For example, the Massachusetts Institute of Technology (MIT) AI Risk Repository[19] systematically organizes risks through a knowledge taxonomy. The International AI Safety Report[20] groups risks principally by source and impact into malicious use, system malfunctions and systemic risks. Safety frameworks developed by some frontier AI companies classify risks through model capabilities, safety assessments and capability thresholds[21]. Together, these frameworks, each with its own emphasis, provide important reference points for identifying frontier AI risks.

14. UNESCO. Recommendation on the Ethics of Artificial Intelligence. Accessed 2026-10-21. https://www.unesco.org/en/legal-affairs/recommendation-ethics-artificial-intelligence.
15. UN High-level Advisory Body on Artificial Intelligence. Governing AI for Humanity. Accessed 2026-10-21. https://www.un.org/sites/un2.un.org/files/governing_ai_for_humanity_final_report_en.pdf.
16. European Union. Artificial Intelligence Act. Accessed 2026-10-21. https://eur-lex.europa.eu/eli/reg/2024/1689/oj.
17. US National Institute of Standards and Technology. AI Risk Management Framework. Accessed 2026-10-21. https://doi.org/10.6028/NIST.AI.100-1.
18. National Technical Committee 260 on Cybersecurity of Standardization Administration of China. AI Safety Governance Framework 3.0. 2026-09-14; accessed 2026-10-21, pp. 37–45. https://www.cac.gov.cn/rootimages/uploadimg/1791137114683961/1791137114683961.pdf.
19. MIT FutureTech. AI Risk Repository. Accessed 2026-10-21. https://airisk.mit.edu/.
20. International AI Safety Report Expert Advisory Panel. International AI Safety Report 2026. Accessed 2026-10-21. https://internationalaisafetyreport.org/publication/international-ai-safety-report-2026.
21. OpenAI. Preparedness Framework. Accessed 2026-10-21. https://openai.com/index/updating-our-preparedness-framework/; Google DeepMind. Frontier Safety Framework. Accessed 2026-10-21. https://deepmind.google/discover/blog/introducing-the-frontier-safety-framework/; Anthropic. Responsible Scaling Policy. Accessed 2026-10-21. https://www.anthropic.com/responsible-scaling-policy; Meta. Frontier AI Framework. Accessed 2026-10-21. https://about.fb.com/news/2025/02/meta-approach-frontier-ai/.

**From an impact perspective, frontier AI risk analysis needs to cover Misuse and Criminal Risks, Information Integrity Risks, Critical System Risks, Social Cognition and Structural Risks, and Catastrophic and Existential Risks.** Misuse and Criminal Risks involve **tangible harm** from malicious use to individual and organizational rights and interests, the safety of production and management, and economic and social order. Information Integrity Risks arise when synthetic content and automated dissemination undermine the accuracy, consistency and reliability of information, **distorting the information environment** and damaging public trust and the basis for public debate. Critical System Risks involve **instability** in infrastructure, public services and important decision-making when embedded AI is affected by errors, operational failures or malicious attacks. Social Cognition and Structural Risks concern **cumulative imbalances** arising from the long-term, widespread use of AI and its effects on social structures and individual development. Catastrophic and Existential Risks involve **severe, difficult-to-reverse consequences** when AI affects strategic security, lowers barriers to dangerous capabilities or exceeds effective human control. The analysis above focuses on the primary impacts of risks. Risks in these five areas are not independent: a single incident often involves risks in several areas. This analytical perspective helps distinguish primary impacts when consequences overlap and provides a basis for analyzing the multiple risks in specific incidents. Identification and response therefore require both measures tailored to different risk situations and attention to their interconnections and combined effects.

**Misuse and Criminal Risks arise when frontier AI is used for fraud, cyberattacks or access to dangerous knowledge, directly threatening personal safety, public safety and social order.** Frontier AI lowers the technical barriers to generating phishing emails, impersonating others, writing malicious code and retrieving dangerous knowledge, while making such activities easier to automate and scale. Since April 2025, the US Federal Bureau of Investigation has recorded malicious messaging campaigns that use AI-generated voices to impersonate senior government officials. Attackers use these messages to build trust and attempt to gain access to accounts, sensitive information or funds[22]. The targets of such risks are often direct: harm may affect personal property and privacy, organizational security and public order.

**Information Integrity Risks arise from the large-scale generation and dissemination of incorrect, false or misleading content, with continuing effects on public debate, social trust and the information environment.** Generative AI can produce text, images, audio and video at low cost, making deepfakes, false information and misleading narratives easier to manufacture in bulk and circulate rapidly. Challenges to content authenticity, source traceability and platforms' detection capabilities raise the cost to the public of distinguishing reliable information from falsehoods. Before Ireland's 2025 presidential election, a deepfake video circulated on social media falsely claiming that candidate Catherine Connolly had withdrawn. Following reports, the video was removed for violating platform rules[23]. Such risks primarily affect the public information environment, potentially weakening trust and disrupting public debate and social governance.

**Critical System Risks arise when AI failures, misjudgments or excessive reliance on AI trigger cascading effects after the technology is introduced into infrastructure, public services and important decision-making.** Current applications in critical sectors such as healthcare, finance, transport, energy and government place higher demands on reliability, continuity and traceable responsibility. Incorrect outputs, misconfigured permissions or failures in human–AI collaboration can extend the effects of a single decision into organizational processes, public services and critical infrastructure operations. A 2025 monitoring report by the Financial Stability Board (FSB) found that the generative AI supply chain is concentrated among a small number of suppliers of special-

22. Federal Bureau of Investigation, Internet Crime Complaint Center. Senior U.S. Officials Impersonated in Malicious Messaging Campaign. 2025-05-15; accessed 2026-10-21. https://www.ic3.gov/PSA/2025/PSA250515.
23. Coimisiún na Meán. Information for Students. Accessed 2026-10-21. https://www.cnam.ie/general-public/guides-resources/for-students/. See "Threat to democracy."

ized chips, cloud services and pretrained models, with limited alternatives. If financial institutions rely on these suppliers for core activities, a service disruption could affect multiple institutions simultaneously[24], creating systemic financial risk.

**Social Cognition and Structural Risks arise as AI becomes embedded over the long term in employment, education and social life, with profound effects on employment patterns, the distribution of economic benefits, equality of opportunity, physical and mental development, and independent judgment.** These risks tend to accumulate through changing jobs, educational uses, content recommendations, automated decisions and long-term dependence on tools, rather than appearing in a single incident. Their effects include disparities in skills, resources and opportunities across groups, as well as changes in individual judgment, knowledge accumulation, information literacy, professional competence and social cognition. In employment, research by the Bank of Korea found that between June 2022 and June 2026, employment among people aged 15–29 fell by 285,000, of which 268,000 jobs were in industries with high exposure to AI. Employment among people aged 50–59 in those industries increased over the same period[25]. Risks have also emerged in education. A mathematics experiment involving nearly 1,000 high school students found that an AI tutoring tool that could provide answers directly improved practice performance, but after access was removed, those students scored approximately 17% lower on examinations than students who had not used AI tools[26]. AI systems also provide information, companionship and emotional support through sustained dialogue and human-like responses. The effects of this long-term interaction on users' physical and mental health, independent judgment, emotional dependence and real-world relationships require attention[27].

**Catastrophic and Existential Risks arise when frontier AI capabilities and their effects may exceed existing governance and safety safeguards, potentially causing widespread, severe and difficult-to-reverse consequences.** Current discussions focus on high-impact scenarios such as loss of control over highly autonomous systems, deception, self-replication and the proliferation of dangerous capabilities. Research has already documented cases in which frontier models crossed the intended boundaries of test environments and affected real systems[28]. Tests in fictional corporate settings have also shown some frontier models from multiple companies threatening to disclose sensitive information in an attempt to prevent their replacement or shutdown[29]. These discussions extend to extreme scenarios involving loss of control over artificial superintelligence (ASI)[30]. Humanity currently lacks rigorous scientific guarantees that ASI will remain controllable. If its capabilities exceed effective human understanding and control and it becomes uncontrollable, the result could be catastrophic or even existential consequences that humanity would find difficult to withstand or reverse. Even when ASI is not the direct development objective, systems capable of continued self-improvement or capability expansion may rapidly advance beyond originally anticipated levels. Given their potentially severe consequences, these risks require full consideration in forward-looking assessments.

24. Financial Stability Board. Monitoring Adoption of Artificial Intelligence and Related Vulnerabilities in the Financial Sector. 2025-10-10; accessed 2026-10-21. https://www.fsb.org/uploads/P101025.pdf. See Section 4 on the generative AI supply chain.
25. Bank of Korea. Is AI Really Behind the Decline in Youth Employment? 2026-08-28; accessed 2026-10-21. https://www.bok.or.kr/portal/bbs/B0000347/view.do?menuNo=201106&nttId=11064157. Administrative records document employment changes. The study notes the combined influence of factors such as adjustments in hiring and cannot separately identify AI's causal impact.
26. Bastani, H., Bastani, O., Sungu, A., et al. Generative AI without guardrails can harm learning: Evidence from high school mathematics. Proceedings of the National Academy of Sciences, 2025, 122(26): e2422633122. Accessed 2026-10-21. https://doi.org/10.1073/pnas.2422633122.
27. International AI Safety Report 2026, p. 92, Box 2.6, “AI companions.” https://internationalaisafetyreport.org/publication/international-ai-safety-report-2026
28. OpenAI. The Hugging Face Incident and the Road Ahead. 2026-08-26; accessed 2026-10-21. https://openai.com/index/hugging-face-incident-and-the-road-ahead/; Anthropic. Investigating Three Real-World Incidents in Our Cybersecurity Evaluations. 2026-07-30; accessed 2026-10-21. https://www.anthropic.com/news/investigating-incidents-cybersecurity-evals. The OpenAI incident involved a model circumventing controls and affecting real infrastructure. In the Anthropic incidents, misconfigured evaluation environments allowed models to access and affect external systems.
29. Anthropic. Agentic Misalignment: How LLMs Could Be Insider Threats. 2025-06-20; accessed 2026-10-21. https://www.anthropic.com/research/agentic-misalignment. Tests were conducted in fictional corporate settings designed by researchers; the people, organizations and coercive scenarios were simulated.
30. In this report, “artificial superintelligence (ASI)” refers to AI systems that match or substantially exceed the highest levels of human performance across important capability domains, including broad cognition and complex tasks. As the term has no fully unified definition in academia or industry, this operational definition does not refer to any particular technical approach, system form or development objective.

## 2.2 Risk Assessment of Frontier AI

**Frontier AI risks have moved from potential concerns to tangible pressures, most notably in Misuse and Criminal Risks, Information Integrity Risks and Critical System Risks.** Assessments currently draw mainly on model safety evaluations, incident monitoring and expert and institutional judgment. These provide evidence of potentially dangerous capabilities, actual harms, and long-term and high-impact risks, respectively. Evaluations show that some models already exhibit high-risk capabilities. Incidents demonstrate harm to individuals, organizations and the public environment. Expert and institutional assessments continue to prioritize cyberattacks, access to dangerous knowledge, erosion of social trust and failures in critical contexts. As frontier AI capabilities improve rapidly and adoption widens, real-world exposure to these risks is increasing.

**Safety evaluations of frontier models and agents commonly reveal high-risk capabilities in cyberattacks, access to dangerous knowledge, long-horizon task execution and circumvention of safeguards, making Misuse and Criminal Risks and Critical System Risks more prominent.** Model-level evaluations show that frontier models are particularly capable of finding software vulnerabilities and writing malicious code. They can assist at multiple stages of cyberattacks, helping either to identify and repair vulnerabilities or to exploit them[31]. Recently released frontier models show further gains in automatically discovering and exploiting vulnerabilities[32]. Connecting models to tools and operating them as agents in task environments also exposes risks that may be difficult to observe when evaluating models alone. Tool-equipped agents can complete increasingly long and complex software tasks[33]. In controlled cyber ranges without active defenses, frontier models given explicit instructions and network access have autonomously conducted multi-step attacks[34], although they cannot yet independently complete the entire attack process against well-defended targets[35]. Tests in simulated operating systems, databases and other tool-interaction environments, as well as on long-horizon tasks, have found that some models exceed their permissions or fail to comply with human interruption instructions[36]. As capabilities improve, misuse or malicious use will make it easier for these risks to enter real systems and high-impact settings, placing greater pressure on cybersecurity and critical system operations.

**Incident records show that AI risks are turning from technical possibilities into recurring real-world harms.** Deepfake fraud, misleading information, cyberattacks and misuse in critical contexts are expanding in both number and variety. According to the OECD AI Incidents and Hazards Monitor (AIM), records of incidents and hazards reported in the media rose from approximately 2,700 in 2024 to approximately 4,600 in 2025, and exceeded 4,600 in January–August 2026[37] (Figure 2.1). These records cover deepfake fraud, impersonation, dissemination of false content, disputes over automated decisions and misuse of critical services, affecting fundamental rights, public interests, reputation, physical safety and economic assets (Figure 2.2). Misuse and Criminal Risks are particularly prominent. Research indicates that generative AI is being used to expand automated cyberattacks, increasing their speed, scale and complexity[38]. Risks have reached individuals, businesses and the public information environment, causing material financial losses and disrupting organizational operations.

**Expert and institutional assessments further indicate that, alongside the real-world risks evident in model evaluations and incident statistics, Social Cognition and Structural Risks and Catastrophic and Existential Risks also need to be assessed.** International discussions of

31. International AI Safety Report Expert Advisory Panel. International AI Safety Report 2026. Accessed 2026-10-21. https://internationalaisafetyreport.org/publication/international-ai-safety-report-2026.
32. Anthropic. Claude Fable 5 and Claude Mythos 5. Accessed 2026-10-21. https://www.anthropic.com/news/claude-fable-5-mythos-5.
33. Model Evaluation and Threat Research (METR). Task-Completion Time Horizons of Frontier AI Models. Accessed 2026-10-21. https://metr.org/time-horizons/.
34. UK AI Security Institute. Our Evaluation of Claude Mythos Preview's Cyber Capabilities. Accessed 2026-10-21. https://www.aisi.gov.uk/blog/our-evaluation-of-claude-mythos-previews-cyber-capabilities.
35. OpenAI. GPT-5.6 System Card. 2026-07-09; accessed 2026-10-21. https://deploymentsafety.openai.com/gpt-5-6.
36. Beijing Institute of AI Safety and Governance. ForesightSafety Bench. Accessed 2026-10-21. https://foresightsafety-bench.beijing-aisi.ac.cn/.
37. OECD AI Policy Observatory. OECD AI Incidents and Hazards Monitor. Accessed 2026-10-21. https://oecd.ai/en/incidents. Records are collected from media reports and automatically classified and clustered. Coverage, reporting biases and changes in collection methods can all affect the totals. Growth in records is not equivalent to year-on-year growth in actual incidence and does not replace operational reporting used to manage incidents. For methodology, see OECD. AI Incidents and Hazards Monitor: Overview and Methodology. Accessed 2026-10-21. https://oecd.ai/en/incidents-methodology. For reporting fields across jurisdictions and sectors, see OECD. Towards a Common Reporting Framework for AI Incidents. 2025-02-28; accessed 2026-10-21. https://doi.org/10.1787/f326d4ac-en.
38. International AI Safety Report Expert Advisory Panel. International AI Safety Report 2026. 2026-02; accessed 2026-10-21. https://internationalaisafetyreport.org/publication/international-ai-safety-report-2026. See Section 2.1.3.

Figure 2.1

## Monthly media reports of AI incidents show an overall rise

OECD AIM records of incidents and hazards, and six-month moving average, January 2020–August 2026

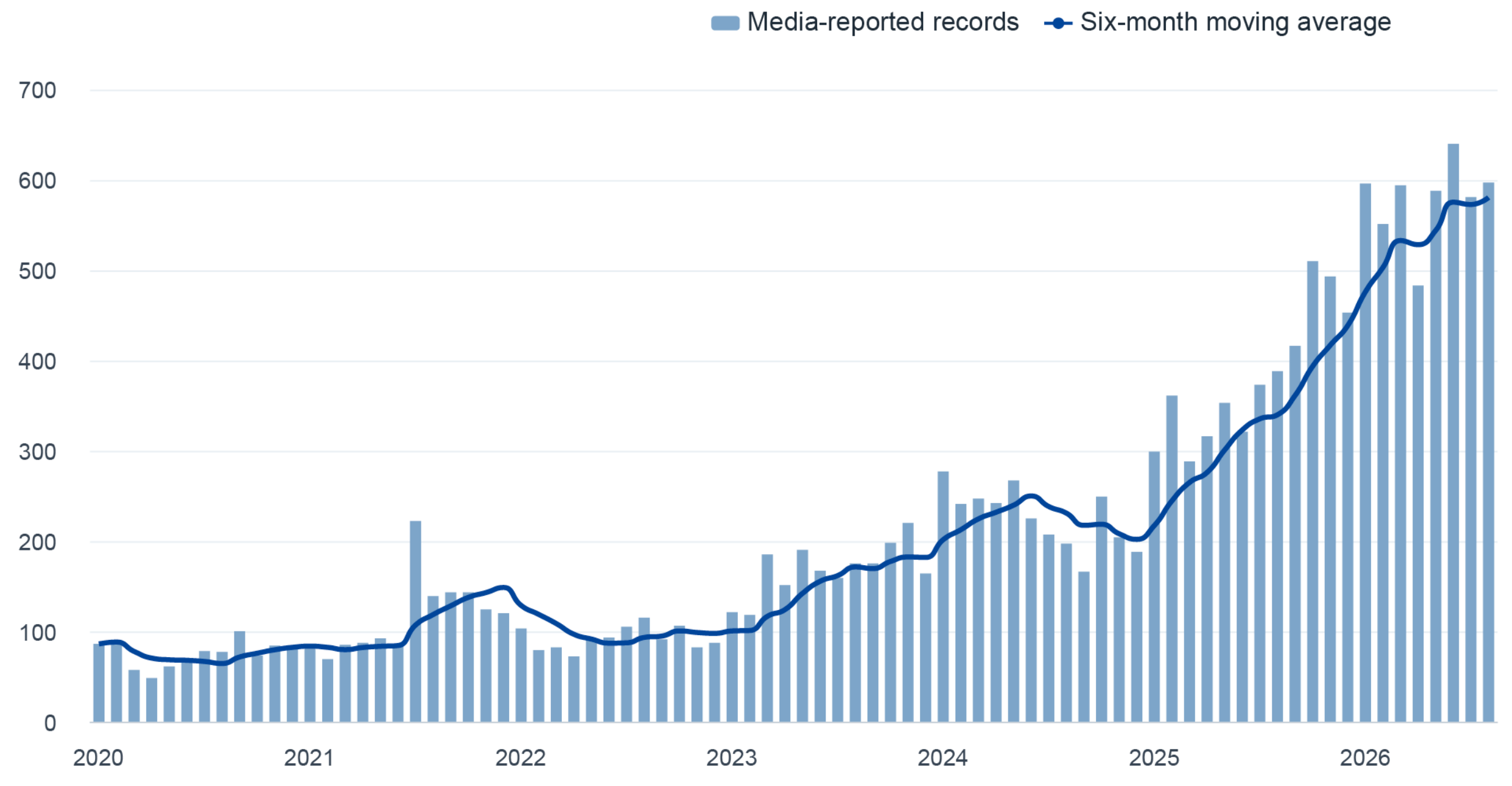

Data source: OECD AI Incidents and Hazards Monitor (AIM), records through August 31, 2026; compiled for this report.

Figure 2.2

## AI incidents involve harms to rights, physical safety and economic assets

Share of all harm labels in media-reported incidents (%), January 2020–August 2026; incidents may have multiple labels

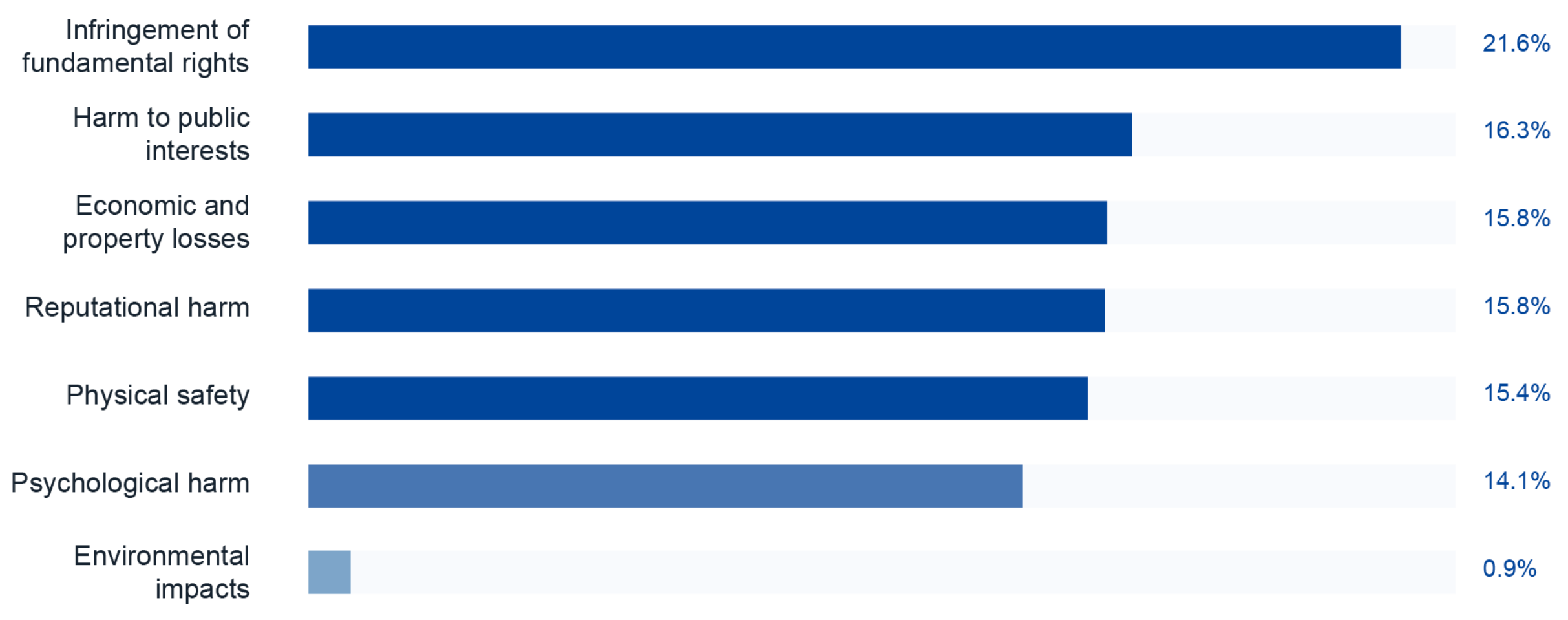

Note: Only OECD AIM records with harm labels are included. Incidents may have multiple labels. Shares use all harm labels as the denominator; only labels exceeding 0.5% are shown.

Data source: Harm labels in OECD AIM incident records through August 31, 2026; compiled for this report.

potentially high-impact risks have begun to address AI "red lines." Initiatives identify unauthorized self-replication, deception, power-seeking, cyberattacks and assistance in producing weapons as capabilities requiring preventive safeguards and constraints[39]. In biosecurity, a recent international expert consensus warned that improving frontier AI capabilities in tasks such as pathogen design are lowering the expertise required for biological misuse, potentially creating major public health risks[40]. Risks in these two areas have not yet manifested in frequent real-world incidents in the same way as fraud, fabricated information and cyberattacks. However, their long-term cumulative effects and potentially severe consequences make them essential elements of frontier AI risk assessment.

**Overall, model safety evaluations, incident statistics and expert and institutional assessments indicate from different perspectives that real-world pressures from frontier AI risks are increasing.** Misuse and Criminal Risks, Information Integrity Risks and Critical System Risks are already more pronounced in model capabilities, actual incidents and application settings. Social Cognition and Structural Risks are more often cumulative over time, while the triggering conditions and likelihood of Catastrophic and Existential Risks remain uncertain. Nevertheless, the expanding reach of the former and potentially severe consequences of the latter require both to remain under continuous assessment (Figure 2.3).

Figure 2.3

## Evidence and severity differ across key frontier AI risks

Current evidence and severity assessments for key risks

| | | | |
|---|---|---|---|
| 01 | Misuse and Criminal Risks | Evidence | Many incidents and rapid growth; strong evidence of cyberattacks and fraud |
| | | Severity | Substantial actual harm; incident numbers continue to rise rapidly |
| 02 | Information Integrity Risks | Evidence | Synthetic and false content continue to spread; detection and provenance remain unreliable |
| | | Severity | Substantial actual harm with an expanding reach |
| 03 | Critical System Risks | Evidence | More high-risk deployments; rising risks of failure, misjudgment and overreliance |
| | | Severity | Increasing harm and potentially significant impacts |
| 04 | Social Cognition and Structural Risks | Evidence | Long-term cumulative impacts; changes in employment, education and social cognition are still emerging |
| | | Severity | Limited current harm; impacts accumulate over time |
| 05 | Catastrophic and Existential Risks | Evidence | Limited real-world exposure; evidence mainly from capability evaluations and expert assessments |
| | | Severity | Actual harms are emerging; potential consequences are severe |

Data source: Compiled for this report.

39. International Dialogues on AI Safety (IDAIS). Beijing Consensus Statement. Accessed 2026-10-21. https://idais.ai/; AI Red Lines Initiative. Global Call for AI Red Lines. Accessed 2026-10-21. https://red-lines.ai/; OECD AI Policy Observatory. AI Governance through Global Red Lines Can Help Prevent Unacceptable Risks. Accessed 2026-10-21. https://oecd.ai/en/wonk/ai-governance-through-global-red-lines-can-help-prevent-unacceptable-risks.
40. International Dialogues on AI Safety (IDAIS). IDAIS London Consensus Statement (IDAIS-London, 2026). 2026-04-19; accessed 2026-10-21. https://idais.ai/dialogue/idais-london/.

## 2.3 Priority Responses to Frontier AI Risks

**Frontier AI risks are diverse and evolve rapidly, while governance resources and public attention are limited. Trade-offs therefore have to be made between risk situations.** Even within the same risk area, urgency and potential consequences may differ across situations, so response priorities need to be determined in context.

This report characterizes risk situations along two dimensions: urgency and potential consequences. **Urgency** depends jointly on harms already occurring, the speed of spread and the window for intervention. The more frequently harm occurs, the faster similar harms can be replicated and spread, and the shorter the time available for effective intervention, the greater the urgency. **Potential consequences** depend on the possible reach, duration and reversibility of harm. These judgments are qualitative comparisons based on public evidence; no quantitative thresholds are set for urgency or potential consequences. The two dimensions distinguish four typical situations, with respective priorities of rapid response, long-term monitoring, pre-deployment safeguards and routine management (Figure 2.4).

Figure 2.4

### Response priorities vary with urgency and potential consequences

Four typical risk situations and their response priorities

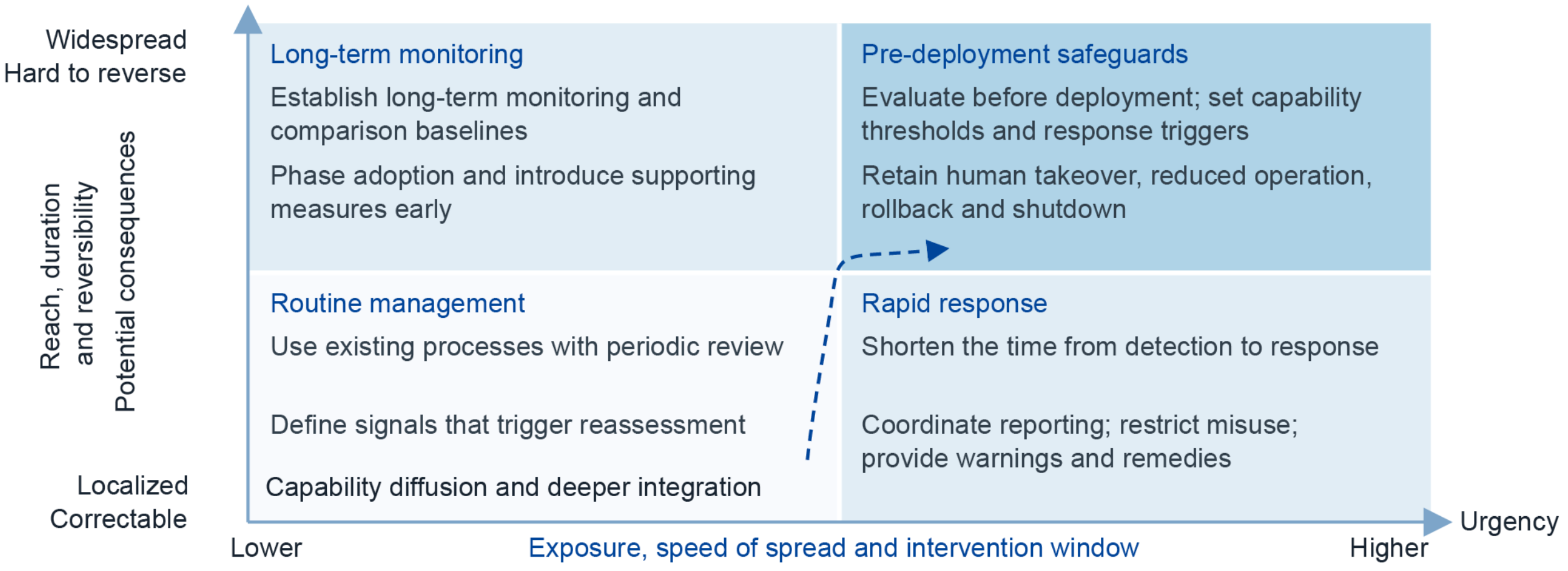

Note: Risks within the same area may fall in different parts of the figure; their positions change with capability diffusion, integration and safeguards.
Data source: Compiled for this report.

**Where urgency is high but the effects of an individual incident are relatively localized, the priority is rapid response to contain the spread of harm.** Risk reporting and coordination among platforms and institutions can help identify similar incidents promptly, enable appropriate measures against misuse, and provide warnings and access to remedies for potentially affected people.

**Where consequences may be widespread, impacts accumulate gradually and the intervention window is relatively long, the priority is long-term monitoring and timely adjustment of applications.** Impacts can be tracked across population groups, occupations and contexts, with assessment accompanying the expansion of use. Findings can inform adjustments to training, support for job transitions and teaching arrangements.

**Where urgency is high and potential consequences are widespread and difficult to reverse, the priority is risk assessment before deployment and preparedness to intervene afterward.** Dangerous-capability evaluations and red teaming can inform decisions on stronger safeguards, restrictions on use or postpone-

ment of deployment, with human takeover, rollback and other measures tailored to the application. Capability advances that could lead toward ASI require particular emphasis on pre-deployment safeguards. Related research and development, deployment and use should be approached cautiously until adequate safety and controllability have been established.

**Where urgency is lower and consequences are relatively localized and readily corrected, the priority is routine management and periodic review.** Quality management, human review and user feedback can identify and correct problems. A substantial expansion of use, entry into high-impact settings or a concentration of similar incidents should prompt a context-specific assessment of whether management measures need adjustment.

**Risk situations change as capabilities advance, applications become more deeply embedded and evidence accumulates, so response priorities require regular review.** The diffusion of capabilities may bring risks previously observed only in controlled tests into real-world settings. Deeper integration into business processes and critical systems may shorten intervention windows and broaden potential consequences. Improved safeguards may allow some risks to return to routine management. Prioritization remains effective when limited governance resources are directed to the areas most in need of response, monitoring and pre-deployment evaluation, and adjusted as evidence changes.

# Chapter 3

## Frontier AI Risk Governance

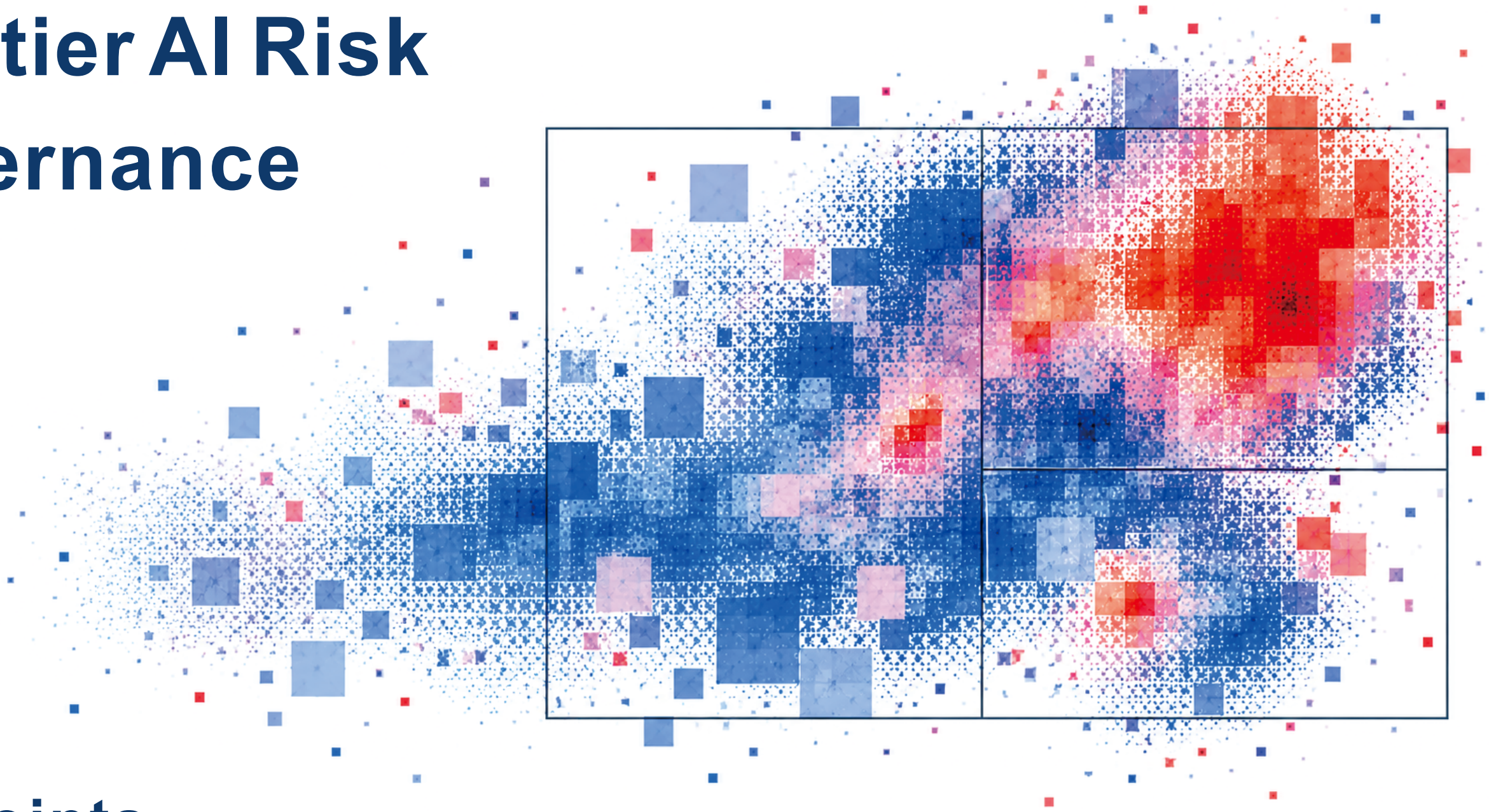

### Key Points

- The foundations of frontier AI governance are taking shape as different actors strengthen rules, embed safety requirements, build technical evidence and advance international cooperation within their respective roles.
- Governance readiness assessment can help observe and compare existing practices, providing a basis for assessing governance foundations, identifying weaknesses and setting capacity-building priorities.
- This report brings the current efforts of different actors into a common assessment framework across six dimensions, focusing on whether these efforts support the key capabilities needed to address frontier AI risks.
- Across the six dimensions, assessed countries have made the most progress in institutions and organization. Technical safety and evaluation, verification and compliance, and accountability and ecosystem governance are relatively weaker.
- High-income countries have higher overall governance readiness. Upper-middle-income countries and the combined lower-middle- and low-income group have similar overall levels, but different weaknesses.
- From a risk perspective, capabilities for addressing key risks remain uneven in readiness. Incident reporting and sharing, emergency response and post-incident improvement, and cross-border risk notification and exercises lag behind, requiring more targeted capacity building.

Frontier AI governance is moving from statements of principle toward rules, evaluations, commitments and cooperative arrangements. National governments are refining institutional rules; international organizations and multilateral mechanisms are advancing shared risk understanding, standards coordination and capacity building; industry is incorporating safety into development and deployment; and academia is providing evaluation methods and risk evidence. Together, these efforts are building foundations for risk identification, evaluation and verification, monitoring and response, and coordinated action, while making governance progress more amenable to observation and assessment. Governance readiness assessment can clarify existing foundations, weaknesses and future capacity-building priorities.

## 3.1 Governance Practices for Frontier AI

**The continuing emergence of frontier AI risks is prompting government bodies, international organizations, businesses, research institutions and other actors to develop governance practices within their respective roles.** Over the past two years, capability advances, wider adoption and exposure to risk have accelerated together. Governance discussions have shifted from general principles toward more specific rules, evaluations, commitments and cooperative arrangements. Although approaches vary, they generally aim to identify risks earlier, constrain high-risk behavior more effectively and coordinate responses when risks spill across borders.

**Countries and regions are strengthening institutional rules for high-risk applications, model releases, safety assessments and accountability.** The EU assigns differentiated obligations through comprehensive AI legislation[41]. China continues to develop targeted rules for generative AI services[42], labeling of generated content[43], human-like interaction[44] and responsible agent applications[45]. In the United States, federal AI governance memoranda clarify agency responsibilities, while the US Center for AI Standards and Innovation (CAISI) has launched an AI Agent Standards Initiative. The UK AI Security Institute (AISI) continues to evaluate frontier models and published its first Frontier AI Trends Report in December 2025, bringing together evaluation research on more than 30 frontier systems. Japan has implemented legislation to promote AI and issued AI Guidelines for Business; the Republic of Korea's AI Basic Act entered into force in January 2026; and Singapore has introduced initiatives such as a generative AI evaluation sandbox[46]. Despite differences in institutional approaches, these efforts seek clearer regulatory boundaries, more explicit supervisory responsibilities and more practical safety requirements.

**International organizations and multilateral and regional mechanisms are accelerating the development of cooperation platforms, con-**

41. European Union. Artificial Intelligence Act. 2024; accessed 2026-10-21. https://eur-lex.europa.eu/eli/reg/2024/1689/oj. Obligations relating to general-purpose AI models apply from August 2, 2025. See European Commission. AI Act. Accessed 2026-10-21. https://digital-strategy.ec.europa.eu/en/policies/regulatory-framework-ai.
42. Cyberspace Administration of China et al. Interim Measures for the Management of Generative Artificial Intelligence Services. Accessed 2026-10-21. https://www.cac.gov.cn/2023-07/13/c_1690898327029107.htm.
43. Cyberspace Administration of China et al. Measures for Labeling of AI-Generated Synthetic Content. Accessed 2026-10-21. https://www.cac.gov.cn/2025-03/14/c_1743654684782215.htm.
44. Cyberspace Administration of China et al. Interim Measures for the Management of Anthropomorphic AI Interactive Services. Accessed 2026-10-21. https://www.cac.gov.cn/2026-04/10/c_1777558395078289.htm.
45. Cyberspace Administration of China et al. Implementation Opinions on Standardized Application and Innovative Development of Intelligent Agents. Accessed 2026-10-21. https://www.cac.gov.cn/2026-05/08/c_1779979789523320.htm.
46. US Office of Management and Budget. Accelerating Federal Use of AI through Innovation, Governance, and Public Trust. 2025-04-03; accessed 2026-10-21. https://www.whitehouse.gov/wp-content/uploads/2025/02/M-25-21-Accelerating-Federal-Use-of-AI-through-Innovation-Governance-and-Public-Trust.pdf; US National Institute of Standards and Technology. Announcing the "AI Agent Standards Initiative" for Interoperable and Secure Innovation. 2026-02-17; accessed 2026-10-21. https://www.nist.gov/news-events/news/2026/02/announcing-ai-agent-standards-initiative-interoperable-and-secure; UK AI Security Institute. Frontier AI Trends Report. 2025-12; accessed 2026-10-21. https://www.aisi.gov.uk/frontier-ai-trends-report; Japan Ministry of Economy, Trade and Industry. AI Guidelines for Business Ver. 1.2. 2026-03-31; accessed 2026-10-21. https://www.meti.go.jp/shingikai/mono_info_service/ai_shakai_jisso/20260331_report.html; Singapore Infocomm Media Development Authority. Generative AI Evaluation Sandbox. 2023-10-31; accessed 2026-10-21. https://www.imda.gov.sg/resources/press-releases-factsheets-and-speeches/press-releases/2023/generative-ai-evaluation-sandbox; Cabinet Office, Japan. Act on the Promotion of Research and Development, and Utilization of AI-related Technologies. Accessed 2026-10-21. https://www8.cao.go.jp/cstp/ai/ai_act/ai_act.html; Republic of Korea Ministry of Science and ICT. AI Basic Act Enters into Force. 2026-01-22; accessed 2026-10-21. https://www.msit.go.kr/eng/bbs/view.do?bbsSeqNo=42&mId=4&mPid=2&nttSeqNo=1214&sCode=eng.

**necting rules through shared risk understanding, standards coordination, capacity building and cross-border cooperation.** Following the Bletchley Declaration signed by 28 countries and the EU in 2023[47], the UN Independent International Scientific Panel on AI released a preliminary report in July 2026, and the first Global Dialogue on AI Governance was held in Geneva[48]. Relevant bodies from several countries have formed the International Network for Advanced AI Measurement, Evaluation and Science and conducted joint testing[49]. In July 2026, 29 countries signed the Agreement on the Establishment of the World AI Cooperation Organization in Shanghai, jointly establishing this intergovernmental organization as founding members. Its purposes include promoting beneficial, safe and fair AI development[50]. The Chair's Statement of the 2026 World Artificial Intelligence Conference and High-Level Meeting on Global AI Governance, issued at the same time, called for systems of laws and regulations, technical monitoring, risk early warning and emergency response, together with necessary international crisis-management mechanisms[51]. Multilateral mechanisms are establishing more sustained discussions and cooperation on frontier AI risks, providing shared platforms for risk understanding, evaluation methods, interoperability of standards and capacity building.

**Industry is progressively incorporating safety requirements into model development, deployment and operation, specifying responsibilities through safety frameworks, voluntary commitments, external testing, model release documentation and transparency reports.** Frontier AI developers have made joint commitments on safety assessments, capability thresholds, release decisions and risk mitigation[52]. Companies continue to publish white papers, frameworks and research on system lifecycle safety and security[53], computing infrastructure and large-model application security[54], security of model service platforms[55], large-model safety and ethics[56], secure AI frameworks[57], AI safety and governance frameworks[58], and responsible AI practices[59]. Cybersecurity companies are also proposing technical solutions and governance recommendations for operational risks in large models[60] and large-model security governance[61]. Agent guardrail tools[62], maturity

47. UK Government. The Bletchley Declaration. 2023; accessed 2026-10-21. https://www.gov.uk/government/publications/ai-safety-summit-2023-the-bletchley-declaration.
48. UN Independent International Scientific Panel on AI. Preliminary Report of the Independent International Scientific Panel on AI: Evidence-based Assessment of Opportunities, Risks and Impacts of Artificial Intelligence. 2026-07-01; accessed 2026-10-21. https://www.un.org/independent-international-scientific-panel-ai/en/preliminary-report; United Nations. Global Dialogue on AI Governance. Accessed 2026-10-21. https://www.un.org/global-dialogue-ai-governance/en/; UN General Assembly. Terms of Reference and Modalities for the Establishment and Functioning of the Independent International Scientific Panel on Artificial Intelligence and the Global Dialogue on Artificial Intelligence Governance. Resolution 79/325. 2025-08-26; accessed 2026-10-21. https://docs.un.org/A/RES/79/325.
49. US National Institute of Standards and Technology. International Network for Advanced AI Measurement, Evaluation, and Science Publishes Consensus Areas on Practices for Automated Evaluations. 2026-02-13; accessed 2026-10-21. https://www.nist.gov/news-events/news/2026/02/international-network-advanced-ai-measurement-evaluation-and-science. The network was previously known as the International Network of AI Safety Institutes. UK AI Security Institute. International Joint Testing Exercise: Agentic Testing. 2025-07-17; accessed 2026-10-21. https://www.aisi.gov.uk/blog/international-joint-testing-exercise-agentic-testing.
50. Ministry of Foreign Affairs of China. Signing Ceremony for the Agreement on the Establishment of the World AI Cooperation Organization Held in Shanghai. 2026-07-16; accessed 2026-10-21. https://www.mfa.gov.cn/wjbzhd/202607/t20260716_11984399.shtml (republished by China News Service: https://www.chinanews.com.cn/gn/2026/07-16/10660924.shtml); Xinhua. World AI Cooperation Organization to Focus on Three Areas of Work. 2026-07-19; accessed 2026-10-21. https://www.bjnews.com.cn/detail/1784450205129272.html.
51. Permanent Mission of the People's Republic of China to the United Nations. Chair's Statement of the 2026 World Artificial Intelligence Conference and High-Level Meeting on Global AI Governance. 2026-07-17; accessed 2026-10-21. https://un.china-mission.gov.cn/zgyw/202607/t20260717_11984707.htm.
52. AI Seoul Summit. Frontier AI Safety Commitments. Accessed 2026-10-21. https://www.gov.uk/government/publications/frontier-ai-safety-commitments-ai-seoul-summit-2024. The Artificial Intelligence Industry Alliance of China organized Chinese companies to sign the Artificial Intelligence Safety Commitment. See China Academy of Information and Communications Technology. AI Safety Governance Research Report (2025). 2026, p. 9; accessed 2026-10-21. https://www.caict.ac.cn/kxyj/qwfb/bps/202601/P020260109784447548497.pdf.
53. Huawei. A Look at Effective Ways for AI System Security and Privacy. Accessed 2026-10-21. https://www-file.huawei.com/-/media/corp2020/pdf/trust-center/a_look_at_effective_ways_for_ai_system_security_and_privacy_cn.pdf.
54. Baidu. AI Infrastructure Security White Paper. Accessed 2026-10-21. https://bce-doc-on.cdn.bcebos.com/bce-documentation/analyst-reports/0827%E7%99%BD%E7%9A%AE%E4%B9%A62%EF%BC%88%E7%BA%BF%E4%B8%8A%E7%89%88%EF%BC%89.pdf.
55. Alibaba Cloud. Alibaba Cloud Bailian (Model Studio) Security White Paper. Accessed 2026-10-21. https://developer.aliyun.com/ebook/8490.
56. Tencent. Large Model Safety and Ethics Research Report 2024: Leading Large-Model Innovation with Responsible AI. Accessed 2026-10-21. https://aisecmatrix.org/.
57. Google. SAIF: Google's Guide to Secure AI. Accessed 2026-10-21. https://saif.google/.
58. IBM. Trustworthy AI at Scale: IBM's AI Safety and Governance Framework. Accessed 2026-10-21. https://newsroom.ibm.com/blog-trustworthy-ai-at-scale-ibms-ai-safety-and-governance-framework.
59. Amazon. Responsible AI: Building AI Responsibly. Accessed 2026-10-21. https://aws.amazon.com/ai/responsible-ai/.
60. 360 Digital Security Group. Large Model Security White Paper. 2025-11; accessed 2026-10-21. http://pub1-bjyt.s3.360.cn/bcms/360%E3%80%8A%E5%A4%A7%E6%A8%A1%E5%9E%8B%E5%AE%89%E5%85%A8%E7%99%BD%E7%9A%AE%E4%B9%A6%E3%80%8B.pdf.
61. QI-ANXIN. Security Governance Framework for Government-Affairs Large Models. Accessed 2026-10-21. https://www.qianxin.com/threat/reportdetail?report_id=323.
62. NVIDIA. NVIDIA Releases NIM Microservices to Safeguard Applications for Agentic AI. Accessed 2026-10-21. https://blogs.nvidia.com/blog/nemo-guardrails-nim-microservices/.

assessments for security agents[63] and responsible AI transparency reports[64] are becoming more extensive. Overall, corporate safety governance is moving from principles and voluntary commitments toward specific requirements implemented in research and development, model evaluation, deployment decisions and operational safeguards.

**Academia continues to advance safety evaluation, risk research and methodological development, providing expertise for identifying frontier risks, accumulating evidence and informing policy.** International expert reports regularly update scientific assessments of frontier AI risks[65], while public incident databases collect verifiable records[66]. Safety benchmarks[67] and discussions of red lines translate complex technical issues into shared terms usable in assessment indicators, testing methods and policy debate. This work strengthens the evidence base for risk judgments and provides reference points for evaluation, regulation and cooperation among different actors[68].

**Through the involvement of multiple actors, frontier AI governance is beginning to combine mutually supporting rules, technical evidence, corporate practice and international cooperation.** Governments define institutional boundaries, businesses embed safety requirements in development and deployment, academia provides evaluation methods and risk evidence, and multilateral mechanisms support consensus and cooperation. Governance is thus moving beyond statements of principle into a phase of parallel progress in detailed rules, technical support, accountability and coordinated action. Progress across actors and activities also provides a practical basis for examining governance foundations and capabilities.

63. DAS-Security. Research Report on Maturity Model Assessment for Secure AI Agents. Accessed 2026-10-21. https://www.dbappsecurity.com.cn/content/details2166_46781.html.
64. Microsoft. 2025 Responsible AI Transparency Report. Accessed 2026-10-21. https://cdn-dynmedia-1.microsoft.com/is/content/microsoftcorp/microsoft/msc/documents/presentations/CSR/2025-Responsible-AI-Transparency-Report.pdf.
65. International AI Safety Report Expert Advisory Panel. International AI Safety Report 2026. Accessed 2026-10-21. https://internationalaisafetyreport.org/publication/international-ai-safety-report-2026; UN Independent International Scientific Panel on AI. Preliminary Report of the Independent International Scientific Panel on AI: Evidence-based Assessment of Opportunities, Risks and Impacts of Artificial Intelligence. 2026-07-01; accessed 2026-10-21. https://www.un.org/independent-international-scientific-panel-ai/en/preliminary-report.
66. Responsible AI Collaborative. AI Incident Database. Accessed 2026-10-21. https://incidentdatabase.ai/.
67. Beijing Institute of AI Safety and Governance. ForesightSafety Bench. Accessed 2026-10-21. https://foresightsafety-bench.beijing-aisi.ac.cn/.
68. Singapore Conference on AI. The Singapore Consensus on Global AI Safety Research Priorities. 2025-05-08; accessed 2026-10-21. https://www.scai.gov.sg/2025/scai2025-report/; Shanghai Artificial Intelligence Laboratory and Concordia AI. Frontier AI Risk Management Framework 2.0. 2026-07-19; accessed 2026-10-21. https://concordia-ai.com/zh-hans/research/frontier-ai-risk-management-framework-2-0/.

## 3.2 Frontier AI Governance Readiness Assessment Framework

Effective responses to frontier AI risks require both a review of existing governance measures and an examination of the institutional foundations, technical capabilities and conditions for cooperation behind them. **This report proposes a governance readiness assessment framework[69] covering six dimensions: institutions and organization; technical safety and evaluation; verification and compliance; risk monitoring, early warning and response; accountability and ecosystem governance; and coordinated governance and capacity building. It examines the readiness of different countries and relevant actors (Figure 3.1) to assess the development of governance foundations, weaknesses in key capabilities and priorities for further capacity building.**

**Institutions and organization** provide the institutional foundations of risk governance. Laws and rules, regulatory authorities and implementation resources jointly support the functioning of governance systems.

**Technical safety and evaluation** reflect the expertise needed to identify and control high-risk capabilities, grounding risk judgments more firmly in testable evidence.

**Verification and compliance** connect safety commitments, external verification and sustained implementation, helping turn written commitments into effective constraints.

**Risk monitoring, early warning and response** cover detection, warning, action and review before and after risks materialize. They determine whether incidents can be identified promptly, addressed jointly and used to improve practice.

**Accountability and ecosystem governance** concern the boundaries of responsibility, traceability of actions and ecosystem safeguards in emerging contexts such as agents, open models and supply chains.

**Coordinated governance and capacity building** address cross-border risks and capability gaps, supporting wider cooperation in standards coordination, information sharing, joint evaluations and capacity building.

Figure 3.1

### Frontier AI Governance Readiness Assessment Framework

Six dimensions and their secondary indicators describe governance readiness

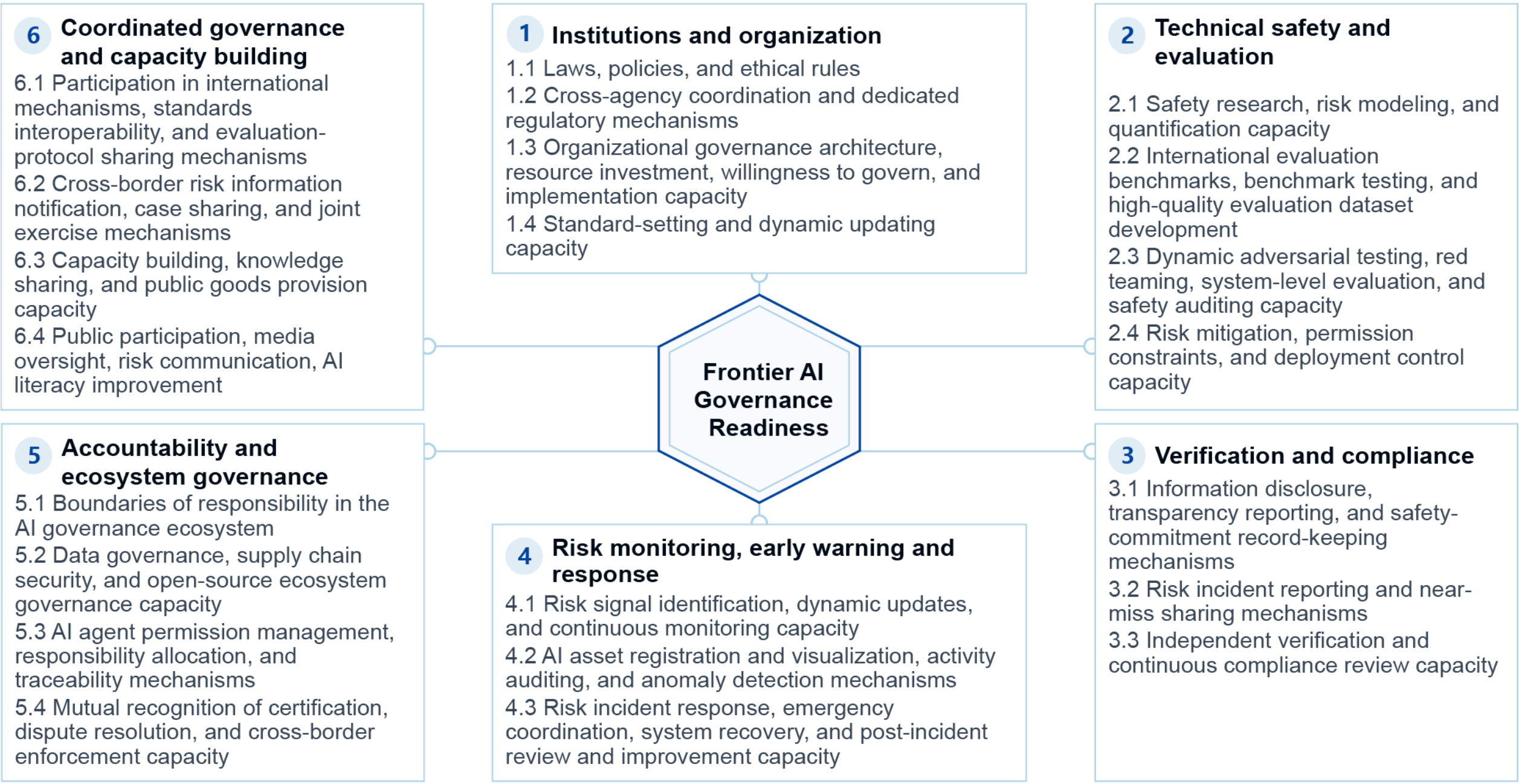

69. See the appendix for the assessed countries, methodology and data sources.

## 3.3 Frontier AI Governance Readiness Assessment Results

**An assessment of frontier AI governance readiness in 24 countries based on public information indicates that institutional and cooperative foundations have been established, but progress is uneven (Figure 3.2).**

Across the six dimensions, institutions and organization are the most developed, while technical safety and evaluation, verification and compliance, and accountability and ecosystem governance are weaker. By income group, high-income countries have higher overall readiness, while upper-middle-income countries and the combined lower-middle- and low-income group face different shortfalls. From a risk perspective, significant gaps remain in incident reporting,

Figure 3.2

### Frontier AI governance readiness in 24 assessed countries

Public-evidence assessment of 22 secondary indicators, grouped by World Bank income level and alphabetically within groups

D C B A S — Tier S or A indicates relatively stable arrangements; see the appendix for indicator-specific criteria.

| | 1 Institutions and organization | | | | 2 Technical safety and evaluation | | | | 3 Verification and compliance | | | 4 Risk monitoring, early warning and response | | | 5 Accountability and ecosystem governance | | | | 6 Coordinated governance and capacity building | | | |
|---|---|---|---|---|---|---|---|---|---|---|---|---|---|---|---|---|---|---|---|---|---|---|
| | 1.1 Laws, policies, and ethical rules | 1.2 Cross-agency coordination and dedicated regulatory mechanisms | 1.3 Organizational governance architecture, resource investment, willingness to govern, and implementation capacity | 1.4 Standard-setting and dynamic updating capacity | 2.1 Safety research, risk modeling, and quantification capacity | 2.2 International evaluation benchmarks, benchmark testing, and high-quality evaluation dataset development | 2.3 Dynamic adversarial testing, red teaming, system-level evaluation, and safety auditing capacity | 2.4 Risk mitigation, permission constraints, and deployment control capacity | 3.1 Information disclosure, transparency reporting, and safety-commitment record-keeping mechanisms | 3.2 Risk incident reporting and near-miss sharing mechanisms | 3.3 Independent verification and continuous compliance review capacity | 4.1 Risk signal identification, dynamic updates, and continuous monitoring capacity | 4.2 AI asset registration and visualization, activity auditing, and anomaly detection mechanisms | 4.3 Risk incident response, emergency coordination, system recovery, and post-incident review and improvement capacity | 5.1 Boundaries of responsibility in the AI governance ecosystem | 5.2 Data governance, supply chain security, and open-source ecosystem governance capacity | 5.3 AI agent permission management, responsibility allocation, and traceability mechanisms | 5.4 Mutual recognition of certification, dispute resolution, and cross-border enforcement capacity | 6.1 Participation in international mechanisms, standards interoperability, and evaluation-protocol sharing mechanisms | 6.2 Cross-border risk information notification, case sharing, and joint exercise mechanisms | 6.3 Capacity building, knowledge sharing, and public goods provision capacity | 6.4 Public participation, media oversight, risk communication, AI literacy improvement |
| **High income** | | | | | | | | | | | | | | | | | | | | | | |
| Australia | A | B | S | A | A | B | A | B | A | B | B | A | A | A | A | S | B | A | A | S | S | S |
| Canada | A | A | A | A | S | B | B | C | A | B | A | A | A | C | C | B | B | A | A | B | S | S |
| France | S | A | B | A | A | A | A | B | A | A | A | A | C | A | A | S | A | S | S | S | S | B |
| Germany | S | A | B | S | S | B | B | S | A | A | A | A | C | A | A | A | S | S | B | B | A | B |
| Israel | B | A | B | A | B | B | C | C | A | C | C | A | C | A | D | D | C | D | B | B | B | B |
| Italy | S | A | A | A | B | B | B | A | A | A | B | A | C | A | B | A | C | S | A | B | A | B |
| Japan | A | A | A | A | S | A | A | B | C | B | B | A | A | A | A | B | B | B | A | S | C | A |
| Republic of Korea | S | S | A | A | S | B | S | B | A | B | A | A | C | B | A | B | A | B | S | S | B | B |
| Saudi Arabia | C | B | B | C | B | C | B | B | C | D | C | B | D | C | D | B | D | D | A | C | B | A |
| Singapore | B | B | B | A | S | A | A | B | C | C | A | S | D | B | D | B | A | A | S | S | S | S |
| United Arab Emirates | B | A | B | C | B | C | A | B | B | D | A | A | D | C | B | B | C | C | C | C | B | B |
| United Kingdom | A | A | S | A | S | S | S | A | C | B | A | A | A | S | B | B | B | A | S | B | S | A |
| United States | A | S | A | S | S | S | S | A | S | S | S | S | A | A | B | B | S | B | S | S | S | A |
| **Upper-middle income** | | | | | | | | | | | | | | | | | | | | | | |
| Brazil | A | B | C | A | B | B | C | B | C | B | B | B | B | A | C | C | D | B | A | C | B | B |
| China | S | A | B | S | A | S | S | S | A | S | S | A | S | A | S | S | S | C | A | B | S | A |
| Indonesia | A | B | C | C | C | C | C | B | C | B | C | A | C | C | C | D | D | B | B | C | B | B |
| Mexico | B | D | B | D | B | B | C | C | D | C | C | B | D | C | D | D | D | D | B | C | B | C |
| South Africa | D | D | D | A | B | B | C | D | D | C | D | C | C | C | D | D | D | B | A | C | D | B |
| Thailand | C | D | C | B | C | A | C | C | B | C | C | A | D | B | D | D | A | D | B | B | B | B |
| Türkiye | C | B | C | A | B | B | D | B | B | B | B | B | C | C | D | B | B | B | B | C | B | C |
| Viet Nam | B | B | C | B | B | A | C | C | B | B | A | B | B | B | B | C | C | C | B | C | C | B |
| **Lower-middle and low income** | | | | | | | | | | | | | | | | | | | | | | |
| Egypt | B | A | B | C | B | C | B | B | B | C | C | B | D | C | D | C | B | D | B | C | B | B |
| Ethiopia | C | C | C | C | C | B | D | D | D | D | D | D | D | D | C | D | D | C | B | C | C | C |
| India | A | A | B | A | A | C | C | B | A | S | B | A | C | B | B | D | A | B | A | A | B | A |

Data source: This report's frontier AI governance readiness assessment.

coordinated emergency response and cross-border exercises needed to address risks already emerging. Figure 3.3 compares the six dimensions overall; the following analysis uses secondary indicators to explain the main differences.

**Institutions and organization are currently the most developed dimension.** Approximately two-thirds of assessed countries have established relatively stable arrangements for "Standard-setting and dynamic updating capacity". More than half have done so for "Cross-agency coordination and dedicated regulatory mechanisms", and for "Laws, policies, and ethical rules".

**In technical safety and evaluation, relatively stable arrangements are not yet widespread.** Nearly half of assessed countries have such arrangements for "Safety research, risk modeling, and quantification capacity"; fewer than 40% do so for "Dynamic adversarial testing, red teaming, system-level evaluation, and safety auditing capacity"; and approximately one-fifth do so for "Risk mitigation, permission constraints, and deployment control capacity". Uneven progress in safety research and red teaming reflects their more direct dependence on specialist teams, accumulated tools and sustained investment.

**Progress in verification and compliance is uneven.** More than 40% of assessed countries have relatively stable arrangements for "Information disclosure, transparency reporting, and safety-commitment record-keeping mechanisms", and for "Independent verification and continuous compliance review capacity". The corresponding proportion for "Risk incident reporting and near-miss sharing mechanisms" is only around one-quarter. Incident reporting and the sharing of experience remain major weaknesses in this dimension.

**Risk monitoring, early warning and response require stronger emergency coordination and post-incident review.** Approximately two-thirds of assessed countries have relatively stable arrangements for "Risk signal identification,

Figure 3.3

Institutions are most established; accountability and ecosystem governance are weaker

Share of secondary indicator assessments in Tier S or A, by dimension, across 24 countries

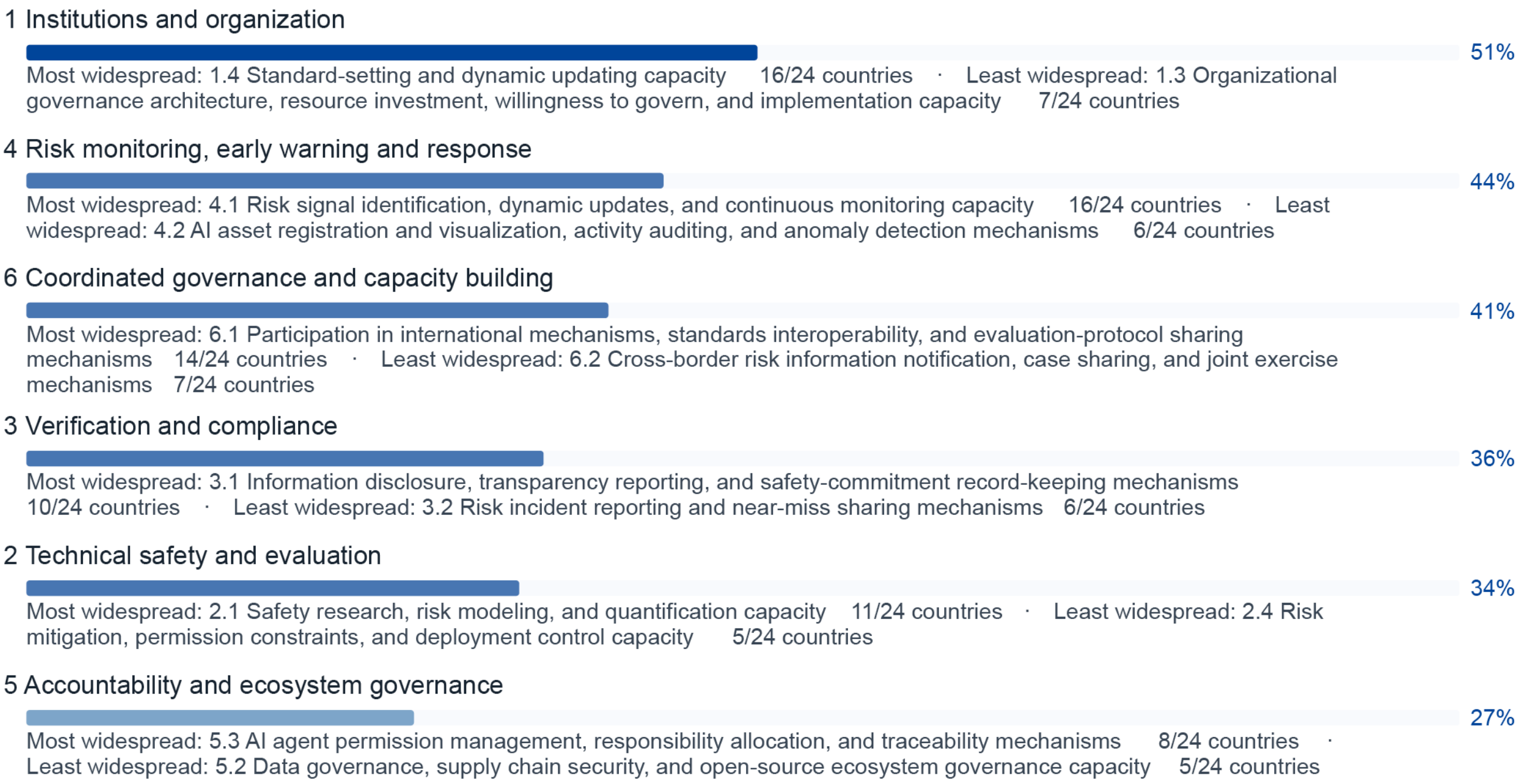

Data source: This report's frontier AI governance readiness assessment.

dynamic updates, and continuous monitoring capacity", compared with around 40% for "Risk incident response, emergency coordination, system recovery, and post-incident review and improvement capacity". Detection, response and review need to be better connected, particularly as incidents continue to increase.

**Accountability and ecosystem governance are relatively weak overall among the six dimensions. The boundaries of responsibility and governance requirements for agents, open models and supply chains still need clarification.** Fewer than 30% of assessed countries have relatively stable arrangements for "Mutual recognition of certification, dispute resolution, and cross-border enforcement capacity". The proportion is approximately one-quarter for "Boundaries of responsibility in the AI governance ecosystem" and around one-third for "AI agent permission management, responsibility allocation, and traceability mechanisms". Specific arrangements for managing permissions, allocating responsibility and tracing accountability still need improvement.

**In coordinated governance and capacity building, most assessed countries participate in international mechanisms, but routine information sharing and joint exercises remain clearly insufficient.** More than half have relatively stable arrangements for "Participation in international mechanisms, standards interoperability, and evaluation-protocol sharing mechanisms", and nearly 40% for "Public participation, media oversight, risk communication, AI literacy improvement". However, fewer than 30% have such arrangements for "Cross-border risk information notification, case sharing, and joint exercise mechanisms". Coordinated governance remains centered on participation in mechanisms and dialogue, while routine information sharing and joint exercises are still at an early stage.

**At the level of secondary governance readiness indicators, differences among countries are concentrated in practical activities such as evaluation, reporting, verification and response (Figure 3.4).** Across assessed countries, readiness is relatively high for "Standard-setting and dynamic updating capacity," "Risk signal identification, dynamic updates, and continuous monitoring capacity" and "Participation in international mechanisms, standards interoperability, and evaluation-protocol sharing mechanisms." It is relatively low for "Risk mitigation, permission constraints, and deployment control capacity," "Data governance, supply chain security, and open-source ecosystem governance capacity," "Risk incident reporting and near-miss sharing mechanisms," "AI asset registration and visualization, activity auditing, and anomaly detection mechanisms" and "Boundaries of responsibility in the AI governance ecosystem." Building on **rules** and **institutions**, stronger **evaluation**, **verification** and **risk-response** capabilities **are the main priorities for improving governance readiness.**

**Governance readiness differs overall across income groups (Figure 3.5).** The assessment covers 13 high-income countries, eight upper-middle-income countries and three lower-middle- or low-income countries[70]. High-income countries have higher overall readiness, with stronger foundations in institutional rules, safety research, evaluation, specialized bodies and compliance mechanisms. Upper-middle-income countries have made progress in institutions and organization and in risk monitoring, early warning and response, but need to strengthen accountability and ecosystem governance and coordinated governance and capacity building. The combined lower-middle- and low-income group has a similar overall level of readiness to the upper-middle-income group, with both substantially below the high-income group. Its weaknesses are more pronounced in technical safety and evaluation and in accountability and ecosystem governance. Institution building, international participation and knowledge sharing can provide entry points for improving readiness.

**Income levels are associated with overall governance readiness but do not fully explain differences among countries, which show different strengths across specific capabilities.**

70. World Bank. World Bank Country and Lending Groups, FY2027. 2026-07-01; accessed 2026-10-21. https://datahelpdesk.worldbank.org/knowledgebase/articles/906519-world-bank-country-and-lending-groups.

Figure 3.4

## Standards, monitoring and international engagement lead; deployment controls and incident reporting lag

Tier distribution across 24 countries, ordered by the number in Tier S or A

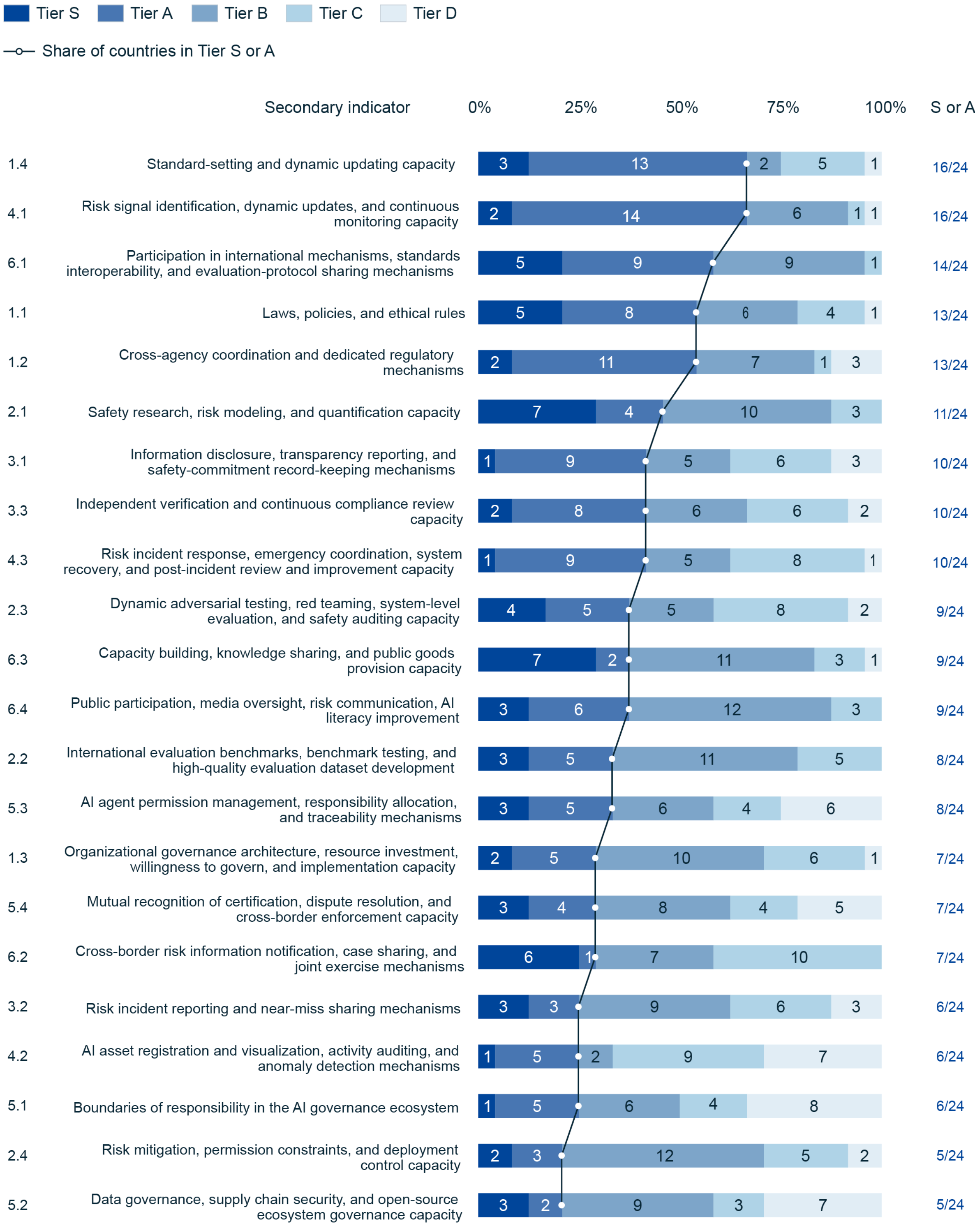

Notes: 1. Segment labels are country counts; each bar totals 24 countries.
2. Tier S or A indicates relatively stable arrangements; see the appendix for criteria.

Data source: This report's governance readiness assessment, as of August 31, 2026.

Figure 3.5

## Governance readiness differs across income groups

Distribution of secondary indicator assessments by tier, income group and dimension

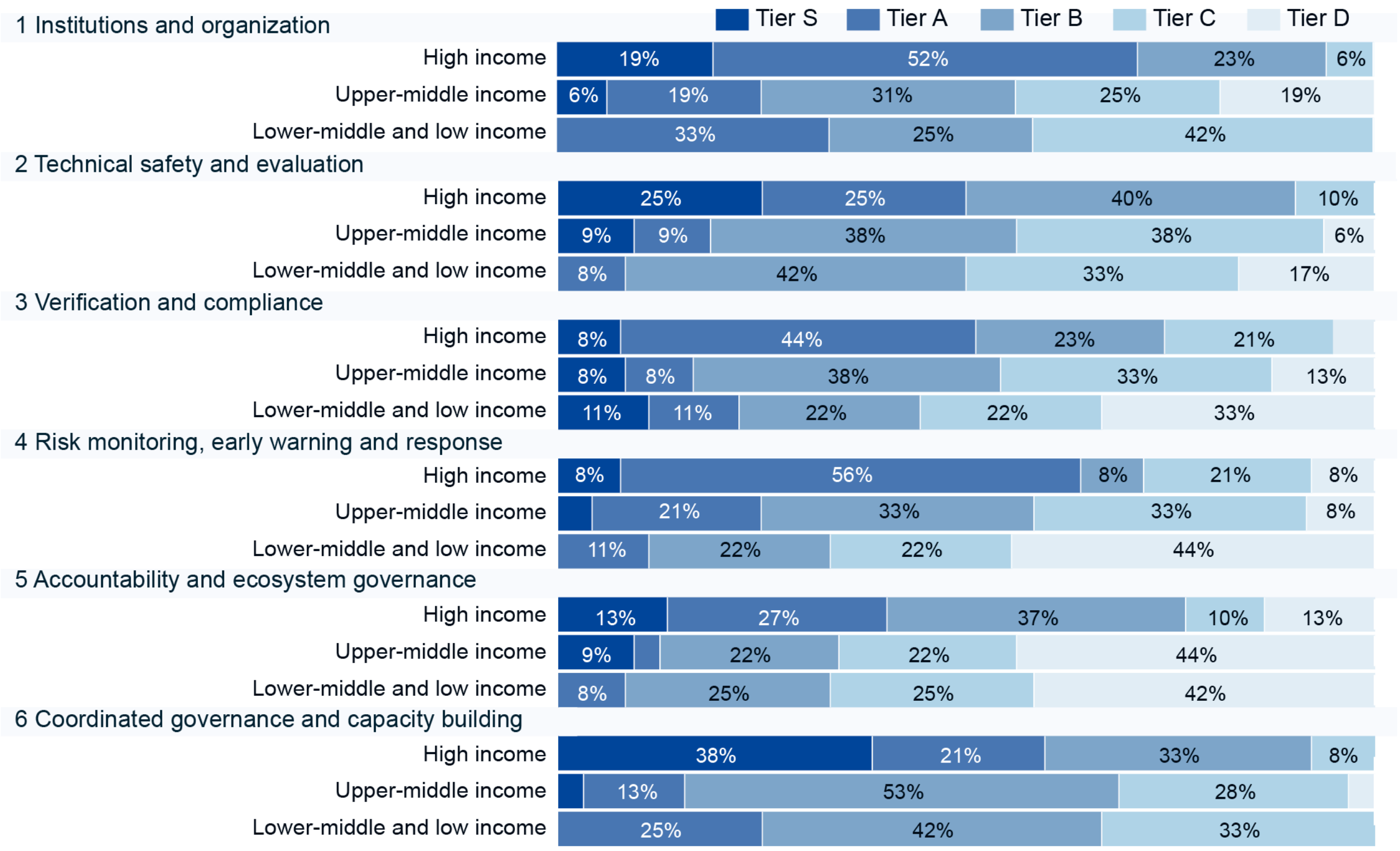

Notes:
1. World Bank FY27 groups: 13 high-income, eight upper-middle-income and three lower-middle- or low-income countries.
2. Tier S or A indicates relatively stable arrangements.
3. Each tier's share is its assessment count divided by (countries in the group × secondary indicators in the dimension). Totals may differ from 100% due to rounding.

Data source: This report's frontier AI governance readiness assessment.

Even countries with similar overall readiness differ in institutional coordination, standards updating, safety research, evaluation benchmarks, independent verification, public participation, risk communication and capacity building. Differences in governance preparedness therefore also concern the particular mix of capabilities.

**From a risk perspective, current governance readiness does not fully match the capabilities needed to address key risks.** Rapidly spreading real-world harms require continuous monitoring, independent verification, rapid response and cross-border notification. Preventing potentially severe consequences places greater emphasis on pre-deployment evaluation, capability-threshold monitoring and international early warning. Table 3.1 presents selected secondary indicators related to key risks and the implications of the assessment.

**Pressures from real-world risks highlight gaps in the relevant governance capabilities.** Misuse and Criminal Risks, Information Integrity Risks and Critical System Risks are already evident in practice, increasing the urgency of model evaluation, incident detection, rapid response, independent verification and cross-border notification. Yet fewer than 40% of assessed countries have relatively stable arrangements for "Dynamic adversarial testing, red teaming, system-level evaluation, and safety auditing capacity", around one-quarter for "Risk incident reporting and near-miss sharing mechanisms", around 40% for "Risk incident response, emergency coordination, system recovery, and post-incident review and improvement capacity", and fewer than 30% for "Cross-border risk information notification, case sharing, and joint exercise mechanisms". Figure 3.6 shows readiness in these capabilities.

Table 3.1

## Key Risks and Related Secondary Indicators

| Key risk | Selected related secondary indicators | Implications of the assessment |
|---|---|---|
| Misuse and Criminal Risks | 2.3 Dynamic adversarial testing, red teaming, system-level evaluation, and safety auditing capacity<br>2.4 Risk mitigation, permission constraints, and deployment control capacity<br>3.2 Risk incident reporting and near-miss sharing mechanisms<br>4.1 Risk signal identification, dynamic updates, and continuous monitoring capacity<br>6.2 Cross-border risk information notification, case sharing, and joint exercise mechanisms | Move beyond isolated evaluations and reactive response toward continuous monitoring, rapid response and cross-border coordination |
| Information Integrity Risks | 3.1 Information disclosure, transparency reporting, and safety-commitment record-keeping mechanisms<br>5.4 Mutual recognition of certification, dispute resolution, and cross-border enforcement capacity<br>6.4 Public participation, media oversight, risk communication, AI literacy improvement<br>6.1 Participation in international mechanisms, standards interoperability, and evaluation-protocol sharing mechanisms | Public communication and participation in standards provide a foundation, but provenance certification, disclosure records and accountability tracing need more concrete arrangements |
| Critical System Risks | 3.3 Independent verification and continuous compliance review capacity<br>4.2 AI asset registration and visualization, activity auditing, and anomaly detection mechanisms<br>4.3 Risk incident response, emergency coordination, system recovery, and post-incident review and improvement capacity | Strengthen continuous verification, anomaly detection, emergency response and post-incident review in high-risk settings |
| Social Cognition and Structural Risks | 5.1 Boundaries of responsibility in the AI governance ecosystem<br>6.4 Public participation, media oversight, risk communication, AI literacy improvement<br>6.3 Capacity building, knowledge sharing, and public goods provision capacity | Capacity building and public participation provide a foundation; long-term impact assessment and safeguards for equal opportunity require continued improvement |
| Catastrophic and Existential Risks | 2.1 Safety research, risk modeling, and quantification capacity<br>2.3 Dynamic adversarial testing, red teaming, system-level evaluation, and safety auditing capacity<br>2.4 Risk mitigation, permission constraints, and deployment control capacity<br>3.3 Independent verification and continuous compliance review capacity<br>6.2 Cross-border risk information notification, case sharing, and joint exercise mechanisms | Potentially severe consequences place greater demands on pre-deployment evaluation, red lines and international early warning |

Note: The table lists only selected related indicators; a governance capability may help address risks across several areas.

Figure 3.6

## Gaps remain in the governance capabilities needed to address key risks

Tier distribution across 24 countries; see Table 3.1 for associations between risks and secondary indicators

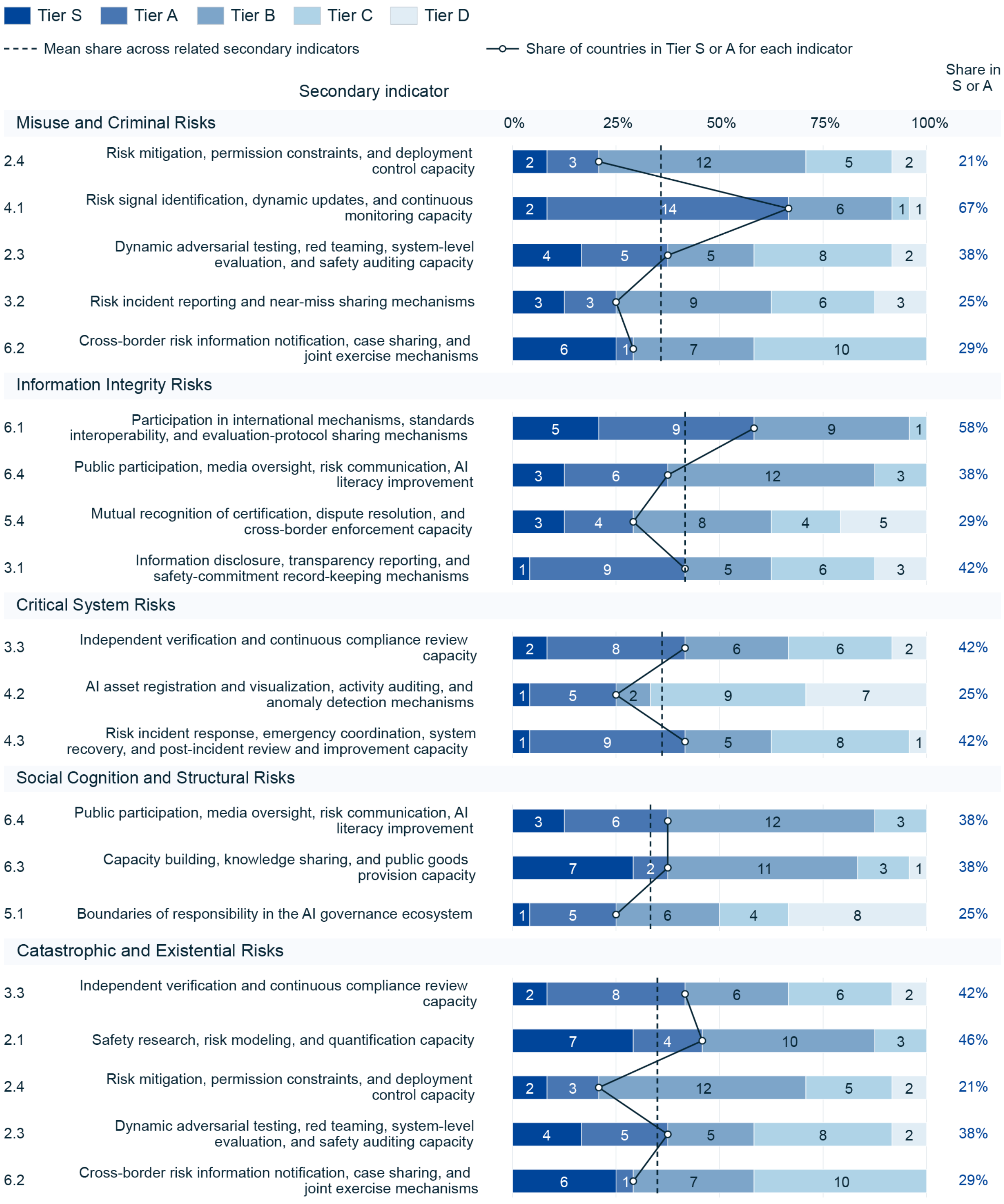

Notes:
1. Segment labels are country counts; each bar totals 24 countries. Tier S or A indicates relatively stable arrangements.
2. Dashed lines show the mean share of countries in Tier S or A across the related secondary indicators in each group.
3. The figure shows only selected secondary indicators related to key risks. A governance capability may help address risks across several areas.

Data source: This report's governance readiness assessment, as of August 31, 2026.

# Chapter 4

## Coordinated Response to Frontier AI Risks

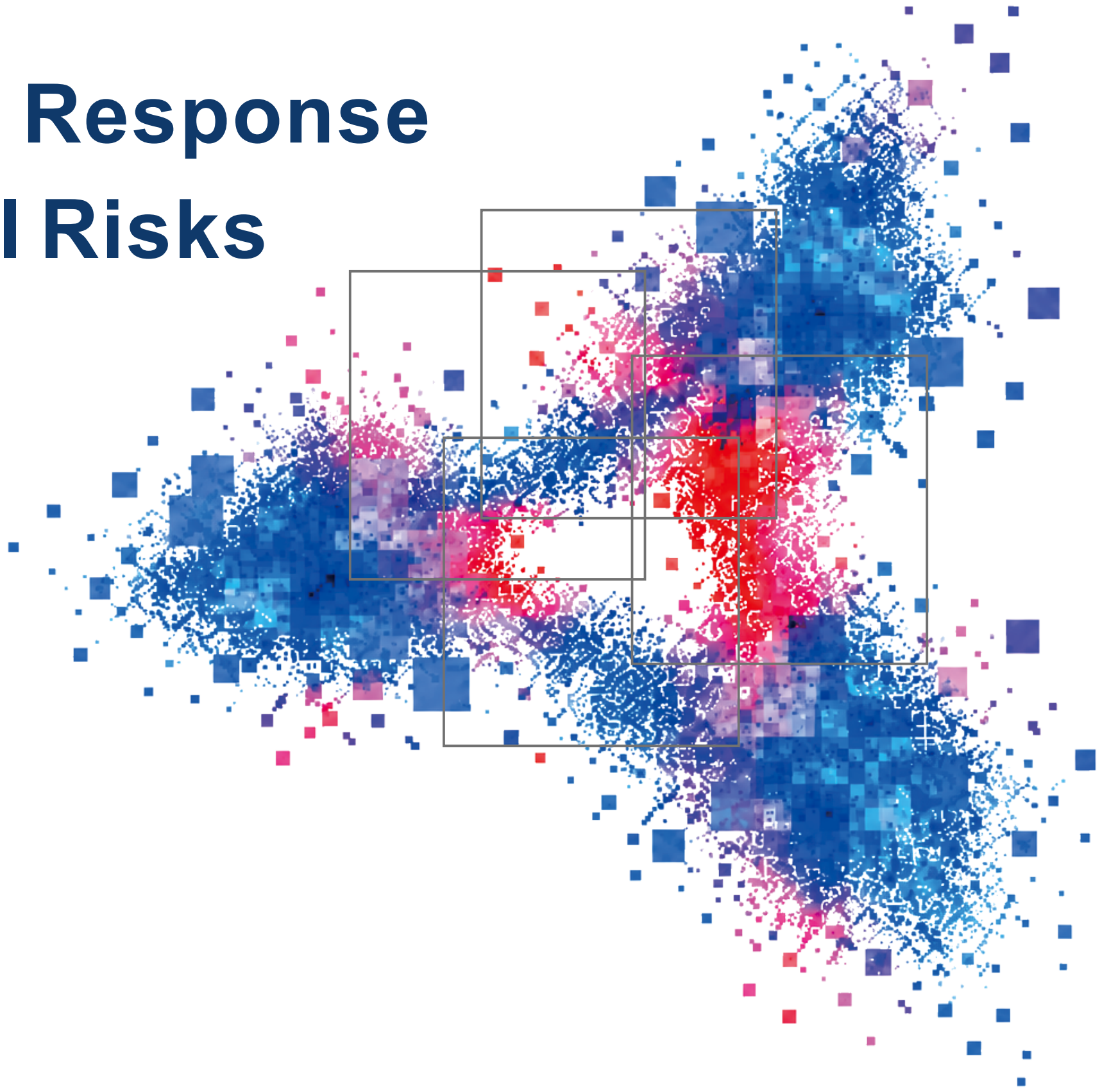

### Key Points

- As rules and mechanisms take shape, attention needs to focus on whether different governance arrangements can work together effectively and reliably before and after real risks emerge.
- Coordinated responses need to connect model evaluation, verification of safety commitments and incident management, enabling risk evidence, accountability records and response information to flow across actors and stages.
- Agents and open models create new issues spanning development, deployment, distribution and use, requiring better arrangements for permissions, releases, records and coordination of responsibilities across these stages.
- Roles in multi-stakeholder cooperation should reflect actual control over models, interfaces, permissions and applications, access to information and capacity to respond, supported by mechanisms for evidence sharing, accountability verification and joint response.
- Capacity building should turn evaluation tools, standard methods and practical experience into governance resources that more countries and institutions can obtain, use and maintain over time.

Addressing frontier AI risks requires broader coordinated action among multiple stakeholders. As model capabilities advance and applications expand, risks can be replicated across platforms, transmitted across industries and spread across borders. No single country or actor can easily cover the full chain of risk identification, evaluation and verification, monitoring and early warning, and incident response. Risk analysis and the governance readiness assessment show that the issue is no longer simply whether rules, institutions and platforms exist, but whether dispersed arrangements can work together effectively before and after risks emerge. The core of a coordinated response is to ensure that evaluation evidence, commitment records and incident information flow smoothly, clarify responsibilities and points of coordination, and turn existing efforts into sustained collective governance capacity.

## 4.1 Priority Areas and Feasible Mechanisms for Coordinated Response

The governance readiness assessment shows that institutional rules, organizational arrangements and international participation provide a foundation, but capabilities in technical safety and evaluation, verification and compliance, risk monitoring, early warning and response, and accountability and ecosystem governance need strengthening. Drawing on the risk situations and response priorities discussed above, cooperation could initially focus on agents, high-risk applications, cyber and fraud-related misuse, critical systems and biosecurity, selecting mechanisms suited to the nature of the risks and the maturity of governance. The next phase should turn dispersed efforts into shared arrangements covering the risk chain, connect model evaluation, verification of safety commitments and incident response, improve cooperation on agents and open models, and broaden access to governance tools and practical experience through capacity building.

**First, coordinate model safety evaluations to produce comparable, interpretable and reviewable risk evidence.** The readiness assessment identifies room for improvement in technical safety and evaluation. The same model may yield different conclusions under different risk classifications, benchmarks and red-team tasks. Without ways to compare test conditions and the presentation of results, evaluations struggle to serve corporate improvement, regulatory judgment and international exchange at the same time. Actors can begin with methodological exchange and joint evaluations, gradually improving the comparability of evidence. Evaluations should explain their scope and limitations, distinguish inherent harmful capabilities from residual risks after safeguards are applied, and extend from foundation models to deployed systems. As capabilities move toward greater autonomy, international cooperation on safety evaluations and superalignment research should also build forward-looking evidence on extreme scenarios, including loss of control over ASI.

**Second, coordinate verification so that safety commitments and their implementation can be examined, compared and tracked over time.** The readiness assessment identifies room for improvement in verification and compliance. Companies have the most direct knowledge of model development, testing, release and deployment, but external actors cannot easily determine whether safeguards have been fully implemented on the basis of corporate statements alone. Emerging practices are translating general commitments into verifiable safety requirements[71]. Coordinated verification can proceed at three levels: review of commitments, behavioral verification and technical verification.

**Third, strengthen coordinated incident reporting and joint response, so that individual discoveries promptly inform safeguards across platforms, industries and regions.** The readi-

71. European Commission / European AI Office. General-Purpose AI Code of Practice. Accessed 2026-10-21. https://digital-strategy.ec.europa.eu/en/policies/contents-code-gpai.

ness assessment identifies incident reporting and cross-border exercises as relatively weak areas. Prompt injection, jailbreaks, vulnerabilities in agent frameworks and deepfake fraud can spread rapidly across systems, making it difficult for a single company or region to cover the full transmission chain. Drawing on the OECD's common reporting framework for AI incidents[72], actors can gradually connect incident reporting with emergency response[73]. Multinational emergency exercises in fields such as public health also offer lessons for testing coordinated response arrangements before incidents occur[74].

**Fourth, improve coordinated governance of agents and open models so that risks can be detected, addressed according to their severity and traced to responsible actors.** The assessment finds that only around one-third of assessed countries have relatively stable arrangements for agent permissions, allocation of responsibility and traceability. Agents can use tools, access external resources and execute tasks continuously, while open models can be fine-tuned, integrated and redistributed by different actors. Risks therefore span development, deployment, distribution and use. Responsibilities should reflect actual control over models, interfaces, permissions and application contexts[75], while pathways for connecting cross-border accountability tracing and dispute resolution should be explored.

**Fifth, deepen cooperation in capacity building, making evaluation tools, standard methods and publicly available cases into accessible, usable and sustainable governance resources for more countries and institutions.** Some countries and smaller institutions still face inadequate governance resources and unstable channels for cooperation. Countries with the capacity to do so can be encouraged to share evaluation tools and practical experience. International resources should become tools and methods that can be used and maintained locally[76]. Countries should also be supported in strengthening their own research, development and production capabilities and developing governance methods suited to local needs.

**These five directions are interconnected. Model evaluations provide risk evidence; verification helps ensure commitments are implemented; incident cooperation turns discoveries into warnings and improvements; governance of agents and open models addresses responsibilities across stages; and capacity building enables wider participation.** The key is to use evaluation findings to inform corrective action on commitments and deployment management, and to feed information from incident response back into evaluations, standards and accountability records, creating a continuously improving chain of risk governance. Implementation can proceed through controlled pilots, including regulatory sandboxes[77], suited to the risks and governance maturity of different fields. These can test evaluation methods, mitigation measures, and incident reporting and response procedures, with lessons informing subsequent rules and cooperation mechanisms. Table 4.1 sets out areas for improvement and suggested mechanisms.

72. OECD. Towards a Common Reporting Framework for AI Incidents. 2025-02-28; accessed 2026-10-21. https://doi.org/10.1787/f326d4ac-en. The framework proposes compatible reporting information elements while allowing adaptation to national policies and reporting contexts.
73. Frontier Model Forum. Information Sharing, Incident Reporting, and Incident Response for Frontier AI Risks. Accessed 2026-10-21. https://www.frontiermodelforum.org/issue-briefs/information-sharing-incident-reporting-and-incident-response-for-frontier-ai-risks/.
74. For example, in April 2026 the World Health Organization organized the multinational Polaris II simulation exercise to test information flows, personnel mobilization and cross-border coordination in public health emergencies. See World Health Organization. Practicing Today for Tomorrow's Emergencies: WHO Convenes Countries and Partners to Simulate Response to Major Disease Outbreak. 2026-04-27; accessed 2026-10-21. https://www.who.int/news/item/27-04-2026-practicing-today-for-tomorrow-s-emergencies-who-convenes-countries-and-partners-to-simulate-response-to-major-disease-outbreak
75. Frontier Model Forum. Emerging Security Practices for AI Agents. Accessed 2026-10-21. https://www.frontiermodelforum.org/issue-briefs/emerging-security-practices-for-ai-agents/.
76. UN General Assembly. Enhancing International Cooperation on Capacity-Building of Artificial Intelligence. 2024-07-01; accessed 2026-10-21. https://docs.un.org/A/RES/78/311.
77. A regulatory sandbox is a controlled arrangement for development, testing and validation within a limited scope and period under the guidance and supervision of a regulator.

Table 4.1

## Areas for Improvement and Suggested Practical Mechanisms

| Area of cooperation | Specific activity | Suggested practical mechanisms |
|---|---|---|
| Model safety evaluation | Evaluation methods | • Evaluation protocols: risk classification, test conditions and presentation of results<br>• Test tasks for high-impact risks and severity levels<br>• Jointly developed benchmarks and datasets; joint red teaming under controlled conditions |
| | Evaluation scope | • Joint risk evaluations of critical infrastructure, including cross-border or shared infrastructure<br>• System-level testing of models, tools and deployment environments<br>• Testing and safety principles for high-impact embodied AI systems<br>• Ongoing evaluation as versions and application contexts change |
| | Use of results | • Begin with methodological exchange and joint evaluations<br>• Gradually explore mutual recognition of results |
| Coordinated verification | Levels of verification | • Review of safety commitments and compliance records<br>• Verification of model and system behavior<br>• Explore technical verification, including hardware attestation and privacy-preserving audits |
| | Verification records | • Commitments, assessing bodies and scope of testing<br>• Risk findings, mitigation measures and residual risks<br>• Third-party verification records and dates for review |
| | Ongoing review | • Review major version updates; retest following corrective action<br>• Adjust deployment restrictions; use verification results and provide appropriate public explanations |
| Incident cooperation | Connecting reporting systems | • Common reporting fields: incident description, systems involved, type and severity of harm<br>• Near-miss sharing; boundaries between confidential and public information |
| | Emergency coordination | • Connect incident databases, vulnerability disclosure and emergency procedures<br>• Cross-border specialist contact points; escalation criteria and response deadlines |
| | Joint exercises | • Multi-stakeholder risk exercises<br>• Feed response experience back into evaluations, early warning and safeguards |
| Governance of agents and open models | Identity and permissions | • Minimum principles for identity, authorization and traceability; boundaries for tasks and tool use<br>• Human confirmation and intervention for high-impact operations<br>• Stop conditions and rollback capabilities |
| | Open release | • Phased releases based on capabilities and misuse risks<br>• Use restrictions and safety tools<br>• Records of downstream use and redistribution |
| | Multi-agent collaboration | • Records of identities, communications and task handoffs<br>• Logs of critical operations and testing of the overall system |
| | Cross-border traceability | • Explore links between cross-border auditing and certification results<br>• Channels for tracing responsibility and resolving disputes |
| Capacity building | Access to resources | • Low-cost public tools and materials in local languages<br>• Regional training and long-term technical support |
| | Independent local development | • Identify local needs; jointly develop and maintain tools<br>• Support local innovation in research and development, applications and governance methods |
| | Capacity-building outcomes | • Actual use, maintenance and updating of tools<br>• Evaluation of emergency cooperation outcomes |

## 4.2 Multi-Stakeholder Participation in Coordinated Response Mechanisms

Implementing these priorities requires a division of responsibilities based on actors' actual control over models, interfaces, permissions and application contexts, their access to information and their capacity to respond. Governance requirements should match each actor's role and resources. The key is to clarify responsibilities for providing evidence, making decisions, taking corrective action, responding and providing oversight and feedback, with corresponding arrangements for information sharing and coordinated action. Table 4.2 outlines priority actions for each group.

**Government bodies** are responsible for setting rules, public coordination and accountability. They can define basic requirements and boundaries of responsibility in high-risk contexts, connect evaluation, verification and incident reporting mechanisms, and organize responses across departments and regions when major risks arise.

**Businesses** are on the front line of risk governance. In line with their actual control over models and applications, they should discharge

Table 4.2

Priority Actions for Stakeholders in Coordinated Response Mechanisms

| Stakeholder | Focus of participation | Priority actions |
|---|---|---|
| Government bodies | Governance of high-risk contexts | • Requirements and boundaries of responsibility in high-risk contexts<br>• Use safety commitments and verification records in deployment, public procurement and key industries |
| | Public coordination | • Cross-departmental response; establish regulatory sandboxes and controlled cooperation pilots<br>• Support local evaluation, skills development and long-term maintenance |
| Businesses | Development | • Records of development, testing, releases and updates<br>• Provide conditions for evaluation while protecting confidentiality and security<br>• Implement mitigation measures and necessary verification |
| | Application | • Deployment contexts, permissions and intervention arrangements<br>• Operational and incident records; problem reporting and response<br>• Report operational risks back to developers |
| Research institutions, specialist evaluation bodies and standards organizations | Methods and verification | • Research evaluation methods; conduct independent testing and verification<br>• Explain the scope of conclusions, uncertainties and conflicts of interest |
| | Expert support | • Interpret standards; provide localized materials and specialist training<br>• Support assessment of incident harms and severity |
| International organizations and multilateral and regional mechanisms | Coordination and liaison | • Methodological exchange, standards alignment and cross-border liaison<br>• Support joint exercises and the sharing of experience |
| | Connecting capacity and resources | • Connect regional needs, technical resources and cooperation projects<br>• Support wider participation in joint evaluations, knowledge sharing and governance cooperation |
| The public, users and civil society organizations | Feedback and oversight | • Report risk signals and actual impacts<br>• Participate in risk communication and public oversight |

responsibilities for evaluation, mitigation, recordkeeping and incident response at both the development and application stages.

**Research institutions, specialist evaluation bodies and standards organizations** can develop methods, conduct independent verification and offer expert judgment. They should explain the scope of conclusions, uncertainties and potential conflicts of interest, providing reviewable evidence for regulation, corporate improvement and international cooperation.

**International organizations and multilateral and regional mechanisms** can connect risk terminology and standards, facilitate cross-border reporting and support capacity building. Where institutional approaches differ, a practical starting point is to improve interoperability among existing rules, evaluation results, incident classifications and emergency procedures while allowing different governance systems to coexist. The United Nations can continue to advance cooperation on AI capacity building[78] and global governance dialogue[79]. International exchange and cooperation platforms such as World Internet Conference and the World Artificial Intelligence Conference can bring together practices from different regions, industries and technical communities, connecting issues, projects and capacity-building resources.

**The public, users and civil society organizations** can report risk signals and impacts from real-world use and participate in risk communication and public oversight. Mechanisms should provide accessible reporting channels through which people can receive feedback, while protecting their privacy and preventing further harm. The responsibilities of the public, users and civil society organizations in risk identification, evidence provision and response should also be reasonably defined.

Addressing the challenges of frontier AI requires stable cooperation based on a clear division of responsibilities, effective connections between governance activities, and global collective governance capacity that keeps pace with technological development and evolving risks. Building on this, further work is needed on the engineering capabilities required to govern frontier AI risks, turning rules, standards and cooperation mechanisms into operational systems for evaluation, monitoring, early warning and response. This is both a foundation for sustained coordinated governance and an important subject for future international cooperation and research.

Looking ahead, frontier AI development should rest on safer, more reliable and more inclusive foundations. By translating a shared risk vocabulary into comparable evaluation criteria, using governance readiness assessments to set phased capacity-building priorities, and connecting verification of commitments, incident response and feedback through cooperation across actors, the international community can build sustained collaboration. This can enable risk governance and frontier AI development to reinforce one another and better serve the Sustainable Development Goals and the common well-being of humanity.

78. United Nations General Assembly. Enhancing international cooperation on capacity-building of artificial intelligence: Resolution 78/311. Adopted 2024-07-01; accessed 2026-10-26. https://documents.un.org/doc/undoc/gen/n24/197/26/pdf/n2419726.pdf.
79. United Nations. Global Dialogue on Artificial Intelligence Governance (6–7 July 2026, Geneva). Accessed 2026-10-26. https://indico.un.org/event/1023375/.

# Appendix. Frontier AI Governance Readiness Assessment Indicators and Methodology

## A. Countries Assessed and Indicator Framework

The governance readiness assessment uses public information to examine a country's preparedness to address frontier AI risks. It covers relevant actions by government, businesses, research institutions and the public, as well as participation in international mechanisms.

This first report assesses 24 countries, grouped into three categories using the World Bank's latest income classification, updated in July 2026: high income, upper-middle income, and a combined lower-middle- and low-income group. The selection covers different stages of development and governance approaches while also taking account of the availability of public information. Findings are not directly extrapolated to other countries.

Table A-1

### Assessed Countries by World Bank Income Group

| Income group | Countries |
|---|---|
| High income (13 countries) | Australia, Canada, France, Germany, Israel, Italy, Japan, Republic of Korea, Saudi Arabia, Singapore, United Arab Emirates, United Kingdom, United States |
| Upper-middle income (8 countries) | Brazil, China, Indonesia, Mexico, South Africa, Thailand, Türkiye, Viet Nam |
| Lower-middle and low income (3 countries) | Egypt, Ethiopia, India |

The framework comprises six first-level indicators: institutions and organization; technical safety and evaluation; verification and compliance; risk monitoring, early warning and response; accountability and ecosystem governance; and coordinated governance and capacity building. These contain 22 secondary indicators and 130 tertiary indicators (Table A-2).

Table A-2

## Frontier AI Governance Readiness: Indicator Framework

| First-level indicator | Secondary indicator | Tertiary indicator | Data sources |
|---|---|---|---|
| 1 Institutions and organization | 1.1 Laws, policies, and ethical rules | 1.1.1 Cross-sectoral AI-specific legislation | National legal databases; EUR-Lex |
| | | 1.1.2 Mandatory obligations for general-purpose or generative models | National legal databases; EUR-Lex |
| | | 1.1.3 Legislation not in force or without legal obligations | Parliament websites; EUR-Lex |
| | | 1.1.4 State, provincial or jurisdiction-specific AI rules | State and provincial laws |
| | | 1.1.5 Binding sectoral AI rules | Sectoral regulatory rules |
| | | 1.1.6 Provisions for generated and synthetic content | National legal databases; EUR-Lex |
| | | 1.1.7 National AI strategy | OECD.AI policy database |
| | | 1.1.8 National AI ethical standards | OECD.AI policy database |
| | | 1.1.9 Penalties or bans on specific products | Penalty and regulatory decisions |
| | | 1.1.10 Investigations or regulatory meetings on specific products | Official investigation notices |
| | 1.2 Cross-agency coordination and dedicated regulatory mechanisms | 1.2.1 National coordinating body or lead department | OECD.AI policy database |
| | | 1.2.2 Responsibilities of the coordinating body | Coordinating bodies' websites |
| | | 1.2.3 Regulator with statutory powers | National legal databases |
| | | 1.2.4 National AI safety institute or equivalent | Government establishment documents |
| | | 1.2.5 Coordinating body meetings and outputs | Coordinating bodies' websites |
| | | 1.2.6 Regulatory guidance, reports or enforcement decisions | Regulatory announcements |
| | | 1.2.7 Joint rules or action across departments | Joint official announcements |
| | 1.3 Organizational governance architecture, resource investment, willingness to govern, and implementation capacity | 1.3.1 Permanent governance body or dedicated position | OECD.AI policy database |
| | | 1.3.2 Mandatory rules for government AI use | Government AI-use policies |
| | | 1.3.3 Dedicated governance budget or funding | Budget and funding documents |
| | | 1.3.4 Register of government AI applications | Government AI registers |
| | | 1.3.5 AI training for civil servants | Civil service training programs |
| | | 1.3.6 Audit or oversight reports | Audit authority reports |
| | | 1.3.7 Adoption of recommendations and corrective action | Audit follow-up notices |
| | 1.4 Standard-setting and dynamic updating capacity | 1.4.1 National AI standards | National standards catalogs |
| | | 1.4.2 Participating membership in international standardization | ISO SC 42 membership list |
| | | 1.4.3 Certification or testing scheme | Accreditation body directories |
| | | 1.4.4 New, revised or reviewed standards | National standards catalogs |
| | | 1.4.5 Standardization plan | Standardization plans |
| 2 Technical safety and evaluation | 2.1 Safety research, risk modeling, and quantification capacity | 2.1.1 National safety research body or program | Government establishment documents |
| | | 2.1.2 Dedicated safety research funding | Research funding announcements |
| | | 2.1.3 Official risk assessment or tiered framework | Official risk frameworks |
| | | 2.1.4 National safety or risk research reports | National institution reports |
| | | 2.1.5 Domestic safety research outputs | DBLP, arXiv |
| | | 2.1.6 International safety research and joint evaluations | AI safety institute network announcements |

| First-level indicator | Secondary indicator | Tertiary indicator | Data sources |
|---|---|---|---|
| | 2.2 International evaluation benchmarks, benchmark testing, and high-quality evaluation dataset development | 2.2.1 Safety benchmarks or testing tools | GitHub, Hugging Face |
| | | 2.2.2 Safety evaluation datasets | Hugging Face dataset cards |
| | | 2.2.3 Benchmark and dataset updates and maintenance | GitHub version history |
| | | 2.2.4 Participation in international joint evaluations | AI safety institute network announcements |
| | | 2.2.5 Official evaluation platform | Official evaluation platforms |
| | | 2.2.6 Overseas adoption of benchmarks or datasets | Foreign model and system cards |
| | 2.3 Dynamic adversarial testing, red teaming, system-level evaluation, and safety auditing capacity | 2.3.1 Red-team results for specific models | Model/system cards; safety institute reports |
| | | 2.3.2 Pre-deployment safety assessment arrangements | National legal databases; EUR-Lex |
| | | 2.3.3 System- or agent-level evaluation results | Safety institute reports; arXiv |
| | | 2.3.4 Official red-teaming or evaluation guidance | Official evaluation guidance |
| | 2.4 Risk mitigation, permission constraints, and deployment control capacity | 2.4.1 Deployment approval, licensing or filing | National legal databases; EUR-Lex |
| | | 2.4.2 Statutory mitigation measures | National legal databases; EUR-Lex |
| | | 2.4.3 Permission and access-control disclosures | Model cards; GitHub |
| | | 2.4.4 Suspension, withdrawal, rollback or takeover arrangements | National legal databases; EUR-Lex |
| | | 2.4.5 Actual suspension or restriction measures | Regulatory decisions |
| | | 2.4.6 Pre-deployment mitigation or testing records | Model and system cards |
| 3 Verification and compliance | 3.1 Information disclosure, transparency reporting, and safety-commitment record-keeping mechanisms | 3.1.1 Statutory disclosure or labeling obligations | National legal databases; EUR-Lex |
| | | 3.1.2 Transparency reports | Corporate transparency reports |
| | | 3.1.3 Safety commitments | AI safety commitment signatories |
| | | 3.1.4 Disclosure of commitment implementation | Commitment progress reports |
| | | 3.1.5 Official disclosure specifications or guidance | Regulatory guidance |
| | | 3.1.6 Supervision of disclosure obligations | Regulatory announcements |
| | 3.2 Risk incident reporting and near-miss sharing mechanisms | 3.2.1 Official incident reporting channel | Emergency response body websites |
| | | 3.2.2 Statutory incident reporting obligations | National legal databases; EUR-Lex |
| | | 3.2.3 Reporting covers near misses | National legal databases; EUR-Lex |
| | | 3.2.4 Incident information referral and notification | National legal databases; EUR-Lex |
| | | 3.2.5 Receipt and handling of specific incidents | Official incident reports |
| | | 3.2.6 Official responses to incidents in databases | AIID, AIAAIC, OECD AIM |
| | 3.3 Independent verification and continuous compliance review capacity | 3.3.1 Statutory pre-deployment assessment | National legal databases; EUR-Lex |
| | | 3.3.2 Official evaluation or certification program | Official evaluation/certification programs |
| | | 3.3.3 National accreditation for management system certification | Accreditation body directories |
| | | 3.3.4 Certification obtained by domestic organizations | Certificate directories of certification bodies |
| | | 3.3.5 Ongoing supervision requirements for deployed systems | National legal databases; EUR-Lex |
| | | 3.3.6 Public official evaluation or certification findings | Official directories and reports |

| First-level indicator | Secondary indicator | Tertiary indicator | Data sources |
|---|---|---|---|
| 4 Risk monitoring, early warning and response | 4.1 Risk signal identification, dynamic updates, and continuous monitoring capacity | 4.1.1 Risk monitoring responsibilities | Emergency response body websites |
| | | 4.1.2 Risk inventory or assessment report | Official risk reports |
| | | 4.1.3 Periodic situation reports | Official risk reports |
| | | 4.1.4 Risk warnings or advisories | Emergency response advisories |
| | | 4.1.5 Updated risk inventories and monitoring rules | Risk inventory versions |
| | | 4.1.6 Risk-information channels | Official information platforms |
| | 4.2 AI asset registration and visualization, activity auditing, and anomaly detection mechanisms | 4.2.1 System registration or filing | National legal databases |
| | | 4.2.2 Public registration information | Registration platforms |
| | | 4.2.3 Logging and auditability requirements | National legal databases; EUR-Lex |
| | | 4.2.4 Audits or inspections | Audit and regulatory reports |
| | | 4.2.5 Detection and handling of anomalous behavior | Official response notices |
| | | 4.2.6 Updated registration and auditing arrangements | Registration and audit documents |
| | 4.3 Risk incident response, emergency coordination, system recovery, and post-incident review and improvement capacity | 4.3.1 Emergency plans covering AI | Emergency management documents |
| | | 4.3.2 Inter-agency emergency coordination | Emergency coordination rules |
| | | 4.3.3 Exercises involving AI scenarios | APCERT and other regional exercise reports |
| | | 4.3.4 Response to actual incidents | AIID; official notices |
| | | 4.3.5 Post-incident review and improvement | Investigation and review reports |
| | | 4.3.6 System recovery and verification of corrections | Official corrective-action notices |
| 5 Accountability and ecosystem governance | 5.1 Boundaries of responsibility in the AI governance ecosystem | 5.1.1 Developer obligations | National legal databases; EUR-Lex |
| | | 5.1.2 Provider obligations | National legal databases; EUR-Lex |
| | | 5.1.3 Deployer obligations | National legal databases; EUR-Lex |
| | | 5.1.4 End-user obligations | National legal databases; EUR-Lex |
| | | 5.1.5 Regulatory responsibilities | National legal databases; EUR-Lex |
| | | 5.1.6 Action under responsibility provisions | Enforcement decisions |
| | 5.2 Data governance, supply chain security, and open-source ecosystem governance capacity | 5.2.1 Training-data governance rules | National legal databases; EUR-Lex |
| | | 5.2.2 Data-protection enforcement or guidance | Data-protection authorities |
| | | 5.2.3 Security requirements for supply chains and third-party components | Procurement and security rules |
| | | 5.2.4 Open-model governance rules | National legal databases; EUR-Lex |
| | | 5.2.5 Investigation and handling of data or supply-chain issues | Regulatory announcements |
| | 5.3 AI agent permission management, responsibility allocation, and traceability mechanisms | 5.3.1 Agent-specific official documents | Agent guidance |
| | | 5.3.2 Agent identity, permission and tool-use limits | Agent guidance |
| | | 5.3.3 Human oversight and authorization for high-risk operations | Agent guidance |
| | | 5.3.4 Distinct developer, deployer and user responsibilities | National legal databases |
| | | 5.3.5 Agent logs and traceability disclosures | Corporate technical documents; GitHub |
| | | 5.3.6 Regulatory action on specific agent products | Emergency response notices |

| First-level indicator | Secondary indicator | Tertiary indicator | Data sources |
|---|---|---|---|
| | 5.4<br>Mutual recognition of certification, dispute resolution, and cross-border enforcement capacity | 5.4.1 Signatory to international accreditation recognition arrangements | IAF and ILAC signatory lists |
| | | 5.4.2 AI-related regulatory cooperation | Cooperation agreements and statements |
| | | 5.4.3 Cross-border disputes and complaints | Complaint-handling mechanisms |
| | | 5.4.4 Cross-border joint investigations or enforcement assistance | Joint cross-border investigation notices |
| | | 5.4.5 Rulings or decisions concerning foreign providers | Judgments and administrative decisions |
| | | 5.4.6 Compliance review or market access for cross-border services | National legal databases; EUR-Lex |
| 6<br>Coordinated governance and capacity building | 6.1<br>Participation in international mechanisms, standards interoperability, and evaluation-protocol sharing mechanisms | 6.1.1 Membership of specialist safety or governance mechanisms | AI safety institute network membership |
| | | 6.1.2 Participation in multilateral governance | OECD and UNESCO meeting documents |
| | | 6.1.3 Participating membership in international standardization | ISO SC 42 membership list |
| | | 6.1.4 Signatory to joint international statements | Joint statement signatory pages |
| | | 6.1.5 International joint evaluations | AI safety institute network announcements |
| | | 6.1.6 Hosting international governance conferences | Official conference websites |
| | 6.2<br>Cross-border risk information notification, case sharing, and joint exercise mechanisms | 6.2.1 Membership of international or regional emergency networks | FIRST and APCERT membership lists |
| | | 6.2.2 Information-sharing arrangements covering AI | Cooperation memoranda |
| | | 6.2.3 Joint safety notices or guidance | Joint guidance signatory pages |
| | | 6.2.4 Multinational exercises involving AI scenarios | APCERT and other regional exercise reports |
| | 6.3<br>Capacity building, knowledge sharing, and public goods provision capacity | 6.3.1 Domestic governance training | Training programs |
| | | 6.3.2 Governance capacity building for other countries | International assistance programs |
| | | 6.3.3 Safety and governance knowledge resources | Official guidance and manuals |
| | | 6.3.4 Safety and governance tools or platforms | GitHub; official platforms |
| | | 6.3.5 Continued resource and tool updates | GitHub and official version histories |
| | 6.4<br>Public participation, media oversight, risk communication, AI literacy improvement | 6.4.1 Public consultation on policies and legislation | Government consultation platforms |
| | | 6.4.2 Explanation of responses to public comments | Consultation response summaries |
| | | 6.4.3 Public risk advisories | Official risk advisories |
| | | 6.4.4 National AI literacy program | OECD.AI policy database |
| | | 6.4.5 Complaints and reporting channels covering AI | Complaints and reporting platforms |

## B. Data and Assessment Methodology

**Data sources and period covered.** The assessment uses public information, principally original laws and regulations, government gazettes, competent authorities' announcements, membership lists of international organizations, standards catalogs and official institutional websites. Research reports and news coverage provide search leads, while judgments are based on primary sources. Specific data sources are listed under each secondary indicator in Section C of this appendix. Institutional facts, including laws, bodies and standards, reflect the position as of August 31, 2026. Records of actions such as enforcement, incident response, exercises and report publication cover January 1, 2024–August 31, 2026.

**Assessment of tertiary indicators.** For each country and tertiary indicator, the assessment records the publishing body, document title, date, URL and evidence supporting the judgment. The following rules apply. EU regulations directly applicable to member states are counted for France, Germany and Italy. State, provincial and other local rules may be counted, with their scope of application noted. AI rules, obligations and channels must specifically address AI or explicitly mention AI systems; general cybersecurity or data-protection rules are not counted. “Regulatory action by a competent authority” includes decisions on specific AI products by data-protection, consumer-protection and competition authorities. Indicators concerning implementation require records of actions such as enforcement, response or exercises; adoption of a rule is not equivalent to implementation. A finding of “No” means that no corresponding evidence was found within the sources listed.

**Assignment of tiers to secondary indicators.** Each secondary indicator is assigned to one of five tiers—S, A, B, C or D—with S the highest. Conditions are checked in descending order from S to D, and the highest qualifying tier is assigned. Classification is based on key arrangements and records of actual actions, rather than counts, weights or sums of conditions met. Tertiary indicators correspond to specific verifiable facts and are assessed individually as “Yes” or “No.” The same fact may support different secondary indicators but is not counted twice within a single indicator.

**Presentation of results.** Results are presented by secondary indicator. They are not aggregated into dimension scores, overall country scores or rankings. Charts in the main report treat Tier S or Tier A as indicating “relatively stable arrangements” and report the corresponding number or proportion of countries.

**Assessment and review.** Supporting tools were used to organize information. The materials and evidence underlying judgments were verified and reviewed by researchers.

**Coverage and timeliness.** The extent of public disclosure and the accessibility of official websites vary among countries. The assessment is based on public information available as of August 31, 2026. Failure to locate evidence does not mean that the relevant arrangements do not exist in a country.

## C. Assessment Scope, Data Sources and Tier Criteria for Secondary Indicators

This section sets out in detail the scope, tertiary indicators, data sources and tier criteria for each secondary indicator. Numbers in the qualifying conditions refer to the tertiary indicators under that indicator.

### 1.1 Laws, policies, and ethical rules

**Assessment scope:** Readiness to manage AI risks through legislation, national policies and ethical standards. The assessment focuses on dedicated AI legislation, legally binding obligations for general-purpose or generative models, national AI strategies and ethical standards, and regulatory action on specific AI prod-ucts. It comprises the following tertiary indicators:

1.1.1 Cross-sectoral AI-specific laws or regulations imposing legal obligations are in force, including administrative regulations and departmental rules;

1.1.2 Mandatory obligations for general-purpose, generative or frontier AI models are in force;

1.1.3 Dedicated AI legislation has been adopted but is not yet in force, has been submitted to the legislature, or is in force without imposing legal obligations, such as a promotion act;

1.1.4 AI-specific regulations are in force at state, provincial or other jurisdictional level;

1.1.5 A sectoral authority has issued binding AI-specific rules for its sector;

1.1.6 Legal provisions specifically addressing AI-generated or synthetic content are in force;

1.1.7 A current national AI strategy, plan or action plan exists;

1.1.8 National AI ethical principles or codes of conduct exist;

1.1.9 A competent authority has issued a penalty, prohibition or other decision concerning a specific AI product;

1.1.10 A competent authority has opened an investigation or held a regulatory meeting concerning a specific AI product.

**Data sources:** Public searches. Search channels and materials include national legislative websites, government gazettes and competent authorities' websites; EUR-Lex; and the OECD.AI national policy database, among others.

**Tier criteria:**

**Tier S:** Comprehensive legal obligations and regulatory practice. Conditions: All of 1.1.1, 1.1.2, 1.1.6, 1.1.7, 1.1.8 and 1.1.9 are met.

**Tier A:** Dedicated rules and regulatory action. Conditions: At least one of 1.1.1–1.1.6 and at least one of 1.1.9–1.1.10 are met.

**Tier B:** Dedicated legislation or rules. Conditions: At least one of 1.1.1–1.1.6 is met.

**Tier C:** A national strategy and ethical framework. Conditions: Both 1.1.7 and 1.1.8 are met.

**Tier D:** The conditions for Tier C are not met.

**1.2 Cross-agency coordination and dedicated regulatory mechanisms**

**Assessment scope:** Coordination and dedicated regulatory arrangements for AI governance, including the establishment and operation of coordinating bodies, statutory regulatory mandates and outputs, dedicated safety institutions, and joint action across departments. Tertiary indicators:

1.2.1 A national AI governance coordinating body exists, or a lead coordinating department is designated by law or national policy;

1.2.2 The coordinating body or lead department has documented responsibilities or working procedures;

1.2.3 A specific authority has been given AI regulatory powers by law or formal regulation;

1.2.4 A national AI safety institute or equivalent specialist body has been established;

1.2.5 Since 2024, the coordinating body or lead department has published records of meetings, decisions or other outputs;

1.2.6 Since 2024, the regulator has issued AI regulatory guidance, supervisory reports or enforcement decisions;

1.2.7 Records exist of multiple departments jointly issuing AI rules, conducting investigations or taking action.

**Data sources:** Public searches. Search channels and materials include government and competent authorities' websites, the OECD.AI national policy database, and membership lists of the International Network of AI Safety Institutes, among others.

**Tier criteria:**

**Tier S:** A complete governance structure with sustained outputs. Conditions: All of 1.2.1–1.2.7 are met.

**Tier A:** Governance bodies are in place and produce substantive outputs. Conditions: Either (1) both 1.2.1 and 1.2.3, plus at least one of 1.2.5–1.2.6; or (2) at least one of 1.2.1 and 1.2.3, plus both 1.2.5 and 1.2.7, are met.

**Tier B:** Governance bodies and some governance outputs. Conditions: At least one of 1.2.1 and 1.2.3, and at least one of 1.2.5–1.2.7, are met.

**Tier C:** Basic governance bodies. Conditions: At least one of 1.2.1, 1.2.3 and 1.2.4 is met.

**Tier D:** The conditions for Tier C are not met.

**1.3 Organizational governance architecture, resource investment, willingness to govern, and implementation capacity**

**Assessment scope:** The organizational, institutional and financial resources committed by government to AI governance, and oversight and correction of the government's own AI use. It covers dedicated bodies, internal rules, earmarked funding, use registers, staff training, audits and corrective action. Commitment and implementation capacity are judged through verifiable arrangements such as funding, internal rules, inspections and corrective action. Legislation applying to society more broadly is assessed under Indicator 1.1. Tertiary indicators:

1.3.1 A permanent body or dedicated position has AI governance responsibilities;

1.3.2 Mandatory rules govern government bodies' own use of AI;

1.3.3 Dedicated AI governance budgets, appropriations or funding programs are public, including earmarked funding for national AI safety institutions;

1.3.4 A public register or inventory of government AI applications exists;

1.3.5 An AI training program for civil servants exists;

1.3.6 Since 2024, an audit or oversight body has published an inspection report on government AI governance or use;

1.3.7 Public records show that recommendations from these inspections have been adopted or followed by corrective action.

**Data sources:** Public searches. Search channels and materials include government mandates and appointment announcements, budget and funding documents, government AI-use policies and regis-ters, and national audit reports, among others.

**Tier criteria:**

**Tier S:** Organizational, institutional and financial resources are in place, and inspection recommendations are implemented. Conditions: All of 1.3.1, 1.3.2, 1.3.3, 1.3.6 and 1.3.7 are met.

**Tier A:** A dedicated body and mandatory internal rules, with supporting investment or oversight. Conditions: Both 1.3.1 and 1.3.2, and at least one of 1.3.3–1.3.6, are met.

**Tier B:** A dedicated body and supporting governance arrangements. Conditions: 1.3.1 and at least one of 1.3.2–1.3.5 are met.

**Tier C:** Basic organizational or internal governance arrangements. Conditions: At least one of 1.3.1, 1.3.2 and 1.3.4 is met.

**Tier D:** The conditions for Tier C are not met.

**1.4 Standard-setting and dynamic updating capacity**

**Assessment scope:** Development and updating of AI standards and participation in international standardization, covering national standards, international participation, conformity assessment, ongoing updates and standardization plans. Tertiary indicators:

1.4.1 Current national AI standards or technical specifications have been published, including identical adoption of international standards;

1.4.2 The country is a participating member of ISO/IEC JTC 1/SC 42 on artificial intelligence;

1.4.3 An AI certification or testing scheme exists;

1.4.4 AI standards have been published, revised or reviewed since 2024;

1.4.5 A public AI standardization strategy or work plan exists.

**Data sources:** Public searches. Search channels and materials include ISO/IEC JTC 1/SC 42 membership lists, standards catalogs and announcements published by national standards bodies, accreditation body directories, official government gazettes and other public materials.

**Tier criteria:**

**Tier S:** Arrangements cover standards development, international participation, certification or testing, and ongoing updates. Conditions: All of 1.4.1–1.4.5 are met.

**Tier A:** National standards and international participation, with supporting implementation arrangements. Conditions: Both 1.4.1 and 1.4.2, and at least one of 1.4.3–1.4.5, are met.

**Tier B:** National standards and supporting arrangements. Conditions: 1.4.1 and at least one of 1.4.2–1.4.5 are met.

**Tier C:** Basic standardization, certification or testing arrangements. Conditions: At least one of 1.4.1, 1.4.2, 1.4.3 and 1.4.5 is met.

**Tier D:** The conditions for Tier C are not met.

**2.1 Safety research, risk modeling, and quantification capacity**

Assessment scope: Capacity for AI safety research and risk assessment, covering national research bod-ies and dedicated funding, official risk assessment frameworks, national-level research outputs, wider domestic research outputs and international cooperation. Tertiary indicators:

2.1.1 A national AI safety research institution, laboratory or dedicated research program exists;

2.1.2 Government has established dedicated funding or a funding program for AI safety research;

2.1.3 Government or a national institution has published an AI risk assessment method, taxonomy or tiered framework;

2.1.4 Since 2024, government or a national institution has publicly released an AI safety or risk research report;

2.1.5 Since 2024, domestic institutions have published AI safety research, in papers or technical reports;

2.1.6 Domestic institutions participate in international AI safety research or joint evaluations.

**Data sources:** Public searches. Search channels and materials include national AI safety institute and national research institution websites, research funding announcements, DBLP and arXiv, and International Network of AI Safety Institutes announcements, among others.

**Tier criteria:**

**Tier S:** Safety research connects institutions, funding, methods and research outputs. Conditions: All of 2.1.1–2.1.4 are met.

**Tier A:** National research bodies and outputs, supported by funding or methods. Conditions: 2.1.1 is met, at least one of 2.1.2–2.1.3 is met, and 2.1.4 is met.

**Tier B:** A national foundation for safety research or risk assessment. Conditions: Any one of (1) 2.1.1, (2) 2.1.3 or (3) 2.1.4 is met.

**Tier C:** Domestic safety research outputs or international research cooperation. Conditions: At least one of 2.1.5–2.1.6 is met.

**Tier D:** The conditions for Tier C are not met.

**2.2 International evaluation benchmarks, benchmark testing, and high-quality evaluation dataset development**

**Assessment scope:** Capacity to develop and maintain AI safety benchmarks, datasets and evaluation platforms, and their international influence. It covers resource development, continued maintenance, official evaluation platforms, international joint evaluations and external adoption. Tertiary indicators:

2.2.1 A domestic actor has developed and publicly released an AI safety benchmark or testing tool;

2.2.2 A domestic actor has publicly released an AI safety evaluation dataset;

2.2.3 Public records of version updates or maintenance exist for these benchmarks or datasets;

2.2.4 Domestic actors participate in international joint evaluations or shared evaluation protocols;

2.2.5 A government or public institution operates an AI evaluation platform or service;

2.2.6 Foreign institutions have used a domestic benchmark or dataset in model evaluations.

**Data sources:** Public searches. Search channels and materials include official GitHub and Hugging Face repositories and dataset cards, technical reports from evaluation bodies, and public model and sys-tem cards from major developers, among others.

**Tier criteria:**

**Tier S:** Benchmarks and datasets are maintained and used abroad, with cooperation or platform support. Conditions: All of 2.2.1, 2.2.2 and 2.2.3 are met, at least one of 2.2.4–2.2.5 is met, and 2.2.6 is met.

**Tier A:** Evaluation resources are maintained, with cooperation, platform support or overseas use. Conditions: At least one of 2.2.1–2.2.2 is met, 2.2.3 is met, and at least one of 2.2.4–2.2.6 is met.

**Tier B:** Evaluation resources exist, with maintenance, cooperation or use records. Conditions: At least one of 2.2.1–2.2.2 and at least one of 2.2.3–2.2.6 are met.

**Tier C:** Basic evaluation resources or an official platform. Conditions: At least one of 2.2.1, 2.2.2 and 2.2.5 is met.

**Tier D:** The conditions for Tier C are not met.

**2.3 Dynamic adversarial testing, red teaming, system-level evaluation, and safety auditing capacity**

**Assessment scope:** Capacity to conduct adversarial testing, system-level evaluation and pre-deployment safety assessment of AI models and systems, covering published red-team findings for specific models, system- or agent-level results, pre-deployment assessment arrangements and official methodological guidance. Tertiary indicators:

2.3.1 Domestic institutions have published red-team or adversarial-testing results for specific models or systems;

2.3.2 Arrangements exist for pre-deployment safety assessment or pre-release review;

2.3.3 Domestic institutions have published system- or agent-level evaluation results;

2.3.4 Official guidance on red teaming or safety evaluation methods exists.

**Data sources:** Public searches. Search channels and materials include reports from AI safety institutes and evaluation bodies, corporate model cards, system cards and safety reports, and regulatory evaluation guidance, among others.

**Tier criteria:**

**Tier S:** Red teaming, system-level evaluation and pre-deployment assessment are in place. Conditions: All of 2.3.1–2.3.3 are met.

**Tier A:** System-level evaluation is conducted, supported by red teaming or institutional and methodological arrangements. Conditions: Either (1) both 2.3.1 and 2.3.3; or (2) all of 2.3.2–2.3.4 are met.

**Tier B:** Red teaming is conducted, or pre-deployment assessment has supporting practice. Conditions: Either (1) 2.3.1; or (2) 2.3.2 and at least one of 2.3.3–2.3.4 are met.

**Tier C:** A foundation for pre-deployment assessment, system-level evaluation or official methods. Conditions: At least one of 2.3.2–2.3.4 is met.

**Tier D:** The conditions for Tier C are not met.

**2.4 Risk mitigation, permission constraints, and deployment control capacity**

**Assessment scope:** Capacity to control AI deployment and operation, covering approval or filing requirements, legally required mitigation, arrangements for suspension or withdrawal, records of actual intervention, and corporate disclosures on access controls and pre-deployment testing. Tertiary indicators:

2.4.1 Deployment approval, licensing or filing requirements apply to high-risk or frontier AI systems;

2.4.2 Risk mitigation measures are legally or officially required;

2.4.3 Domestic actors publish descriptions of model or system permissions and access controls;

2.4.4 Arrangements exist for suspension, withdrawal, rollback or human takeover;

2.4.5 Since 2024, a competent authority has actually initiated suspension, withdrawal or restriction measures;

2.4.6 Domestic actors publish pre-deployment mitigation or safety-testing records for specific systems.

**Data sources:** Public searches. Search channels and materials include legislation and regulatory guidance, corporate model cards, system cards and technical documentation, and regulatory announcements and decisions, among others.

**Tier criteria:**

**Tier S:** Deployment, mitigation, permission and intervention arrangements are in place, with actual intervention records. Conditions: All of 2.4.1–2.4.5 are met.

**Tier A:** Intervention arrangements have led to suspension, withdrawal or restrictions. Conditions: Either (1) all of 2.4.1, 2.4.4 and 2.4.5; or (2) all of 2.4.2, 2.4.4, 2.4.5 and 2.4.6 are met.

**Tier B:** Deployment or mitigation requirements have supporting disclosures or action records. Conditions: At least one of 2.4.1–2.4.2 and at least one of 2.4.3, 2.4.5 and 2.4.6 are met.

**Tier C:** Basic deployment or operational controls. Conditions: At least one of 2.4.1–2.4.4 is met.

**Tier D:** The conditions for Tier C are not met.

**3.1 Information disclosure, transparency reporting, and safety-commitment record-keeping mechanisms**

**Assessment scope:** Rules and practices for disclosure of AI use, risks and governance information, covering statutory disclosure requirements, official guidance, transparency reports, safety commitments and implementation reports, and supervision of disclosure obligations. Tertiary indicators:

3.1.1 Laws or regulations impose AI disclosure or labeling obligations;

3.1.2 Domestic actors publish AI transparency or responsible AI reports;

3.1.3 Domestic actors sign or publish AI safety commitments;

3.1.4 Domestic actors publicly report implementation of or progress on safety commitments;

3.1.5 Government or regulators issue AI disclosure specifications, templates or guidance;

3.1.6 Since 2024, a competent authority has supervised, verified or taken action concerning disclosure obligations.

**Data sources:** Public searches. Search channels and materials include laws and regulatory guidance, transparency reports and model cards on corporate websites, signatory lists for AI safety commitments, and regulatory announcements, among others.

**Tier criteria:**

**Tier S:** Disclosure, implementation reporting, guidance and supervision are in place. Conditions: All of 3.1.1–3.1.6 are met.

**Tier A:** Statutory disclosure and disclosure practice, supported by guidance or supervision. Conditions: 3.1.1, at least one of 3.1.2–3.1.3, and at least one of 3.1.5–3.1.6 are met.

**Tier B:** Statutory disclosure with supporting practice or specifications. Conditions: 3.1.1 and at least one of 3.1.2, 3.1.3 and 3.1.5 are met.

**Tier C:** Basic disclosure requirements, practice or guidance. Conditions: At least one of 3.1.1, 3.1.2, 3.1.3 and 3.1.5 is met.

**Tier D:** The conditions for Tier C are not met.

**3.2 Risk incident reporting and near-miss sharing mechanisms**

**Assessment scope:** Arrangements and practices for receiving, handling and sharing information on AI incidents, covering official reporting channels, statutory obligations, coverage of near misses, information sharing, and records of actual receipt and response. Tertiary indicators:

3.2.1 An official AI incident reporting channel exists;

3.2.2 Laws, regulations or formal rules impose AI incident reporting obligations;

3.2.3 Reporting explicitly covers near misses, hazardous situations or precursors of major risks;

3.2.4 Formal arrangements exist for referring or sharing incident information among institutions or notifying the public;

3.2.5 Since 2024, official records have been published of the receipt, handling or reporting of specific AI incidents;

3.2.6 AI incidents in the country recorded in public databases have received an official response or action.

**Data sources:** Public searches. Search channels and materials include regulators' and national computer emergency response bodies' websites, original legislation, AIID, the AIAAIC Repository and OECD AIM, among others.

**Tier criteria:**

**Tier S:** Reporting, information exchange and actual response are connected. Conditions: All of 3.2.1, 3.2.2, 3.2.4 and 3.2.5, and at least one of 3.2.3 and 3.2.6, are met.

**Tier A:** Reporting channels and statutory obligations, with actual receipt or handling of incidents. Conditions: All of 3.2.1, 3.2.2 and 3.2.5 are met.

**Tier B:** Reporting or sharing arrangements, or public records of handling and official responses. Conditions: Either (1) at least one of 3.2.1, 3.2.2 and 3.2.4; or (2) both 3.2.5 and 3.2.6 are met.

**Tier C:** Public handling of or official responses to specific incidents. Conditions: At least one of 3.2.5–3.2.6 is met.

**Tier D:** The conditions for Tier C are not met.

### 3.3 Independent verification and continuous compliance review capacity

**Assessment scope:** Arrangements and practices for independent verification of AI systems and organizations and continued compliance review after deployment. It covers statutory pre-deployment assessment, official evaluation and certification schemes, national accreditation, ongoing supervision and public verification findings. Tertiary indicators:

3.3.1 Laws or regulations require specified AI systems to undergo third-party assessment, certification or safety assessment by a competent authority before use;

3.3.2 A government-established or government-recognized AI evaluation or certification scheme exists, including voluntary schemes;

3.3.3 The national accreditation body includes AI management system certification within its accreditation scope;

3.3.4 Specific domestic organizations have obtained AI management system or AI product certification;

3.3.5 Laws, regulations or official schemes require ongoing supervision or periodic reassessment of deployed AI systems;

3.3.6 Since 2024, a competent authority or official program has published assessment, verification or certification conclusions on specific AI systems or organizations.

**Data sources:** Public searches. Search channels and materials include original legislation, regulatory announcements, announcements and directories of national accreditation bodies and certification authorities, and official evaluation program websites, among others.

**Tier criteria:**

**Tier S:** Pre-deployment assessment, certification and accreditation, and ongoing supervision are connected. Conditions: 3.3.1, at least one of 3.3.2–3.3.3, and both 3.3.5 and 3.3.6 are met.

**Tier A:** Formal assessment arrangements, with ongoing supervision or published findings. Conditions: At least one of 3.3.1–3.3.2 and at least one of 3.3.5–3.3.6 are met.

**Tier B:** Formal verification arrangements, or certification and ongoing supervision. Conditions: Either (1) at least one of 3.3.1–3.3.3; or (2) both 3.3.4 and 3.3.5 are met.

**Tier C:** Domestic organizations have obtained AI-related certification. Conditions: 3.3.4 is met.

**Tier D:** The conditions for Tier C are not met.

### 4.1 Risk signal identification, dynamic updates, and continuous monitoring capacity

**Assessment scope:** Capacity to identify and update AI risk signals continuously, covering monitoring mandates, risk assessments or situation reports and their regularity, early-warning records, updates to risk inventories or monitoring rules, and public risk-information channels. Tertiary indicators:

4.1.1 An official institution has responsibility for AI risk monitoring;

4.1.2 An official AI risk inventory, assessment or situation report exists;

4.1.3 These monitoring or situation reports are published periodically;

4.1.4 Official AI risk warnings, notices or advisories have been issued since 2024; 4.1.5 Updates or version records exist for risk inventories or monitoring rules; 4.1.6 An AI risk-information channel is available to the public or industry.

**Data sources:** Public searches. Search channels and materials include national computer emergency response bodies, regulators and AI safety institute websites, and official risk reports and notices, among others.

**Tier criteria:**

**Tier S:** Monitoring, periodic reporting, early warning and updating are in place. Conditions: All of 4.1.1–4.1.5 are met.

**Tier A:** Monitoring responsibilities and risk reports, with periodic publication or warnings. Conditions: Both 4.1.1 and 4.1.2, and at least one of 4.1.3–4.1.4, are met.

**Tier B:** Monitoring responsibilities with supporting outputs or information channels. Conditions: 4.1.1 and at least one of 4.1.2, 4.1.4 and 4.1.6 are met.

**Tier C:** Basic risk monitoring or information provision. Conditions: At least one of 4.1.1, 4.1.2, 4.1.4 and 4.1.6 is met.

**Tier D:** The conditions for Tier C are not met.

**4.2 AI asset registration and visualization, activity auditing, and anomaly detection mechanisms**

**Assessment scope:** Capacity to track AI deployments and detect anomalies, covering system registration or filing and public access, logging requirements, audit and inspection records, detection and handling of anomalous behavior, and updates to registration coverage. Tertiary indicators:

4.2.1 An AI system or algorithm registration or filing scheme exists;

4.2.2 Registration or filing information is publicly searchable;

4.2.3 Laws or regulations require AI systems to retain logs or auditable records;

4.2.4 AI systems have been audited, inspected or verified since 2024;

4.2.5 Records exist of detecting and addressing anomalous AI system behavior since 2024;

4.2.6 Registration or auditing arrangements have been updated or expanded.

Data sources: Public searches. Search channels and materials include government algorithm registers and AI use-case platforms, original legislation, and audit and regulatory reports, among others.

**Tier criteria:**

**Tier S:** Registration, disclosure, logging, auditing and anomaly response are in place. Conditions: All of 4.2.1–4.2.5 are met.

**Tier A:** Registration or filing is public and audits or inspections have been conducted. Conditions: All of 4.2.1, 4.2.2 and 4.2.4 are met.

**Tier B:** Registration with supporting oversight or response arrangements. Conditions: 4.2.1 and at least one of 4.2.2–4.2.5 are met.

**Tier C:** Basic registration, auditing or anomaly-detection arrangements. Conditions: At least one of 4.2.1–4.2.6 is met.

**Tier D:** The conditions for Tier C are not met.

### 4.3 Risk incident response, emergency coordination, system recovery, and post-incident review and improvement capacity

**Assessment scope:** Capacity to respond to AI incidents, covering emergency plans, inter-agency coordination, exercises involving AI scenarios, response to actual incidents, post-incident review and improvement, and system recovery. Tertiary indicators:

4.3.1 An emergency plan or response procedure covers AI;

4.3.2 Explicit inter-agency emergency coordination arrangements exist;

4.3.3 Emergency exercises involving AI scenarios have been conducted since 2024;

4.3.4 Actual AI incidents have been handled since 2024;

4.3.5 Post-incident reviews, investigation reports or improvement measures exist;

4.3.6 Records exist of system recovery, reactivation or verification of corrective action.

**Data sources:** Public searches. Search channels and materials include national computer emergency response bodies and competent authorities' websites, exercise reports from regional organizations such as APCERT, official incident notices and investigation reports, and AIID, among others.

**Tier criteria:**

**Tier S:** Plans, practical response, review, and coordination or recovery arrangements are in place. Conditions: Both 4.3.1 and 4.3.5, at least one of 4.3.2 and 4.3.6, and at least one of 4.3.3–4.3.4 are met.

**Tier A:** Emergency plans and practical records, with supporting coordination or improvement. Conditions: 4.3.1, at least one of 4.3.2, 4.3.5 and 4.3.6, and at least one of 4.3.3–4.3.4 are met.

**Tier B:** Emergency plans with coordination, exercise or response records. Conditions: 4.3.1 and at least one of 4.3.2–4.3.4 are met.

**Tier C:** Basic emergency plans or practical records. Conditions: At least one of 4.3.1, 4.3.3 and 4.3.4 is met.

**Tier D:** The conditions for Tier C are not met.

### 5.1 Boundaries of responsibility in the AI governance ecosystem

**Assessment scope:** The extent to which laws allocate responsibilities among AI developers, providers, deployers, end users and regulators, and the practice of holding actors accountable under these provisions. Tertiary indicators:

5.1.1 Laws or regulations specify AI developers' obligations;

5.1.2 Laws or regulations specify AI providers' or service operators' obligations;

5.1.3 Laws or regulations specify AI deployers' or user organizations' obligations;

5.1.4 Laws or regulations specify end users' obligations or codes of conduct;

5.1.5 Laws or regulations specify regulators' duties and procedures;

5.1.6 Since 2024, a competent authority has enforced accountability, taken action or issued formal guidance under these provisions.

**Data sources:** Public searches. Search channels and materials include national legal databases and government gazettes, regulatory announcements and enforcement decisions, among others.

**Tier criteria:**

**Tier S:** Responsibilities are specified for all five roles, with implementation records. Conditions: All of 5.1.1–5.1.5 and 5.1.6 are met.

**Tier A:** Responsibilities are specified for at least four roles, with implementation records. Conditions: At least four of 5.1.1–5.1.5 are met, and 5.1.6 is met.

**Tier B:** Responsibilities are clearly specified for at least three roles. Conditions: At least three of 5.1.1–5.1.5 are met.

**Tier C:** Responsibilities are clearly specified for at least one role. Conditions: At least one of 5.1.1–5.1.5 is met.

**Tier D:** The conditions for Tier C are not met.

**5.2 Data governance, supply chain security, and open-source ecosystem governance capacity**

**Assessment scope:** Rules and implementation for AI training data, supply chains and open models, covering training-data governance, data-protection enforcement, supply-chain security requirements, open-model rules, and actual inspections and action. Tertiary indicators:

5.2.1 Governance rules apply to AI training data;

5.2.2 Since 2024, the data-protection authority has undertaken AI-related enforcement or issued formal guidance;

5.2.3 Security requirements apply to AI supply chains, procurement or third-party components;

5.2.4 Rules, guidance or restrictions address open models or open-source components;

5.2.5 Since 2024, a competent authority has investigated or addressed AI data or supply-chain issues.

**Data sources:** Public searches. Search channels and materials include documents from data-protection and AI authorities, government procurement and cybersecurity rules, and regulatory enforcement decisions, among others.

**Tier criteria:**

**Tier S:** Rules cover data, supply chains and the open-source ecosystem, with implementation records. Conditions: All of 5.2.1–5.2.5 are met.

**Tier A:** Data rules, regulatory guidance and inspections or action, extending to related ecosystem governance. Conditions: Both 5.2.1 and 5.2.2 are met, at least one of 5.2.3–5.2.4 is met, and 5.2.5 is met.

**Tier B:** Data rules with supporting regulatory or ecosystem arrangements. Conditions: 5.2.1 and at least one of 5.2.2–5.2.4 are met.

**Tier C:** Basic data or ecosystem governance arrangements. Conditions: At least one of 5.2.1–5.2.4 is met.

**Tier D:** The conditions for Tier C are not met.

**5.3 AI agent permission management, responsibility allocation, and traceability mechanisms**

**Assessment scope:** Governance arrangements for AI agents, covering dedicated official rules or guidance, permission and tool-use limits, human oversight and authorization, allocation of responsibility, behavioral records and traceability, and regulatory action on specific agent products. Tertiary indicators:

5.3.1 An official document sets governance requirements specifically for AI agents;

5.3.2 Official documents impose restrictive requirements on agent identification, permission boundaries or tool use;

5.3.3 Official documents require human oversight, human takeover or prior authorization for high-risk agent operations;

5.3.4 Laws or regulations distinguish responsibilities among AI developers, deployers and users;

5.3.5 Domestic actors publish agent behavior logs, action records or technical descriptions of traceability;

5.3.6 Since 2024, a competent authority has issued risk advisories, conducted investigations or required corrective action concerning specific agent products or services.

**Data sources:** Public searches. Search channels and materials include regulatory and national computer emergency response announcements, original legislation and official guidance, and corporate technical documentation and system cards, among others.

**Tier criteria:**

**Tier S:** Agent rules, permissions, oversight and traceability are in place, with regulatory action. Conditions: All of 5.3.1, 5.3.2, 5.3.3, 5.3.5 and 5.3.6 are met.

**Tier A:** Agent-specific rules and permission or oversight requirements, with regulatory action. Conditions: 5.3.1 is met, at least one of 5.3.2–5.3.3 is met, and 5.3.6 is met.

**Tier B:** Agent-specific policy arrangements. Conditions: 5.3.1 is met.

**Tier C:** A foundation for human oversight, differentiated responsibilities or behavioral traceability. Conditions: At least one of 5.3.3–5.3.5 is met.

**Tier D:** The conditions for Tier C are not met.

**5.4 Mutual recognition of certification, dispute resolution, and cross-border enforcement capacity**

**Assessment scope:** Capacity for cross-border AI cooperation and enforcement, covering international recognition of certification, regulatory cooperation, handling of disputes, joint investigations, enforcement against foreign AI service providers, and compliance review of cross-border services. Tertiary indicators:

5.4.1 The national accreditation body is a signatory to an IAF or ILAC mutual recognition arrangement;

5.4.2 A regulatory cooperation arrangement, memorandum or joint statement involving AI has been signed with another country;

5.4.3 A mechanism accepts AI-related cross-border disputes or complaints;

5.4.4 Since 2024, AI-related joint investigations, enforcement assistance or information cooperation have taken place with other countries;

5.4.5 Since 2024, judicial rulings or administrative decisions concerning foreign AI service providers have been issued;

5.4.6 Compliance review or market-access arrangements cover cross-border AI services.

**Data sources:** Public searches. Search channels and materials include IAF and ILAC signatory lists, regulatory cooperation agreements and joint announcements, court judgments and administrative decisions, among others.

**Tier criteria:**

**Tier S:** Mutual recognition, cooperation, dispute resolution and cross-border enforcement arrangements are in place. Conditions: All of 5.4.1–5.4.6 are met.

**Tier A:** Cross-border investigations and adjudication, supported by dispute or market-access mechanisms. Conditions: At least one of 5.4.3 and 5.4.6, and both 5.4.4 and 5.4.5, are met.

**Tier B:** Cross-border joint investigations or action against foreign providers. Conditions: At least one of 5.4.4–5.4.5 is met.

**Tier C:** Cross-border dispute resolution or compliance and market-access mechanisms. Conditions: At least one of 5.4.3 and 5.4.6 is met.

**Tier D:** The conditions for Tier C are not met.

**6.1 Participation in international mechanisms, standards interoperability, and evaluation-protocol sharing mechanisms**

**Assessment scope:** The depth of participation in international AI governance, covering membership of specialist mechanisms, engagement in multilateral governance, international standardization, joint statements, joint evaluations and hosting governance conferences. Tertiary indicators:

6.1.1 The country belongs to a specialist international AI safety or governance mechanism;

6.1.2 Since 2024, it has participated in AI governance through multilateral mechanisms such as the UN, OECD or UNESCO;

6.1.3 It is a participating member of ISO/IEC JTC 1/SC 42;

6.1.4 It is a signatory to an international joint AI statement, common methodology or consensus document;

6.1.5 It participates in international joint evaluations or shared evaluation protocols;

6.1.6 It has hosted an international AI governance conference or summit since 2024.

**Data sources:** Public searches. Search channels and materials include membership lists and meeting documents on international mechanisms' websites, ISO/IEC JTC 1/SC 42 membership lists and original joint international statements, among others.

**Tier criteria:**

**Tier S:** Participation spans international mechanisms, standards, statements, evaluations and conferences. Conditions: All of 6.1.1–6.1.6 are met.

**Tier A:** Participation in multilateral governance and international standardization, with other substantive engagement. Conditions: At least one of 6.1.1, 6.1.5 and 6.1.6, and both 6.1.2 and 6.1.3, are met.

**Tier B:** Multilateral governance participation with a foundation for international cooperation. Conditions: At least one of 6.1.1, 6.1.3, 6.1.4 and 6.1.6 is met, and 6.1.2 is met.

**Tier C:** Basic international governance participation. Conditions: At least one of 6.1.2–6.1.4 is met.

**Tier D:** The conditions for Tier C are not met.

**6.2 Cross-border risk information notification, case sharing, and joint exercise mechanisms**

**Assessment scope:** Capacity to exchange AI risk information and conduct joint exercises with other countries, covering membership in international emergency cooperation networks, information-sharing arrangements explicitly covering AI, joint safety notices or guidance, and multinational exercises involving AI scenarios. Tertiary indicators:

6.2.1 The national computer emergency response body belongs to an international or regional emergency cooperation organization;

6.2.2 A risk-information sharing agreement, memorandum or joint-action arrangement explicitly covering AI has been signed with another country;

6.2.3 Since 2024, AI safety or security notices, warnings or guidance have been jointly issued with foreign institutions;

6.2.4 Since 2024, the country has participated in multinational joint exercises involving AI scenarios.

**Data sources:** Public searches. Search channels and materials include membership lists of FIRST, APCERT and similar organizations, national computer emergency response and cybersecurity agency websites, signatory pages of original joint guidance, and exercise plans and completion reports, among others.

**Tier criteria:**

**Tier S:** Cross-border sharing arrangements, joint outputs and practical exercises. Conditions: All of 6.2.2–6.2.4 are met.

**Tier A:** Participation in multinational exercises, with sharing arrangements or joint outputs. Conditions: At least one of 6.2.2–6.2.3 is met, and 6.2.4 is met.

**Tier B:** Cross-border sharing, joint notices or exercises. Conditions: At least one of 6.2.2–6.2.4 is met.

**Tier C:** Membership of an international or regional emergency cooperation organization. Conditions: 6.2.1 is met.

**Tier D:** The conditions for Tier C are not met.

**6.3 Capacity building, knowledge sharing, and public goods provision capacity**

**Assessment scope:** Capacity to provide training, knowledge and public tools for AI safety and governance, covering domestic governance training, international capacity-building programs, public knowledge resources, safety and governance tools available to others, and continued updates. Tertiary indicators:

6.3.1 Government or public institutions provide domestic AI safety or governance training;

6.3.2 The national government or government-funded institutions host, implement or fund AI capacity-building programs for other countries that include governance, safety or ethics;

6.3.3 Government or public institutions publish AI safety or governance resources, such as guidance, manuals or teaching materials;

6.3.4 Domestic institutions publicly provide AI safety or governance tools, evaluation datasets or platforms for use by others;

6.3.5 Records show continued updates or iteration of these training programs, resources or tools.

**Data sources:** Public searches. Search channels and materials include government and international organization project pages, UNESCO, ITU and OECD capacity-building program lists, and public resources on GitHub and Hugging Face, among others.

**Tier criteria:**

**Tier S:** Domestic training, international support and public resources are in place and continuously updated. Conditions: All of 6.3.1–6.3.5 are met.

**Tier A:** Domestic training, international support, knowledge resources and tools. Conditions: All of 6.3.1–6.3.4 are met.

**Tier B:** Domestic training with resources or updates. Conditions: 6.3.1 and at least one of 6.3.3–6.3.5 are met.

**Tier C:** Basic training or public resources. Conditions: At least one of 6.3.1, 6.3.3 and 6.3.4 is met.

**Tier D:** The conditions for Tier C are not met.

**6.4 Public participation, media oversight, risk communication, AI literacy improvement**

**Assessment scope:** Channels for public participation in AI governance and access to risk information, covering public consultation on policies and legislation and official responses, public risk communication and literacy education, and complaints and reporting channels covering AI. Media and public oversight are principally examined through channels by which the public can report AI-related problems. Tertiary indicators:

6.4.1 Public consultation on AI policy or legislation has taken place since 2024;

6.4.2 An official summary of public comments or explanation of how they were handled has been published;

6.4.3 Official AI risk advisories or prevention materials are available to the public;

6.4.4 A national AI literacy, education or skills-training program exists;

6.4.5 Public complaints or reporting channels cover AI-related issues.

**Data sources:** Public searches. Search channels and materials include government consultation platforms and response documents, and materials from regulators, consumer-protection bodies and public education institutions, among others.

**Tier criteria:**

**Tier S:** Public participation, responses to comments, risk communication and literacy building are in place. Conditions: All of 6.4.1–6.4.5 are met.

**Tier A:** Public consultation with an explanation of responses or a complaints channel. Conditions: Either (1) both 6.4.1 and 6.4.2; or (2) both 6.4.1 and 6.4.5 are met.

**Tier B:** Public consultation with risk advisories, education or complaints arrangements. Conditions: 6.4.1 and at least one of 6.4.3–6.4.5 are met.

**Tier C:** Basic public participation or risk education. Conditions: At least one of 6.4.1, 6.4.3 and 6.4.4 is met.

**Tier D:** The conditions for Tier C are not met.

# World Internet Conference Specialized Committee on Artificial Intelligence

The Specialized Committee on Artificial Intelligence, as the first professional branch of World Internet Conference, was established in 2024, including three programs: Standards Program, AI Safety and Governance Program, and Industry Program. The Committee brings together leading experts and professionals in the field of AI from international organizations, think tanks, research institutes, professional associations and industry, which is dedicated to fostering international cooperation and coordinated development, and aims to facilitate global sharing of AI achievements. Through thematic seminars, outcome sharing, and initiative releases, it continuously consolidates international consensus and supports inclusive and sustainable AI development.

# Frontier AI Safety and Governance 2026
# Risk Understanding and Coordinated Governance

Follow us on Facebook: @wicinternet

Follow us on X: @wicinternet

世界互联网大会
World Internet
Conference